\documentclass[aps,prb,twocolumn,showpacs,superscriptaddress]{revtex4}
\usepackage{amssymb,amsmath}
\usepackage{graphicx}
\usepackage{multirow}
\usepackage{hyperref}
\usepackage{appendix}
\usepackage{mathtools}
\usepackage{amsmath}
\usepackage{xcolor}
\usepackage{siunitx}
\usepackage{yfonts}
\usepackage{esint}
\usepackage{subcaption}
\graphicspath{{./IMG/}}
\newcommand{\occf}{f}

\begin{document}

\title{Variational Density Functional Perturbation Theory for Metals}

\author{Xavier Gonze*}
\affiliation{European Theoretical Spectroscopy Facility, Institute of Condensed Matter and Nanosciences, Universit\'{e} catholique de Louvain, Chemin des \'{e}toiles 8, bte L07.03.01, B-1348 Louvain-la-Neuve, Belgium}

\author{Samare Rostami}
\affiliation{European Theoretical Spectroscopy Facility, Institute of Condensed Matter and Nanosciences, Universit\'{e} catholique de Louvain, Chemin des \'{e}toiles 8, bte L07.03.01, B-1348 Louvain-la-Neuve, Belgium}

\author{Christian Tantardini*}
\affiliation{Hylleraas center, Department of Chemistry, UiT The Arctic University of Norway, PO Box 6050 Langnes, N-9037 Troms\o, Norway.}
\affiliation{Department of Materials Science and Nanoengineering, Rice University, Houston, Texas 77005, United States of America.}

\email{xavier.gonze@uclouvain.be, \\
christiantantardini@ymail.com}

\date{\today}

\begin{abstract}
Density functional perturbation theory is a well-established method to study responses of molecules and solids, especially responses to atomic displacements or to different perturbing fields
(electric, magnetic). 
Like for density functional theory, the treatment of metals is delicate, due to the Fermi-Dirac statistics and electronic bands crossing the Fermi energy. 
At zero temperature, there is an abrupt transition from occupied states to unoccupied ones, usually addressed with smearing schemes. 
Also, at finite temperature, fractional occupations are present, and the occupation
numbers may vary in response to the perturbation.

The present work establishes the characteristics of density functional perturbation theory stemming from the underlying variational principle, in the case of metals.
After briefly reviewing variational density functional theory for metals, the convexity of  the entropy function of the occupation number is analyzed, and, at finite temperature, the benefit of resmearing the
Fermi-Dirac broadening with the Methfessel-Paxton one is highlighted. 
Then the variational expressions for the second-order
derivative of the free energy are detailed, exposing the different possible gauge choices. 
The influence of the inaccuracies in the unperturbed wavefunctions from the prior  
density functional theory calculation is studied.
The whole formalism is implemented in the ABINIT software package.
\end{abstract}

\maketitle

\section{Introduction}
Density functional perturbation theory (DFPT) has been 
implemented and used 
for decades for the study of responses of molecules, solids, and nanostructures to different types of perturbations, 
including atomic displacements, applied electric field or magnetic field, or cell parameter changes.
\cite{Baroni1987,Gonze1992,Gironcoli1995,Gonze1995a, Gonze1997,Gonze1997a, Baroni2001,Gonze2005a,Wu2005,Ricci2019}
It proves a method of choice for the computation of
phonon band structures,
\cite{Petretto2018a},
linear dielectric response\cite{He2014}, Born effective charges\cite{Ghosez1998}, 
thermal expansion\cite{Fleszar1990,Rignanese1996},
piezoelectricity\cite{Wu2005}, Raman tensors\cite{Veithen2005},
electro-optic effect\cite{Veithen2004},
electron-phonon\cite{Giustino2017,Miglio2020} and phonon-phonon couplings \cite{Gonze1989,Debernardi2000}, flexoelectricity\cite{Royo2019}, 
thermodynamical\cite{Lee1995a,Rignanese1996}
and many other properties.
The list of applications of DFPT 
continues to increase regularly. 

Many basic concepts and theorems of DFPT
have been established a long time ago\cite{Baroni1987,Gonze1992,Gonze1995,Gonze1995a, Gonze1997,Gonze1997a, Baroni2001}. 
DFPT stems from the Taylor expansion of 
quantities present in density functional theory (DFT) 
when an external parameter is changed by a small amount. 
The above-mentioned properties are directly connected to the (possibly high-order) derivatives of the energy with respect to  
such small parameters characterizing the strength of the perturbations.
Since the first-order derivatives of the energy with respect to atomic displacement, electric field, magnetic fields and cell parameter changes, respectively, 
are forces, electric dipole  or electric polarization, magnetic dipole or magnetic polarization, and stress, respectively, their linear response to additional applied fields are linked to second-order derivatives of the energy.

It is well-known that DFT is based on a variational principle: the energy is minimized with respect to trial
Kohn-Sham wavefunctions. 
DFPT inherits also from this property of DFT a variational
principle for the second-order derivative of the energy 
with respect to trial first-order wavefunctions\cite{Gonze1992}.
While the linear-response formalism can be derived without
making explicit usage of this variational property, the quantities computed numerically, determined using iterative solvers with some stopping criterion, are
more accurate with the variational formulation than 
with alternative, possibly simpler, non-variational
formulations.
Also, algorithms to determine the optimal first-order
wavefunctions can benefit 
from the variational character of the second-order energy.
In addition, the variational principle is crucial for establishing higher-order DFPT, thanks to the so-called ``2n+1'' theorem\cite{Gonze1989,Gonze1995,Gonze1995a}.

The specificities of the treatment of metals within DFPT have been established by de Gironcoli\cite{Gironcoli1995} in 1997, 
based on the treatment of metals in DFT.
At variance with the DFT theory for
finite systems and insulators, in the DFT theory of metals, the occupation numbers, usually fractional, have to be determined. 
The electronic entropy appears,
and the internal energy is replaced by the free energy.
Such varying occupation numbers are present when dealing with finite temperature, 
but also appear in practice even at zero temperature, in order to deal with the abrupt transition between occupied states to unoccupied state at the Fermi energy.
Such case is tackled using smearing schemes, that allow to reduce the numerical burden of the integration of a discontinuous occupation function in the Brillouin Zone.
The most efficient high-order smearing schemes\cite{Methfessel1989} have their own problems, as described by dos Santos \textit{et al.}\cite{DosSantos2023}
since the occupation function of the energy becomes 
non-monotonic.

For DFPT, de Gironcoli described the
specificities of linear responses due to varying occupation numbers and due to entropy, and provided phonon band structures for Al, Pb and Nb.  However, he did not present a variational formulation of
the second-order derivative of the free energy.
This result is still lacking in the literature. However,
it had been derived, implemented (at least in the ABINIT package)\cite{Gonze2002,Gonze2020,Romero2020}, and used
for many studies of metals, e.g.
for computing the phonon band structure of lead\cite{Verstraete2008}, bismuth\cite{Diaz-Sanchez2007,Diaz-Sanchez2007a} and polonium\cite{Verstraete2010}, all three with spin-orbit coupling, or the 
electronic transport properties of lithium,\cite{Xu2014} and osmium and osmium silicide\cite{Xu2013} among others.

Motivated by the interest to fill this gap, but also by some recent publications related to 
the response properties of metals by Canc\`es \textit{et al.}\cite{Cances2022} as well as improved treatment 
in high-order smearing schemes in DFT, by dos Santos \textit{et al.}\cite{DosSantos2023},
the present publication has the aim to lay down the
variational treatment of DFPT for metals.
The second-order derivative of the free energy
is formulated as a variational functional of trial first-order wavefunctions
and trial first-order density matrix. The second-order entropy is present in the second-order free energy, and depends on the
first-order and selected second-order changes of the occupation numbers,
both derived from the
first-order density matrix. 
Non-variational expressions are also presented.

The invariance of DFT for metals with respect to unitary transformations inside the wavefunction space is more intricate than 
in the case of DFT for gapped systems at 0 K.\cite{Gonze1995a}
In the latter, a unitary transformation of wavefunctions inside the occupied space leaves the density, total energy, and Kohn-Sham potential invariant.
In the case of metals, a unitary transformation 
of the wavefunctions must be accompanied by a simultaneous transformation of the (one-body) density matrix.
The wavefunctions might not be eigenstates of the Hamiltonian, and the density matrix might not be diagonal. 
Still, such transformed trial wavefunctions and trial density matrix correspond to the same free energy,
and hence minimize the free energy functional.
This has been developed in 1997 by Marzari, Payne and Vanderbilt (MPV),\cite{Marzari1997} in their variational formulation of DFT for metals, on which we will rely to derive the variational formulation of DFPT for metals. 

As outlined for DFT above, in DFPT also, several sets of first-order wavefunctions (and first-order density matrix elements for the metal case) minimize the energy,
related by some well-defined transformation rule.
A choice among such possibilities is referred to fixing the gauge, and the invariance of the second-order free energy with respect to the gauge choice is called the gauge freedom.
The gauge freedom in the DFPT of metals is more complicated than the one in the DFPT of gapped systems at 0 K, and this is described as well in the present work. The parallel and diagonal gauges are defined,
extending to the metal case the well-known results obtained for gapped systems. 
However, more freedom is allowed, due to the added variability of the density matrix. This will be described as well, as different formulations might be of interest in different contexts. The connection 
with the article of Canc\`es \textit{et al.}\cite{Cances2022} will be made.

Concerning smearing schemes, we first remark that 
the non-monotonic behavior of the occupation number as a function of the energy not only induces problems at the level of the determination of the Fermi energy, as outlined recently by dos Santos et al,\cite{DosSantos2023} but also makes the one-level entropy function of the occupation number multivalued and non-convex. 
DFPT is also impacted,
as the second-order free energy might not be an extremum, because of the non-positive-definiteness of the contribution of the second-order entropy.
When a finite temperature is considered, it is shown that the resmearing procedure\cite{Verstraete2001} is a procedure in which, possibly, higher-order smearing might be used without sacrificing the 
monotonic behavior of the occupation function, provided the resmearing parameter is not too large.
For the resmearing using the Methfessel-Paxton (MP) scheme,\cite{Methfessel1989}
it is shown that a smearing parameter smaller or equal to twice the
physical electronic temperature can be used. 
For such range of parameters, the occupation function is monotonic, the one-level entropy is univalued and convex.

Coming to applications, in addition to the results already available in the literature, the convergence
of phonon frequencies of copper, with respect to the wavevector grid and to smearing schemes is provided.
For this case, one can distinguish two regimes, 
a first one, ``medium precision", in which the
target numerical precision is requested at the
level of the absolute value of
phonon frequencies, and a second one, ``high precision",
in which the target is the study of the temperature
dependence of the phonon frequencies. 

Finally, the impact of the precision requirement (or lack of precision) for the unperturbed wavefunctions, on the precision of the second-order free energy, is examined. 
Indeed, when dealing with metals, in practice, the preliminary DFT calculation of the density also includes wavefunctions with vanishing occupations. 
Depending on the stopping criterion, the highest energy ones might possibly not be well converged, as they do not influence the density anyhow. 
Thus, the impact of lack of precision of such 
wavefunctions in DFT is negligible. 
By contrast, it is found that such lack of numerical convergence might have an impact in the subsequent
DFPT calculations, in agreement with the recent observation by Canc\`es \textit{et al}.\cite{Cances2022}
This impact is analyzed thanks to a simple three-level model. The error in the second-order free energy is found to be proportional to the norm of the
residual of such wavefunctions. 
Canc\`es \textit{et al.}\cite{Cances2022} propose a Schur complement technique to deal with such problem. 
Actually, increasing the number of states in the underlying DFT calculation, then filtering less-converged states to start subsequent DFPT calculations solves the problem, if their occupation is really negligible.

The structure of this article is as follows. After the present introduction, Sec.~\ref{Sec:VarDFTMetals} deals with variational DFT for metals: the MPV\cite{Marzari1997} variational DFT for metals is reviewed, some considerations on 
the  space of potentially occupied wavefunctions are introduced, and then smearing schemes are detailed.
In the latter it is shown that resmearing\cite{Verstraete2001} the
Fermi-Dirac distribution with the MP smearing\cite{Methfessel1989} at finite temperature does not break the monotonic behavior of the occupation function, for a range of resmearing parameter. Related to section Sec.~\ref{Sec:VarDFTMetals}, Sec.~S1 in the Supporting Information fixes notation problems and typos present in Ref.~\onlinecite{Verstraete2001}.

In Sec.~\ref{Sec:F2} the variational second-order free energy within DFPT, that includes the treatment of the second-order entropy, is presented.
The gradient of the second-order free energy 
is written, and linked with the de Gironcoli 
linear-response DFPT approach for metals\cite{Gironcoli1995}.
In the Supporting Information, the Sec.~S2 gives a detailed  derivation of the variational second-order free energy, while the Secs.~S3 and S4 give some technical details
to obtain the gradients, also related to the 
non-hermiticity freedom for the first-order off-diagonal density matrix elements. 

Sec.~\ref{Sec:different_gauges} focuses on the choice of gauge. The gauge freedom is first presented, followed by the definition and properties of the parallel gauge, as well as the definition and properties of the diagonal gauge. 
The section finishes with the complete suppression of first-order occupation matrix elements. In the Supporting Information,  the covariance of first-order wavefunctions and first-order density matrix elements is presented in Sec.~S5. Then the derivation of the first-order density expression with modified first-order wavefunctions is explained in Sec.~S6, and finally non-variational expressions are written down, for the case of the parallel gauge, in Sec.~S7.

While the previous sections neglected the
Bloch characteristics of the first-order wavefunctions and energies, as well as the 
presence of a Brillouin Zone, Sec.~\ref{Sec:Periodic} upgrades such results for explicitly periodic systems. 

Sec.~\ref{Sec:Applications} presents the study 
of some phonon frequencies of copper, especially
focusing on the wavevector grid sampling and 
its interplay with the smearing parameter.
The ``medium-precision'' and ``high-precision" regimes are distinguished.
In the Supporting Information, Sec.~S8 provides additional figures.

The influence of underconverged unoccupied states on the second-order free energy is quantified, and analyzed using a simple model
in Sec.~\ref{Sec:underconvergedGSwfs},
with details of the mathematical treatment given in the Supporting Information, Sec.~S9.

Sec.~\ref{Sec:Conclusion} summarizes the results.


\section{Variational DFT for metals}
\label{Sec:VarDFTMetals}


In this section, first, 
the variational approach to DFT 
of metals~\cite{Marzari1997} is reviewed, with notations that will then be used to treat the DFPT case. 
The need to define a space of potentially occupied wavefunctions is highlighted. 
Smearing schemes are the focus of the last part of this section on DFT. In particular, the
monotonic behavior of the occupation function is linked to the
convexity and singlevaluedness of the entropy function of the occupation number.

\subsection{Variational formulation of DFT with varying occupation numbers at finite temperature}
\label{sec:VarDFPTvarocc}

MVP~\cite{Marzari1997} introduced in 1997 a variational free energy for the density functional theory with varying occupation numbers at finite temperature, especially relevant to treat metals.
This approach will also be a basis for variational DFPT. 
For simplicity, the formalism is presented for non-spin-polarized systems ($n_{\textrm{s}}=2$ accounts for the spin degeneracy). Generalization to spin-polarized systems, including the non-collinear case is trivial.
In this section, as well as Secs.~\ref{Sec:F2} and~\ref{Sec:different_gauges}, one considers finite systems (with $N$ being the total number of electrons).
Periodic systems are treated in Sec.~\ref{Sec:Periodic}. Atomic (Hartree) units are used throughout.

The MVP electronic free energy $F[T;\{\psi_i\},\{\rho_{ij}\}]$, for a given temperature $T$, is a functional of the (trial) wavefunctions $\{\psi_i\}$
that form an orthonormal basis set, and of the (trial) matrix representation $\{\rho_{ij}\}$ of the one-particle density matrix operator $\hat{\rho}$ in this orthonormal set. Explicitly:
\begin{eqnarray}
F[T;\{\psi_i\},\{\rho_{ij}\}]
&=& 
n_{\textrm{s}} \sum_{ij}\rho_{ji} \langle \psi_i | \hat{K} + \hat{v}_{\textrm{ext}} | \psi_j \rangle
\nonumber \\
&+&
E_{\textrm{Hxc}}[\rho] - TS[\{\rho_{ij}\}].
\label{eq:free_energy}
\end{eqnarray}
In this expression, the sums over $i$ and $j$ extend to infinity, $\hat{K}$ is the kinetic energy operator, $\hat{v}_{\textrm{ext}}$ is the external
potential (e.g. created by the nuclei, as well as any other  additional external potential), $E_{\textrm{Hxc}}$ is the DFT Hartree and exchange-correlation energy functional of the density $\rho(\textbf{r})$, 
which is defined as
\begin{equation}
\rho(\textbf{r}) = 
n_{\textrm{s}} \sum_{ij}\rho_{ji} \psi_i^*(\textbf{r})
     \psi_j(\textbf{r}).
\label{eq:density}
\end{equation}
The one-particle density matrix is hermitian, with all its eigenvalues $\occf_{\gamma}$ - actually occupation numbers of the corresponding state - being between 0 and 1 for the Fermi-Dirac entropy (see later for the behavior of occupation numbers with high-order smearing schemes). 
$S[\{\rho_{ij}\}]$ is the entropy, considered as a functional of the density matrix elements. Explicitly,
\begin{equation}
S[\{\rho_{ij}\}] = n_{\textrm{s}} \sum_{\gamma} ks(\occf_{\gamma})
=n_{\textrm{s}} \textrm{Tr}[ks(\hat{\rho})],
\label{eq:S}
\end{equation}
where $s(f)$, the one-level entropy function (adimensional), is to be specified, and $k$ is Boltzmann's constant.
The usual physical situation corresponds to the Fermi-Dirac 
entropy function $s_\textrm{FD}(f)$, given by
\begin{equation}
s_\textrm{FD}(f) = -\Big(f \ln(f) + (1-f) \ln(1-f) \Big),
\label{eq:sFD}
\end{equation}
Smearing techniques, introduced for numerical reasons, will modify such entropy function. 
In what follows, equations are presented
in terms of a generic $s(f)$ function, with examples using the Fermi-Dirac entropy function. 
The formulas for other entropy functions
are presented in Sec.~\ref{sec:smearing_schemes}.

The trace of the occupation matrix is constrained to $N$, the number of electrons,
possibly taking into account the spin degeneracy,
\begin{equation}
n_{\textrm{s}} \sum_{i} \rho_{ii}=
n_{\textrm{s}} \sum_{\gamma} \occf_{\gamma}=
N.
\label{eq:constraintN}
\end{equation}
\\
Following MVP~\cite{Marzari1997},
one defines the Hamiltonian matrix,
with elements

\begin{equation}
\label{eq:H}
H_{ij}[\rho] = 
\langle \psi_i | \hat{K} + \hat{v}_{\textrm{ext}} 
+ \hat{v}_{\textrm{Hxc}}[\rho] | \psi_j \rangle,
\end{equation}

\indent where $\hat{v}_{\textrm{Hxc}}[\rho]$  is a local operator with 

\begin{equation}
v_{\textrm{Hxc}}[\rho](\textbf{r}) = \frac{\delta E_{\textrm{Hxc}}[\rho]}{\delta \rho(\textbf{r})}.
\end{equation}

MVP introduce 
the Lagrange multiplier $\mu$ (identified to the chemical potential) that enforces the constraint Eq.~(\ref{eq:constraintN}). It is such that
\begin{equation}
H_{ij}[\rho] - kT[s'(\hat{\rho})]_{ij}=\mu \delta_{ij}
,
\end{equation}
where the notation $[s'(\hat{\rho})]_{ij}$ is used in place of $d (\textrm{tr} [s(\hat{\rho})])/d\rho_{ji}$.
MVP also obtain that at the
minimum, the Hamiltonian and occupation matrices can be simultaneously diagonalized.

Working with diagonal Hamiltonian and occupation matrices is convenient, but
one is free to avoid diagonalizing them, the so-called ``gauge freedom" that MVP exploit indeed.
Arbitrary unitary transformations between
the wavefunctions can be accompanied by adequate unitary transformation of the occupation matrix, such that the density, 
Eq.~(\ref{eq:density}), the entropy, Eq.~(\ref{eq:S}), and the free energy,
Eq.~(\ref{eq:free_energy}), are invariant.

If the wavefunctions are chosen such as to diagonalize both Hamiltonian and density matrices, one has
\begin{equation}
H_{ij}[\rho] =\epsilon_i \delta_{ij}
,
\end{equation}
and
\begin{equation}
\epsilon_i 
- kTs'(\occf_i)=\mu.
\label{eq:chemicalpotential}
\end{equation}
In particular, for the Fermi-Dirac entropy,
\begin{equation}
s'_\textrm{FD}(f) = \frac{ds_\textrm{FD}}{df} 
=\ln \Big(\frac{1}{f} - 1 \Big)
.
\label{eq:11}
\end{equation}

Eq.~\ref{eq:chemicalpotential} can be inverted, to deliver the occupation number as a function of the eigenenergy,
\begin{equation}
\occf_i=[s']^{-1}(\frac{\epsilon_i-\mu}{kT}),
\label{eq:f_i_from_spm1}
\end{equation}
where the notation $[s']^{-1}$ is for the reciprocal of the
$s'$ function.

For the Fermi-Dirac entropy function 
Eq.~(\ref{eq:sFD}), the Eq.~(\ref{eq:f_i_from_spm1}) delivers the usual Fermi-Dirac occupations, where
\begin{eqnarray}
\occf_{i} &=& f_{\textrm{FD}}\big( (\mu - \epsilon_i )/kT \big)
\label{eq:fi_FD}
\\
f_{\textrm{FD}}(x)&=& \big( \exp(-x)+1 \big)^{-1}.
\label{eq:FD}
\end{eqnarray}
Note that, for consistency with the usual definitions
for smearing schemes, we choose the $f_\textrm{FD}(x)$
function
to monotonically increase from 0 to 1.
Then, Eq.~(\ref{eq:fi_FD}) is such that for high-energy states (large $\epsilon_i$), the occupation number tends rapidly to zero, exponentially.

In what follows, the choice to diagonalize the Hamiltonian (together with the occupation matrix) will be referred to as the ``diagonal gauge''. 
In the diagonal gauge,

\begin{equation}
\hat{H} | \psi_i \rangle =
\big( \hat{K} + \hat{v}_{\textrm{ext}} 
+ \hat{v}_{\textrm{Hxc}}[\rho] 
\big)
| \psi_i \rangle
= \epsilon_i | \psi_i \rangle.
\label{eq:KohnSham}
\end{equation}

For the derivation of DFPT equations, done later, this variational formulation of DFT for metals at finite temperature is reformulated
as an
unconstrained minimization,
based on Lagrange multipliers, as follows.
The free energy 
is augmented with
the Lagrange contributions from both types of constraints, namely,

\begin{eqnarray}
F^+[T;\{\psi_i\},\{\rho_{ij}\}]
&=& 
F[T;\{\psi_i\},\{\rho_{ij}\}]
\nonumber \\
&-&
\sum_{ij} \Lambda_{ji} 
n_{\textrm{s}} 
\Big( \langle \psi_i | \psi_j \rangle
- \delta_{ij} \Big)
\nonumber \\
&-&
\mu \Big( n_{\textrm{s}} \sum_{i}  \rho_{ii} - N \Big).
\label{eq:augmented_free_energy}
\end{eqnarray}
The optimal electronic free energy $F_{\textrm{el}}(T)$ is obtained by minimizing such
augmented free energy, 
\begin{eqnarray}
F_{\textrm{el}}[T]=
\min_{\{\psi_i\},\{\rho_{ij}\}}
F^+[T;\{\psi_i\},\{\rho_{ij}\}],
\label{eq:F_el_min}
\end{eqnarray}
without specific constraints on  $\{\psi_i\},\{\rho_{ij}\}$
during the minimization,
but where the Lagrange multipliers $\Lambda_{ji}$ 
and $\mu$ are chosen to enforce them afterwards.
In this formulation as well, unitary transforms between the wavefunctions
and occupation matrix leave the density, entropy and free energy invariant.
One can check that, at the minimum,
\begin{equation}
\label{eq:42.7}
\Lambda_{ki} = \sum_j H_{kj} \rho_{ji}.
\end{equation}
In the diagonal gauge, this becomes
\begin{equation}
\label{eq:42}
\Lambda_{ji} = \epsilon_i \delta_{ji}\rho_{ii} .
\end{equation}

\subsection{The space of potentially occupied wavefunctions}
\label{sec:Space_pocc}

In practice, first-principles calculations for metals at finite temperature (or finite smearing) only treat explicitly 
a finite number of eigenstates, among which, some are (nearly) fully occupied, some have intermediate occupation numbers,
and some have vanishing occupation numbers. Occupation numbers of the states outside of this space of potentially occupied states are so small that they can be set to zero and ignored. 
Thus, there is an ``active space" of potentially occupied wavefunctions. This space plays the same role than the occupied state space for the first-principles treatment of semiconductors.

This approach might not be practical when the temperature is quite large, yielding a large number of wavefunctions in the active space. 
However, even a temperature as high as 6000 Kelvin (corresponding to about 0.5 eV, that is beyond melting of all known materials at ordinary pressure), does not induce an unreasonable increase of the number of states, compared to the number of bands strictly needed at 0 Kelvin.
\\

The number of potentially occupied wavefunctions,
$pocc$, is defined at the start of the computation. 
It must exceed sufficiently the number of electrons $N$,
in order for the highest states in this space to have vanishing occupations.  
So, instead of minimizing Eq.~(\ref{eq:augmented_free_energy}) with definition Eq.~(\ref{eq:free_energy}),
that implicitly suppose that the set of $\{ \psi_{i} \}$ spans the whole Hilbert space, the functional to be considered is defined in terms of a finite number of orthonormal functions, and the corresponding finite occupation matrix elements, with $i$ and $j$ running from 1 to $pocc$. This gives the following modified definition:

\begin{align}
    & F^{+}[T, \{ \psi_{i} \}, \{ \rho_{ij} \}] = n_\textrm{s} \sum_{ij}^{{pocc}} \rho_{ji} h_{ij} + E_\textrm{Hxc}[\rho] \nonumber \\
    & -kT n_{\textrm{s}} \sum_{\gamma}^{{pocc}} s(\occf_{\gamma}) - \sum_{ij}^{{pocc}} \Lambda_{ji}
    n_\textrm{s} \Big(\langle \psi_{i} | \psi_{j} \rangle - \delta_{ij}\Big) \nonumber \\
    & - \mu \bigg( \bigg( \sum_{i}^{{pocc}}n_\textrm{s}\rho_{ii} \bigg) - N \bigg).\label{eq:14x}
\end{align}

\indent  
The orthonormalization constraint only applies between the functions belonging to the potentially occupied wavefunctions. 
The notation ``S$_{pocc}$" will later denote that space of functions.

\subsection{Smearing schemes}
\label{sec:smearing_schemes}

Smearing schemes have the goal to decrease the numerical effort
needed to deal with rapidly varying occupation numbers
when the Brillouin Zone of metals is sampled. 
They allow to rely on fewer wavevectors to obtain the
same numerical precision than without smearing.
The difficulty to reach numerical convergence is especially
acute for a vanishing temperature, since the occupation
of levels discontinuously changes from 1 to 0 at the Fermi level.
Generally speaking, the occupation of an energy level is defined through an
occupation function $f(x)$ whose argument is the difference
between the Fermi energy and the energy of the level, 
rescaled by either $kT$, for the Fermi-Dirac
case, or by a smearing energy $\sigma$, for pure numerical
smearing schemes.
The occupation function $f(x)$ vanishes for infinitely negative $x$ and
tends to 1 for infinite positive $x$.

As mentioned in Refs.~\onlinecite{Methfessel1989, DosSantos2023}, all such occupation functions can be generated from
an associated smearing function $\Tilde{\delta}(\varepsilon)$, which is normalized to 1. The related occupation function is 
\begin{equation}
f(x) = \int_{- \infty}^{x} \Tilde{\delta}(\varepsilon) d\varepsilon. \label{eq:185}
\end{equation}
where $x = \frac{\mu - \epsilon}{\sigma}$.
Note that $x$ is adimensional, as well as the integrand 
$\varepsilon$, while $\epsilon$, $\mu$, 
$\sigma$ and $kT$ have the dimension of an energy. $x$ and $\varepsilon$ are actually ``rescaled" energies, without dimensions.
For the Fermi-Dirac (FD) case, the smearing function is
\begin{align}
    \Tilde{\delta}_{\textrm{FD}}(x) =
    \frac{1}{\big( \exp(x)+1\big)\big(\exp(-x)+1\big)}.
    \label{eq:187}
\end{align} 
The occupation function deduced from this smearing function is  given by Eqs.~(\ref{eq:fi_FD}) and ~(\ref{eq:FD})
as expected.

In order to obtain the entropy $s(f)$ as a function of the occupation, one defines first the
entropy $\Tilde{s}(x)$ as a function of the adimensional $x$,
\begin{equation}
\Tilde{s}(x) = - \int_{- \infty}^{x} \varepsilon \Tilde{\delta}(\varepsilon) d\varepsilon. \label{eq:186} 
\end{equation}
Note the slight change of notation for the entropy function of the rescaled energy, $\Tilde{s}$, with respect to the one in Ref.~\onlinecite{DosSantos2023}, ``$s$'' for the same quantity. Indeed, the notation ``$s$" is already used in the present context for the entropy as a function of $f$, see Eq.~(\ref{eq:S}).
The latter had not been examined in 
Ref.~\onlinecite{DosSantos2023}.

Such function $s(f)$ is deduced 
from Eq. (\ref{eq:186})
by using the reciprocal of Eq. (\ref{eq:185}), denoted $[f]^{-1}$ so that
\begin{align}
&    s(f)= \Tilde{s}\big([f]^{-1}(f)\big).
\end{align}
Indeed, with definitions Eqs.~(\ref{eq:185}) and ~(\ref{eq:186}), and the same definition of $x$ as in the text before Eq.~(\ref{eq:185}), the relation 
Eq.~(\ref{eq:f_i_from_spm1}),
that links the occupation number to the energy through the
derivative of the entropy as a function of the occupation number,
is fulfilled.
Note however that $s(f)$ might be a multivalued function, in the case where the $f(x)$ function is not 
monotonically decreasing. This is encountered
for advanced smearing schemes. Also, the 
one-level contribution
to the free energy, $-Ts(f)$,
might not be convex.

For the Fermi-Dirac case,
the corresponding $-kTs_{\textrm{FD}}(f)$, Eq.~(\ref{eq:sFD}), is univalued and convex.

Beyond the Fermi-Dirac case, the Gaussian and MP smearing functions are often encountered. We will not analyze so-called ``cold smearings'',
\cite{Marzari1999} that for the purpose of the
present analysis, exhibit the same problematic feature as the 
MP smearing, namely the non-monotonicity of the occupation function.

For the Gaussian (G) case, the smearing function is
\begin{align}
    \Tilde{\delta}_{\textrm{G}}(x) = \frac{1}{\sqrt{\pi}} \exp(-x^{2}) \label{eq:tilde_delta_G}.
\end{align}
The occupation function $f_{\textrm{G}}(x)$ is $1+\textrm{erf}(x)/2$. 
Its reciprocal cannot be expressed easily.
The entropy function of $x$ is half the broadening function,
\begin{align}
    \Tilde{s}_{\textrm{G}}(x) = \frac{1}{2\sqrt{\pi}} \exp(-x^{2}) \label{eq:tilde_s_G}.
\end{align}
This entropy function of the occupation is univalued.

For the MP broadening,

\begin{align}
    \Tilde{\delta}_{\textrm{MP}}(x) = \frac{1}{\sqrt{\pi}} \bigg( \frac{3}{2} - x^{2} \bigg) \exp(-x^{2}). \label{eq:188}
\end{align}
$f_{\textrm{MP}}(x)$ can be larger than 1 and smaller than 0,
and is not easily expressed.
It is not monotonically increasing, 
hence the $s_{\textrm{MP}}(f)$ function is multivalued. 
The corresponding entropy function of the scaled energy is
\begin{align}
    \Tilde{s}_{\textrm{MP}}(x) = \frac{1}{2\sqrt{\pi}} 
    \bigg( \frac{1}{2} - x^{2} \bigg) \exp(-x^{2}) \label{eq:tilde_s_MP}.
\end{align}

By convention, for the Gaussian and MP scheme,
one replaces $kT$ by the smearing parameter $\sigma$
in the definition of the individual occupations
in terms of the distribution function, Eq. (\ref{eq:fi_FD}). 
As mentioned earlier, the goal of the Gaussian and MP smearings is to 
provide the $T=0$ properties of metals 
with less numerical effort than with the sudden change of the occupation from 1 to 0 at the Fermi energy, albeit with some loss of precision, nevertheless under control.
The advantage of the MP smearing function comes from vanishing low-order Taylor terms up to and including the third order with respect to the
smearing parameter $\sigma$, in the expansion of the correction to the free energy due to the smearing.
In the Gaussian smearing, the second order does not vanish, while it vanishes for MP and cold smearing.
However, as mentioned above, the MP occupation function becomes non-monotonic.

At finite temperatures, a smearing 
methodology can also help. The so-called ``resmearing'' scheme has been introduced to obtain 
physical finite-temperature quantities
with decreased numerical effort.\cite{Verstraete2001, Verstraete2004}
The resmeared delta function $\Tilde{\delta}_{\textrm{rsm}}$ is defined as
\begin{align}
    \Tilde{\delta}_{\textrm{rsm}}(y,R) = \int  \Tilde{\delta}_{\textrm{FD}}(y-Rz) \Tilde{\delta}_{2}(z) dz, \label{eq:41}
\end{align}
where $R=\frac{\sigma}{kT}$ is the ratio between
the smearing parameter and the physical electronic temperature,
and $\delta_2$ is the broadening function (either the Gaussian broadening or the MP broadening in the present work)
that is convoluted with the Fermi-Dirac broadening
function $\Tilde{\delta}_\textrm{FD}$. 
While the broadening function $\delta_2$
might indeed be such generic function, 
in the remaining of this work, we will focus on the case $\delta_2=\delta_{\textrm{MP}}$.

The actual broadening function corresponding to some physical electronic temperature $T$,
denoted as the ``total broadening'' in Refs.~\onlinecite{Verstraete2001}
and ~\onlinecite{Verstraete2004}, is obtained as

\begin{align}
    \Tilde{\delta}_{\textrm{tot}}
    (\mu-\epsilon,kT,\sigma) = \frac{1}{kT} \Tilde{\delta}_{\textrm{rsm}} \bigg( \frac{\mu-\epsilon}{kT}, \frac{\sigma}{kT} \bigg), \label{eq:42appendix}
\end{align}
where the explicit dependence of such function on three arguments having the dimension
of an energy has been made clear.

While preparing the present publication, it became clear that the notation in the original reference\cite{Verstraete2001}, to which one of us contributed, was fuzzy.
Also, several typos were present.
In order to bypass such problems, the Sec.~ S1 of the Supporting Information contains a mathematically rigorous rewriting of the key equations found in the original reference.

Depending on the ratio $R$, the total  broadening
resembles the original Fermi-Dirac broadening (small $R$) or the other broadening function $\Tilde{\delta}_2$ (large $R$), although in the latter case, its argument is rescaled by $R$, and its value is inversely rescaled by $R$, to keep the integral unity.

For the specific resmearing of the Fermi-Dirac broadening with MP 
broadening function, the value $R=2$ is critical, since it is the largest $R$ value for which the total broadening is positive for the entire
range of its argument, as will be shown later.
For this reason, the resmeared function with $R=2$ will be illustrated: in the forthcoming figures, results are presented with 
a smearing parameter $\sigma$ twice bigger than the energy corresponding to the physical electronic temperature, $kT$.
Note that the authors of Ref.~\onlinecite{DosSantos2023} used 
$R$ of about 2.565 in comparing the FD and Gaussian cases. On the one hand, they 
were interested to work with the FD and G functions, while we are interested in the FD and MP functions, and, on the other hand, they based their study on another criterion (the similarity between the Gaussian and Fermi-Dirac occupation function).
In practice, the resmearing parameter $\sigma$ is however taken to be a constant, irrespective of the physical electronic temperature.

\begin{figure*}
\begin{subfigure}{0.45\textwidth}
\centering
 \includegraphics[width=\linewidth]{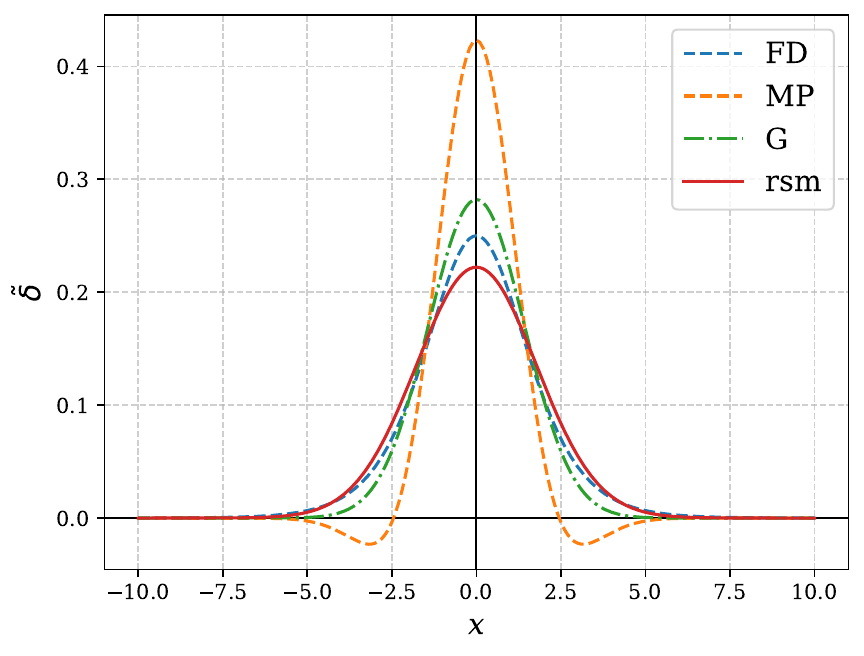}
 \caption{Comparative representation of the Fermi-Dirac broadening function $\Tilde{\delta}_{\textrm{FD}}(x)$ with the rescaled Methfessel-Paxton one $R\Tilde{\delta}_{\textrm{MP}}(x/R)$, 
 the rescaled Gaussian one $R\Tilde{\delta}_{\textrm{G}}(x/R)$, and the
 rescaled smearing one $R\Tilde{\delta}_{\textrm{rsm}}(x/R,R)$,
 with $R$=2, see text.}
 \label{fig:subfig1}
 \end{subfigure}
 \hfill
\begin{subfigure}{0.45\textwidth}
\centering
 \includegraphics[width=\linewidth]{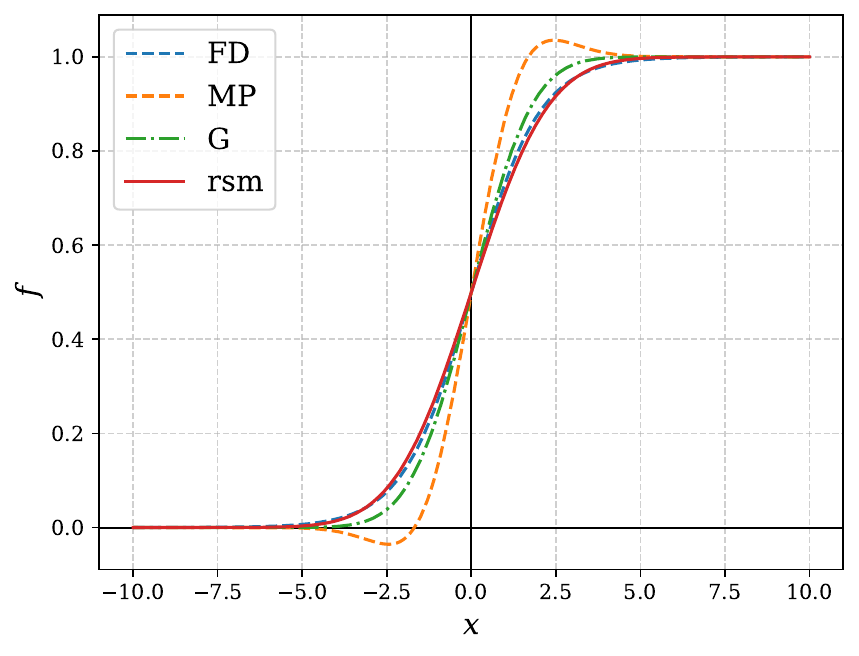}
 \caption{Comparative representation of various occupation functions $f_{\textrm{FD}}(x)$, $f_{\textrm{MP}}(x/R)$, $f_{\textrm{G}}(x/R)$, and $f_{\textrm{rsm}}(x/R,R)$,
 with $R$=2, see text.}
 \label{fig:subfig2}
 \end{subfigure}
 \hfill
 \begin{subfigure}{0.45\textwidth}
 \centering
 \includegraphics[width=\linewidth]{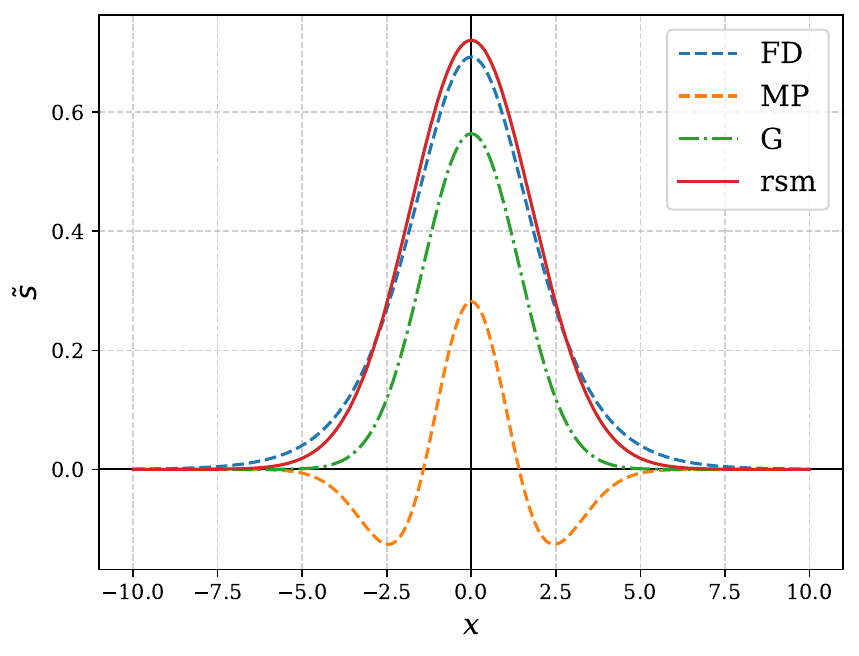}
 \caption{Comparative representation of various entropy functions $\tilde{s}_{\textrm{FD}}(x)$,
 $\tilde{s}_{\textrm{MP}}(x/R)$,
 $\tilde{s}_{\textrm{G}}(x/R)$
 and
 $\tilde{s}_{\textrm{rsm}}(x/R,R)$,
 with $R$=2, see text.}
 \label{fig:subfig3}
 \end{subfigure}
 \hfill 
 \begin{subfigure}{0.45\textwidth}
 \centering
 \includegraphics[width=\linewidth]{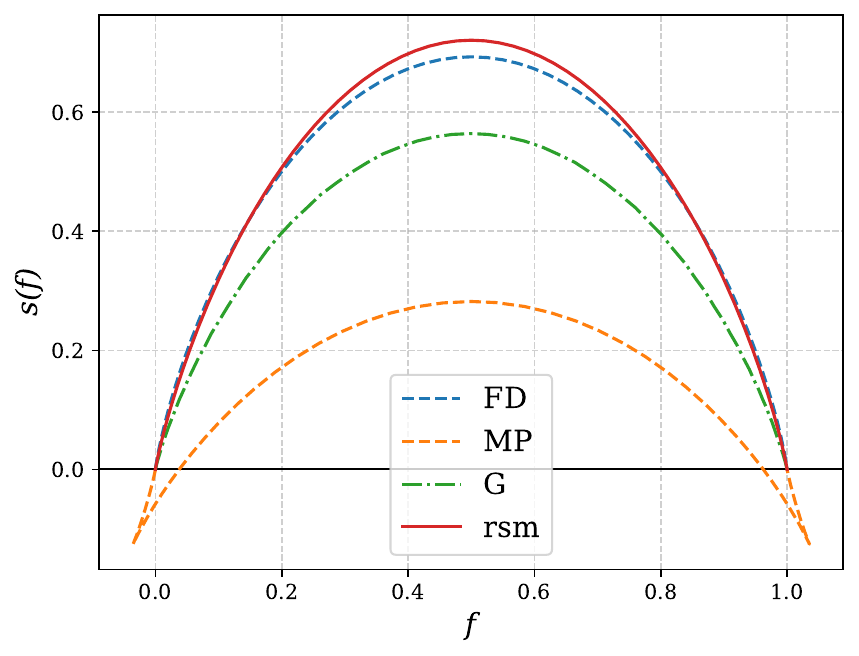}
 \caption{Comparative analysis of various entropy functions $s(f)$ of the occupation $f$.}
 \label{fig:subfig4}
 \end{subfigure}
 \caption{Comparative representation of the broadening function, occupation function, and entropy functions 
 ($\tilde{s}$~and~$s$) in the Fermi-Dirac (FD),
 Methfessel-Paxton (MP), Gaussian (G) and resmeared cases. The last three are displayed with rescaling factor $R=2$.
 Also, $\delta_2=\delta_{\textrm{MP}}$ for the resmearing case.}
\label{fig:fig1ch}
\end{figure*}

Fig. \ref{fig:subfig1} presents the different broadening functions
mentioned above (with $R=2$ and $\delta_2=\delta_{\textrm{MP}}$ for $\Tilde{\delta}_{\textrm{rsm}})$.
All functions, except MP, are positive, going smoothly from 0 to their maximum, then back to 0. Only the MP broadening exhibits negative values for some range of its argument.
The asymptotic behavior of $\Tilde{\delta}_{\textrm{rsm}}$
is analyzed in 
Sec.~S1 of the Supporting Information. It is shown there that the exponentially decreasing tail of $\Tilde{\delta}_{\textrm{rsm}}$ changes sign at $R=2$.

Similarly, Fig.~\ref{fig:subfig2} presents the different associated occupation functions, with the characteristic 
monotonically increasing behavior, except for the MP
scheme, and Fig.~\ref{fig:subfig3} presents the different one-level entropy functions $\Tilde{s}(x)$, all positive everywhere except the MP one.
In the context of the variational DFT (or DFPT), the shape 
of the entropy function with the occupation as argument, $s(f)$, is crucial.
Indeed, it enters the free energy to be minimized through the $-kTs(f)$ contribution of each one-electron level. 
Non-convexity of this term might induce non-convexity of the global free energy functional. 

The positive monotonic behavior of $\frac{\partial f}{\partial x}$ is directly linked to the convexity of the $-kTs(f)$ function.
Indeed, using the chain rule followed by Eqs.
\ref{eq:185} and \ref{eq:186}, one finds
\begin{align}
\frac{\partial s}{\partial f}=
\frac{\partial \Tilde{s}}{\partial x}
\frac{\partial x}{\partial f}=
-x \Tilde{\delta}(x)
\big(\Tilde{\delta}(x)   \big)^{-1}=
-x
\end{align}
This equation is derived with respect to $f$, to give
\begin{align}
\frac{\partial^{2} s}{\partial f^{2}}
=
-\frac{\partial x}{\partial f}
=
-\Bigg( \frac{\partial f}{\partial x} \Bigg)^{-1}
\label{eq:d2sdf2}
\end{align}

So, if $\frac{\partial f}{\partial x} \geq 0$ everywhere, then also everywhere

\begin{align}
    \frac{\partial^{2} s}{\partial f^{2}} \leq 0, \label{eq:182}
\end{align}

\noindent and $-kTs(f)$ is convex.
$\frac{\partial f}{\partial x} \geq 0$ everywhere is also the criterion to avoid multiple chemical potentials, as described in Ref.~\onlinecite{DosSantos2023}. This criterion is violated by the MP scheme, but fulfilled in the other schemes.

The $s(f)$ function is represented in Fig.~\ref{fig:subfig4} for the different smearing schemes.
The MP one is particularly interesting.
Its domain of definition extends beyond
the 0 to 1 range, and, outside of this range, the function is multivalued, with two branches.
This multivalued function has characteristics singularities at the smallest values that it can reach, where two branches merge with common tangent. 
This happens at the critical rescaled energy $x^*$ at
which the broadening MP function vanishes,
$x^*=\pm \sqrt{3/2}$.
Indeed, at that value, both $f_{\textrm{MP}}(x)$ and 
$\Tilde{s}_{\textrm{MP}}(x)$ reach an extremum.
Their curvature is identical on the left and right of $x^*$, which explains the common tangent.

\section{Second-order free energy for metals}
\label{Sec:F2}

\subsection{Variational formulation of DFPT for metals}
\label{sec:VarDFPTmetals}

Ref.~\onlinecite{Gonze1995} describes a general framework for the perturbation theory of variational principles, including the case of constraints. 
In Ref.~\onlinecite{Gonze1995a},
such framework is applied
to DFT in case of discretized levels and fixed occupation numbers at zero Kelvin. 
We follow the variational framework of Ref.~\onlinecite{Gonze1995}, including its notations, and generalize Ref.~\onlinecite{Gonze1995a} to varying, metallic occupations.
The details of the derivation are presented in the Sec.~S2 of the Supporting Information.

For the unperturbed wavefunctions 
and occupation matrix, one works in the diagonal gauge: the starting wavefunctions $|\psi_i^{(0)}\rangle$, belonging to S$_{pocc}$, fulfill Eq.~(\ref{eq:KohnSham}), and the unperturbed occupation numbers are obtained from  
Eq.~(\ref{eq:f_i_from_spm1}) - or its equivalent for
smearing schemes other than FD. 
Due to Eq.~(\ref{eq:KohnSham}), all off-diagonal elements of the unperturbed density matrix vanish.  

The augmented variational second-order free energy, a functional of the first-order wavefunctions and first-order density matrix elements, including Lagrange multipliers terms, is obtained as
\begin{widetext}
\begin{eqnarray}
F^{+(2)}[T,\{\psi_i^{(1)}\},\{\rho_{ij}^{(1)}\}]
&=& 
n_{\textrm{s}} \sum_{i}^{{pocc}} \occf_i^{(0)}
F^{(2)}_{i}[\psi_i^{(1)}]
+
n_{\textrm{s}} \sum_{ij}^{{pocc}} \rho_{ji}^{(1)}
F_{ij}^{(1)}[\psi_i^{(1)}]
\nonumber \\
&+&
\frac{1}{2} \int \int K_{\textrm{Hxc}}(\textbf{r},\textbf{r}')
\rho^{(1)}(\textbf{r})
\rho^{(1)}(\textbf{r}')
d\textbf{r}
d\textbf{r}'
\nonumber\\
&-&
TS^{(2)}
[T;\{ \rho_{ij}^{(1)} \}]
-
n_{\textrm{s}} \sum_{ij}^{{pocc}} \Lambda_{ji}^{(1)} 
\Big( 
\langle \psi_i^{(1)} | \psi_j^{(0)} \rangle
+
\langle \psi_i^{(0)} | \psi_j^{(1)} \rangle
 \Big)
-
n_{\textrm{s}} \mu^{(1)} 
\sum_{i}^{{pocc}} \rho_{ii}^{(1)},
\label{eq:F+2}
\end{eqnarray}
with the shorthand notations

\begin{equation}
    F^{(2)}_{i}[\psi_i^{(1)}] = 
\langle \psi_i^{(1)} | \hat{H}^{(0)} - \epsilon_i^{(0)}| \psi_i^{(1)} \rangle
+ \langle \psi_i^{(0)} | \hat{v}^{(2)}_{\textrm{ext}} | \psi_i^{(0)} \rangle +\Big( \langle \psi_i^{(1)} | \hat{v}^{(1)}_{\textrm{ext}}  | \psi_i^{(0)} \rangle + (\textrm{c.c.}) \Big) \label{eq:F2i}
\end{equation}

\indent and 

\begin{align}
F_{ij}^{(1)}[\psi_i^{(1)}]
= \langle \psi_i^{(0)} | \hat{v}_{\textrm{ext}}^{(1)} | \psi_j^{(0)} \rangle 
+ \langle \psi_i^{(1)} | \hat{H}^{(0)} | \psi_j^{(0)} \rangle
+ \langle \psi_i^{(0)} | \hat{H}^{(0)} | \psi_j^{(1)} \rangle. \label{eq:F1ij}
\end{align}
\end{widetext}

The dependence of this second-order free energy $F^{+(2)}$ on the zero-order quantities \{$|\psi_i^{(0)}\rangle$\} and $\{f^{(0)}_{i}\}$ is not mentioned explicitly, for sake of compactness. Similarly, the dependence of the second-order entropy $S^{(2)}$ on  $\{f^{(0)}_{i}\}$ is not mentioned.
This choice is made because the unperturbed system is considered known, 
and one is focusing on the effect of perturbations on the system. 
In Sec.~\ref{Sec:underconvergedGSwfs},
we will study the effect of underconverged \{$|\psi_i^{(0)}\rangle$\}.

The temperature is explicitly mentioned as an argument of $F^{+(2)}$ and $S^{(2)}$.
They indeed depend on it, directly. 
Moreover, note
that a change of $T$ also affects 
$|\psi_i^{(0)}\rangle$ and $\{f^{(0)}_{i}\}$, and thus, indirectly,
$F^{+(2)}$ and $S^{(2)}$.

The Hartree and exchange-correlation kernel is defined as
\begin{equation}
K_{\textrm{Hxc}}[\rho](\textbf{r},\textbf{r}') = \frac{\delta^2 E_{\textrm{Hxc}}[\rho]}{\delta \rho(\textbf{r})\delta \rho(\textbf{r}')}.
\end{equation}
The first-order density is computed from

\begin{eqnarray}
\label{eq:57d}
\rho^{(1)}(\textbf{r}) &=& 
n_{\textrm{s}} \sum_{i}^{pocc}\occf_i^{(0)} \Big( \psi_i^{(1)*}(\textbf{r})
     \psi_i^{(0)}(\textbf{r})
+
\psi_i^{(0)*}(\textbf{r})
 \psi_i^{(1)}(\textbf{r})
\Big)
\nonumber\\
&+&
n_{\textrm{s}} \sum_{ij}^{pocc}\rho_{ji}^{(1)} \psi_i^{(0)*}(\textbf{r})
     \psi_j^{(0)}(\textbf{r})
.
\end{eqnarray}

As in the case of the unperturbed situation, the 
minimization of the augmented second-order free energy delivers the
optimal electronic second-order free energy
\begin{eqnarray}
F^{(2)}_{\textrm{el}}[T]=
\min_{\{\psi_i^{(1)}\},\{\rho_{ij}^{(1)}\}}
F^{+(2)}[T;\{\psi_i^{(1)}\},\{\rho_{ij}^{(1)}\}].
\end{eqnarray}
The Lagrange parameters  
$\Lambda_{ji}^{(1)}$ and $\mu^{(1)}$
in Eq.~(\ref{eq:F+2}) must be tuned, after minimization, so that
the constraints
\begin{equation}
\langle \psi_i^{(1)} | \psi_j^{(0)} \rangle
+
\langle \psi_i^{(0)} | \psi_j^{(1)} \rangle
=0,
\label{eq:81}
\end{equation}
for $i$ and $j$ in S$_{pocc}$,
and 
\begin{equation}
\sum_{i}^{{pocc}} \rho_{ii}^{(1)}=0
\label{eq:trf1}
\end{equation}
are enforced.

The second-order entropy term, evaluated with zero- and first-order elements of the density matrix (no second-order elements, see Ref.~\onlinecite{Gonze1995,Gonze1995a}) needs to be worked out carefully.
Indeed, although none of the second-order elements of the density matrix should be taken into account (following  Ref.~\onlinecite{Gonze1995}), the eigenvalues of the density matrix will be modified up to second order from first-order variations of the density matrix, and this will have an effect on the evaluation of the trace present in the second-order entropy term.
From Eq.~(\ref{eq:S}), \begin{widetext}
\begin{equation}
\Big( S
[\{\rho^{(1)}_{ij}\}] \Big)^{(2)}
= n_{\textrm{s}} \, \sum_{\gamma} 
k\Big( s(\occf_\gamma) \Big)^{(2)} 
= n_{\textrm{s}} \, \sum_{\gamma} 
k\Bigg( 
s'(\occf_\gamma^{(0)})\occf_\gamma^{(2)}
+
s''(\occf_\gamma^{(0)})
\frac{
\big(
\occf_\gamma^{(1)}
\big)^2}{2}
\Bigg).
\label{eq:S2}
\end{equation}
\end{widetext}

In the Fermi-Dirac case,
the first-order derivative of $s$ with respect to  its argument is given by
Eq.~(\ref{eq:11}), while
the second-order derivative is
\begin{equation}
\label{eq:12}
s''_\textrm{FD}(f) = \frac{d^2s_\textrm{FD}}{df^2} 
=-\frac{1}{(1-f)f},
\end{equation}
a function that is negative for all values of $f$ between 0 and 1, with negative curvature in this range, and that diverges at both 0 and 1. 
Taking into account the $-kT$ prefactor of the second-order entropy in the augmented second-order free energy, Eq.~(\ref{eq:F+2}), the $s''$ term gives a positive contribution to that second-order free energy.
The derivatives of occupation matrix eigenvalues $\occf_\gamma^{(1)}$
and $\occf_\gamma^{(2)}$
are to be computed from
$\occf^{(0)}_{i}$
and 
$\rho_{ij}^{(1)}$, excluding any higher-order contribution from the occupation matrix, in line with the general DFPT formalism.\cite{Gonze1995,Gonze1995a}
The eigenvalues $\occf_\gamma$ are computed by diagonalizing the $\rho$ matrix, and similarly for their perturbation expansion, expressed in terms of Sternheimer equations of different orders.
The first-order eigenvalues are found easily using the Hellmann-Feynman theorem
~\cite{Hellmann1937, Feynman1939},
\begin{equation}
\occf_{\gamma}^{(1)}=\rho_{\gamma\gamma}^{(1)},
\end{equation}
while the second-order eigenvalues are obtained
as 
\begin{equation}
\occf_{\gamma}^{(2)}=\rho_{\gamma\gamma}^{(2)}-
\sum^{pocc'}_{i} \frac{|\rho_{i\gamma}^{(1)}|^2}{f_i^{(0)}-f_\gamma^{(0)}},
\label{eq:f2}
\end{equation}
where the prime superscript to the summation sign means that the sum over $i$ excludes the vanishing denominator case. 
The latter equation is valid in the non-degenerate case, but might be generalized to the degenerate case through degenerate perturbation theory.
Eq.~(\ref{eq:f2}) contains the second-order 
$\rho_{\gamma\gamma}^{(2)}$ that must be discarded in the context of
the computation of Eq.~(\ref{eq:S2}) and its contribution to Eq.~(\ref{eq:F+2}), as mentioned previously. 
Thus the second-order entropy contribution is
\begin{widetext}
\begin{eqnarray}
-TS^{(2)}
[\{\rho_{ij}^{(1)}\}]
&=&
-kTn_{\textrm{s}} 
\Bigg[
-\sum_{ij}^{pocc'}s'(\occf_{j}^{(0)})
\frac{|\rho_{ij}^{(1)}|^2}{\occf_{i}^{(0)}-\occf_{j}^{(0)}}
+
\sum_{i}^{pocc}s''(\occf_{i}^{(0)})
\frac{(\rho_{ii}^{(1)})^2}{2}
\Bigg],
\label{eq:64bis+78}
\end{eqnarray}
where the prime superscript to the summation sign means that the double sum over $i$ and $j$ excludes the vanishing denominator case.
It can be further worked out, using  Eq.~(\ref{eq:chemicalpotential}) and 
~(\ref{eq:d2sdf2}), eliminating the $s$ function and its derivatives, then using the hermiticity of the $\hat{\rho}^{(1)}$ operator:

\begin{equation}
-TS^{(2)}[\{\rho_{ij}^{(1)}\}]
=
- n_{\textrm{s}} 
\Bigg[
\sum^{pocc'}_{ij}
\frac{\epsilon_i^{(0)}-\epsilon_j^{(0)}}
{\occf_{i}^{(0)}-\occf_{j}^{(0)}}
\frac{|\rho_{ij}^{(1)}|^2}{2} + \sum_{i}^{pocc}
\Bigg( \frac{\partial f}{\partial x}\Big|_{x=\epsilon^{(0)}_i-\mu^{(0)}}
\Bigg)^{-1}
\frac{(\rho_{ii}^{(1)})^2}{2}
\Bigg]. \label{eq:87}
\end{equation}
\end{widetext}

With this explicitation of the second-order entropy, the expression of the second-order variational free energy 
Eqs.~(\ref{eq:F+2})-(\ref{eq:F1ij}) is complete.

Quadratic terms in $\{ \psi_{i}^{(1)} \}$ appear in 
Eq.~(\ref{eq:F2i}) and in the Hxc contribution, third term of 
Eq.~(\ref{eq:F+2}).
Quadratic terms in $\{ \rho_{ij}^{(1)} \}$ appear in the entropy contribution, Eq.~(\ref{eq:87}), as well as in the 
Hxc contribution. 
Also, bilinear terms in $\{ \psi_{i}^{(1)} \}$ and $\{ \rho_{ij}^{(1)} \}$ appear in the second term of Eq.~(\ref{eq:F+2}), 
and in the Hxc contribution.

The whole expression must be definite positive with respect to changes of $\{ \psi_{i}^{(1)} \}$ and $\{ \rho_{ij}^{(1)} \}$ taken in their quadratic/bilinear contribution.
For a monotonically decreasing $f(x)$ function, the prefactor of $\rho_{ij}^{(1)}$ or $\rho_{ii}^{(1)}$ in Eq.~(\ref{eq:64bis+78}) (or Eq.~(\ref{eq:87})) is positive, and these contributions are convex.
The situation is also clear for the $K_{\textrm{H}}$ contribution to Eq.~(\ref{eq:F+2}), but not so for the whole $K_{\textrm{Hxc}}$.
Indeed, while $K_{\textrm{H}}$ is a positive-definite kernel, the kernel $K_{\textrm{xc}}$ is not (even, $K_{\textrm{xc}}$ is definite-negative in the LDA).
In order to finalize the analysis of the extremal character
of Eq.~(\ref{eq:F+2}), we need also to address the term quadratic in $\psi_{i}^{(1)}$. This will be done when discussing the gauge choices.

\subsection{Minimization of the second-order free energy}
\label{sec:minimization_F2}

The second-order free energy $F^{+(2)}$, Eq.~(\ref{eq:F+2}), can now be minimized, by computing
the gradients with respect to the variables $\{\psi_i^{(1)}\}$ and $\{\rho_{ij}^{(1)}\}$.
The need to impose the hermitian character of $\{\rho_{ij}^{(1)}\}$ might seem to yield some complication. However this can be bypassed by generalizing Eq.~(\ref{eq:F+2}) to non-hermitian $\{\rho_{ij}^{(1)}\}$,
as it is done in the Supporting Information Sec.~S3.
The gradients are explicitly written in Supporting Information Sec.~S4.
At the minimum, the gradients vanish, and one finds the following equations, that are independent of the choice of gauge.
Depending on the gauge, such expressions might further simplify. This will be seen in the next Sec.~\ref{Sec:different_gauges}.

Imposing zero diagonal occupation gradient  delivers
\begin{equation}
\label{eq:111}
    \rho_{ii}^{(1)} = \frac{\partial f}{\partial \epsilon} \bigg|_{\epsilon_{i}^{(0)}-\mu^{(0)}}(\epsilon_{i}^{(1)}-\mu^{(1)}),
\end{equation}
while for the case of off-diagonal occupation gradients, one gets 
\begin{equation}
\label{eq:114}
\begin{split}
\rho_{ij}^{(1)} & = \frac{\occf_{i}^{(0)} - \occf_{j}^{(0)}}{\epsilon_{i}^{(0)} - \epsilon_{j}^{(0)}} \langle \psi_{i}^{(0)} | \hat{H}^{(1)} | \psi_{j}^{(0)} \rangle \\
& - (\occf_{i}^{(0)} - \occf_{j}^{(0)}) ( \langle \psi_{i}^{(1)} | \psi_{j}^{(0)} \rangle - \langle \psi_{i}^{(0)} | \psi_{j}^{(1)} \rangle),
\end{split}
\end{equation}
\indent where
\begin{equation}
\label{eq:H1}
\hat{H}^{(1)}=
\hat{v}_{\textrm{ext}}^{(1)}
+
\int
K_{\textrm{Hxc}}[\rho](\textbf{r},\textbf{r}')
\rho^{(1)}(\textbf{r}')
d\textbf{r}'.
\end{equation}

Imposing zero projected gradient of $F^{+(2)}$ with respect to $\langle \psi_{i}^{(1)} |$ in
the S$_{pocc}$ space gives an expression for the first-order Lagrange multipliers

\begin{equation}
\label{eq:116}
\begin{split}
    \Lambda_{ki}^{(1)} & = \occf_{i}^{(0)} [ (\epsilon_{k}^{(0)} - \epsilon_{i}^{(0)}) \langle \psi_{k}^{(0)} | \psi_{i}^{(1)} \rangle + \langle \psi_{k}^{(0)} | \hat{H}^{(1)} | \psi_{i}^{(0)} \rangle ] \\
    &  + \rho_{ki}^{(1)} \epsilon_{k}^{(0)}.
\end{split}
\end{equation}

The diagonal elements are

\begin{equation}
    \Lambda_{ii}^{(1)} = \occf_{i}^{(0)} \epsilon_{ii}^{(1)} + \rho_{ii}^{(1)} \epsilon_{i}^{(0)}.
\end{equation}

Imposing zero projected gradient of $F^{+(2)}$ with respect to $\langle \psi_{i}^{(1)}|$ out of the S$_{pocc}$ space gives the usual Sternheimer equation of DFPT~\cite{Gonze1995a},
\begin{equation}
\label{eq:118b}
\hat{P}_{\perp} (\hat{H}^{(0)} - \epsilon_{i}^{(0)}) \hat{P}_{\perp} | \psi_{i}^{(1)} \rangle = - \hat{P}_{\perp} \hat{H}^{(1)} | \psi_{i}^{(0)} \rangle.
\end{equation}
This is also directly connected to a key equation in the work of de Gironcoli,\cite{Gironcoli1995} the projection of his Eq.(11)
in the space perpendicular to the active space of unperturbed wavefunctions.
\\

\section{The different gauges}
\label{Sec:different_gauges}

\subsection{The gauge freedom}
\label{sec:gauge_freedom}

From the very start, the diagonal gauge has been chosen for the unperturbed wavefunctions and occupations, namely Eq.~ (\ref{eq:KohnSham}), giving Eq.~(\ref{eq:42}) and 
\begin{equation}
\label{eq:119}
\rho_{ij}^{(0)} = \delta_{ij} \occf_i^{(0)},
\end{equation}

\begin{equation}
\label{eq:120}
\hat{H}^{(0)} | \psi_{j}^{(0)} \rangle = \epsilon_{j}^{(0)} |\psi_{j}^{(0)} \rangle,
\end{equation}
and
\begin{equation}
\label{eq:121}
    \Lambda_{kj}^{(0)} = \occf_j^{(0)} \delta_{kj} \epsilon_{j}^{(0)}.
\end{equation}

However, no gauge choice has been made for the first-order quantities, while there is indeed a gauge freedom, originating from the possibilities of a unitary transform in the starting problem.
The constraints (to be fullfilled whatever the gauge) are Eqs.~(\ref{eq:81})
 and ~(\ref{eq:trf1}).
 Eq.~(\ref{eq:81}) fixes the symmetric part of the
 scalar product between the 
zero-order and first-order wavefunctions.
However, the asymmetric part of the scalar product between the zero-order and first-order wavefunctions is not fixed:
\begin{equation}
    \langle \psi_{i}^{(1)} | \psi_{j}^{(0)} \rangle - \langle \psi_{i}^{(0)} | \psi_{j}^{(1)} \rangle = A_{ij}
\end{equation}

We first examine the consequences of choosing $A_{ij} = 0$, that is called parallel gauge for the first-order wavefunctions, then examine other possibilities.
Note that $\hat{H}^{(1)}$, $\rho^{(1)}$ and $F^{(2)}$ must be invariant under such choice.
Sec.~S5 of the Supporting Information shows how the first-order wavefunctions and occupation matrix elements
change concurrently.
\newline\\

\subsection{The parallel gauge}
\label{sec:parallel-gauge}

First-order wavefunctions in the parallel gauge are noted $| \psi_{||,i}^{(1)} \rangle$, and similarly for the first-order density matrix elements.
One imposes 
\begin{equation}
\label{eq:124}
\langle \psi_{i}^{(0)} | \psi_{||,j}^{(1)} \rangle = 0
\end{equation}
\indent when $i$ and $j \in \textrm{S}_{pocc}$.
The second-order free energy $F^{+(2)}$, 
Eqs.~(\ref{eq:F+2})-(\ref{eq:F1ij}), simplifies: the two last contributions to  Eq.~(\ref{eq:F1ij})
vanish, as well as the fifth term of Eq.~(\ref{eq:F+2}).

\begin{widetext}
One gets
\begin{align}
F^{+(2)}[T; \{ \psi_{||,i}^{(1)} \}; \{ \rho_{||,ij}^{(1)} \} ]
& = n_\textrm{s} \sum_{i}^{{pocc}} \occf_{i}^{(0)} [ \langle \psi_{||,i}^{(1)} | \hat{H}^{(0)} - \epsilon_{i}^{(0)} | \psi_{||,i}^{(1)} \rangle  + \langle \psi_{i}^{(0)} | v_{\textrm{ext}}^{(2)} | \psi_{i}^{(0)} \rangle + \big( \langle \psi_{||,i}^{(1)} | v_{\textrm{ext}}^{(1)} | \psi_{i}^{(0)} \rangle + (\textrm{c.c.}) \big) ]  \label{eq:125a} \\
& + n_\textrm{s} \sum_{ij}^{pocc} \rho_{||,ji}^{(1)} \langle \psi_{i}^{(0)} | v_{\textrm{ext}}^{(1)} | \psi_{j}^{(0)} \rangle
 + \frac{1}{2} \int \int K_\textrm{Hxc}(\textbf{r},\textbf{r'}) \rho^{(1)}(\textbf{r}) \rho^{(1)}\textbf{(r')} d\textbf{r} d\textbf{r'} \label{eq:125c} \\
& - \frac{n_\textrm{s}}{2}  \sum_{ij}^{pocc'} \frac{\epsilon_{i}^{(0)}-\epsilon_{j}^{(0)}}{\occf_{i}^{(0)} - \occf_{j}^{(0)}} |\rho_{||,ij}^{(1)}|^{2} - \frac{n_\textrm{s}}{2} \sum_{i}^{{pocc}} \frac{\partial \epsilon}{\partial f} \Bigg|_{\occf_{i}^{(0)}} \cdot 
\big( \rho_{||,ii}^{(1)}
\big)^{2}  -n_\textrm{s} \mu^{(1)} \sum_{i}^{{pocc}} \rho_{||,ii}^{(1)} .\label{eq:125f}
\end{align}
\end{widetext}

The analysis of the extremal character of $F^{+(2)}$, started at the end of Sec.~\ref{sec:VarDFPTmetals}, can be pursued.
Indeed, the term quadratic in $\psi_{||,i}^{(1)}$ in Eq.~(\ref{eq:125a}),
\begin{align}
\langle \psi_{||,i}^{(1)} | \hat{H}^{(0)} - \epsilon_{i}^{(0)} | \psi_{||,i}^{(1)} \rangle,
\label{eq:184}
\end{align}
is obviously convex, since
 $\psi_{||,i}^{(1)}$ can be decomposed in the basis of eigenvectors of $\hat{H}^{(0)}$, and has only components with eigenenergies $\epsilon_{j}^{(0)}$ higher (or equal) to $\epsilon_{i}^{(0)}$ (see details in Sec.~S7 of the Supporting Information).

The combination of $K_{\textrm{Hxc}}$ with the positive-definiteness of Eq.~(\ref{eq:184}), and the ones of Eq.~(\ref{eq:125c}) and Eq.~(\ref{eq:125f}) (discussed at the end of Sec.~\ref{sec:VarDFPTmetals}) allows one to better understand the $F^{(2)}$ extremal character.
In any case, this property is also linked to the extremal character of the unperturbed $F$.

Let us now examine the equations at the minimum, in the parallel gauge. Some of them do not change:
Eq.~(\ref{eq:111}) is unchanged and $\rho^{(1)}$ is still obtained from Eq.~(\ref{eq:57d}).
The off-diagonal first-order density matrix gradients, see Eq.~(\ref{eq:114}), are simplified and deliver at the minimum : 
\begin{equation}
\label{eq:126}
    \rho_{||,ij}^{(1)} = \frac{\occf_{i}^{(0)}-\occf_{j}^{(0)}}{\epsilon_{i}^{(0)}-\epsilon_{j}^{(0)}} \langle \psi_{i}^{(0)} | \hat{H}^{(1)} | \psi_{j}^{(0)} \rangle.
\end{equation}

The projected gradient of $1^{st}$ order wavefunctions in the S$_{pocc}$ space, Eq.~(\ref{eq:116}), becomes

\begin{equation}
\label{eq:127}
    \Lambda_{ki}^{(1)} = \occf_{i}^{(0)} \langle \psi_{k}^{(0)} | \hat{H}^{(1)} | \psi_{i}^{(0)} \rangle + \rho_{||,ki}^{(1)} \epsilon_{k}^{(0)}.
\end{equation}

\subsection{The diagonal gauge}
\label{sec:diagonal-gauge}

Is it possible to choose a gauge where all the matrix elements $\rho^{(1)}_{ij}$ vanish ? Indeed, this would bring back the formalism for metals to the one found for gapped systems, 
without modification of the occupations.

Unfortunately, it is not possible to adjust the diagonal values of $\rho^{(1)}$ thanks to a choice of gauge. 
Indeed, whatever the gauge,
    \begin{equation}
    \label{eq:111bis}
        \rho_{ii}^{(1)} = \rho_{||,ii}^{(1)} = \frac{\partial f}{\partial \epsilon} \Bigg|_{\epsilon_{i}^{(0)}-\mu^{(0)}} (\epsilon_{i}^{(1)}-\mu^{(1)}).
    \end{equation}
If some states are partially occupied, 
$\frac{\partial f}{\partial \epsilon}$ does not vanish, and thus also $\rho_{ii}^{(1)}$ does not vanish (except possibly
due to symmetry reasons).
By contrast, for the non-diagonal elements, it is possible to impose
\begin{equation}
\label{eq:141}
    0 = \rho_{||,ji}^{(1)} - \frac{1}{2} A_{ji} (\occf_{j}^{(0)} - \occf_{i}^{(0)}).
\end{equation}
This choice will be called the diagonal gauge.
The relation between the diagonal and parallel gauge wavefunctions is

\begin{equation}
    \label{eq:145}
    | \psi_{\textrm{d}i}^{(1)} \rangle = | \psi_{||,i}^{(1)} \rangle - \sum_{j}^{pocc'} \frac{\langle \psi_{j}^{(0)} | \hat{H}^{(1)} | \psi_{i}^{(0)} \rangle}{\epsilon_{j}^{(0)} - \epsilon_{i}^{(0)}} | \psi_{j}^{(0)} \rangle.
\end{equation}

Then, $|\psi_{\textrm{d}i}^{(1)} \rangle$ fulfills

\begin{align}
\hat{P}_{\perp,i} \big(\hat{H}^{(0)}  - \epsilon_{i}^{(0)} \big) \hat{P}_{\perp,i} | \psi_{\textrm{d}i}^{(1)} \rangle = -  \hat{P}_{\perp,i} \hat{H}^{(1)} | \psi_{i}^{(0)} \rangle, \label{eq:146b}
\end{align} 

\indent that is, the Sternheimer equation, in the diagonal gauge. The notation $\hat{P}_{\perp,i}$
is for the projector on the space perpendicular to the unperturbed state $i$. 
Also, 
\begin{equation}
\label{eq:148}
\langle \psi_{\textrm{d}i}^{(1)} | \psi_{j}^{(0)} \rangle + \langle \psi_{i}^{(0)} | \psi_{\textrm{d}j}^{(1)} \rangle = 0.
\end{equation}

The second-order free energy can be computed in the diagonal gauge, and simplifies due to the constraint
Eq.~(\ref{eq:148}).
Eqs.~(\ref{eq:F+2})-(\ref{eq:F1ij}) become
\begin{widetext}
\begin{align}
F^{+(2)} [T, \{ \psi_{\textrm{d}i}^{(1)} \}, \{ \rho_{\textrm{d}ii}^{(1)} \} ] & = n_\textrm{s} \sum_{i}^{{pocc}} \occf_{i}^{(0)} \Bigg[ \langle \psi_{\textrm{d}i}^{(1)} | \hat{H}^{(0)} - \epsilon_{i}^{(0)} | \psi_{\textrm{d}i}^{(1)} \rangle \nonumber + \langle \psi_{i}^{(0)} | \hat{v}_{\textrm{ext}}^{(2)} | \psi_{i}^{(0)} \rangle + \bigg( \langle \psi_{\textrm{d}i}^{(1)} | \hat{v}_{\textrm{ext}}^{(1)} | \psi_{i}^{(0)} \rangle + (\textrm{c.c.}) \bigg) \Bigg] \nonumber \\
& + n_\textrm{s} \sum_{i}^{{pocc}} \rho_{\textrm{d}ii}^{(1)}  \langle \psi_{i}^{(0)} | \hat{v}_{\textrm{ext}}^{(1)} | \psi_{j}^{(0)} \rangle + \frac{1}{2} \int \int K_\textrm{Hxc}(\textbf{r},\textbf{r'}) \rho^{(1)}(\textbf{r}) \rho^{(1)}(\textbf{r'}) d\textbf{r} d\textbf{r'} \nonumber\\
& - \frac{n_\textrm{s}}{2} \sum_{i}^{{pocc}} \frac{\partial \epsilon}{\partial f} \Bigg|_{f_{i}^{(0)}} \big(
\rho_{\textrm{d}ii}^{(1)}
\big)^{2}   
-n_\textrm{s} \mu^{(1)} \sum_{i}^{{pocc}} \rho_{\textrm{d}ii}^{(1)} , \label{eq:147}
\end{align}
\end{widetext}

\indent with

\begin{align}
\rho^{(1)}(\textbf{r}) & = n_\textrm{s} \bigg[ \sum_{i}^{{pocc}} \rho_{\textrm{d}ii}^{(1)} \psi_{i}^{*(0)}(\textbf{r}) \psi_{i}^{(0)}(\textbf{r})  \nonumber \\
& + \occf_{i}^{(0)} \big( \psi_{\textrm{d}i}^{*(1)}(\textbf{r}) \psi_{i}^{(0)}(\textbf{r}) + \psi_{i}^{*(0)}(\textbf{r}) \psi_{\textrm{d}i}^{(1)}(\textbf{r}) \big) \bigg]. \label{eq:149}
\end{align}

Note the presence of only the diagonal elements of $\rho^{(1)}$ in both Eqs.~(\ref{eq:147}) and (\ref{eq:149}).

\subsection{Complete suppression of first-order occupation matrix elements}
\label{sec:suppress_f1}

The diagonal gauge is numerically inconvenient, because of the presence of the denominator $\epsilon_{j}^{(0)}-\epsilon_{i}^{(0)}$ in Eq.~(\ref{eq:145}), so that the corresponding term can become very large for small differences, while the contribution of pairs $ij$ and $ji$ will nearly cancel each other in Eq.~(\ref{eq:148}) and in Eqs.~(\ref{eq:149}). 
Also, one would prefer to use the same formula (hence the same coding) to build $\rho^{(1)}(\textbf{r})$ as in the case of insulator, with the only modification being the presence of occupation numbers:

\begin{align}
\rho^{(1)}(\textbf{r}) & = n_\textrm{s} \sum_{i}^{{pocc}} \occf_{i}^{(0)} 
\nonumber \\  & \bigg( \psi_{\textrm{mod},i}^{*(1)}(\textbf{r}) \psi_{i}^{(0)}(\textbf{r}) + \psi_{i}^{*(0)}(\textbf{r}) \psi_{\textrm{mod},i}^{(1)}(\textbf{r}) \bigg) \label{eq:153}
\end{align}

This can be achieved as follows.
Instead of Eq.~(\ref{eq:145}) one defines

\begin{align}
| \psi_{\textrm{mod},i}^{(1)} \rangle & = |\psi_{||,i}^{(1)} \rangle + \sum_{j}^{{pocc}} \Theta(\occf_{i}^{(0)},\occf_{j}^{(0)}) \nonumber \\
& \frac{\occf_{j}^{(0)} - \occf_{i}^{(0)}}{\occf_{i}^{(0)}} \frac{\langle \psi_{j}^{(0)} | \hat{H}^{(1)} | \psi_{i}^{(0)} \rangle}{\epsilon_{j}^{(0)} - \epsilon_{i}^{(0)}} \cdot | \psi_{j}^{(0)} \rangle, \label{eq:154}
\end{align}

\indent where $\Theta(\occf_{i}^{(0)},\occf_{j}^{(0)})$, to be defined later, is such that

\begin{align} 
& \Theta(\occf_{i}^{(0)},\occf_{j}^{(0)}) + \Theta(\occf_{j}^{(0)},\occf_{i}^{(0)}) = 1 \label{eq:155} \\
& \Theta(\occf_{i}^{(0)} = 0,\occf_{j}^{(0)}) = 0 \label{eq:156}
\end{align}

\indent This allows one to avoid the divergence in Eq.~(\ref{eq:154}). Note that  when $\occf_{i}^{(0)} = \occf_{j}^{(0)}$,

\begin{align}
\Theta(\occf_{i}^{(0)},\occf_{j}^{(0)}) = \frac{1}{2}. 
\end{align}

Also, in Eq.~(\ref{eq:154}), one has to understand that
\begin{align}
\frac{\occf_{j}^{(0)} - \occf_{i}^{(0)}}{\epsilon_{j}^{(0)} - \epsilon_{i}^{(0)}} = \frac{\partial f}{\partial \epsilon} \bigg|_{\epsilon_{i}^{(0)}}
\end{align}
when $\epsilon_{j}^{(0)} = \epsilon_{i}^{(0)}$.

In principle, the occupation numbers are positive, but this is broken in case of advanced smearing schemes.
So, the function $\Theta$ should be defined also outside of the $0 \leq \occf^{(0)} \leq 1$ range.

In the Supporting Information, Sec.~S6, it is
checked that
the condition expressed by Eq.~(\ref{eq:155}) insures that the computation of Eq.~(\ref{eq:153}) delivers the correct $\rho^{(1)}$, equal to the one obtained in the parallel gauge.
Similarly, one can show that the terms linear in $\psi_{i}^{(1)}$ and $\rho_{ji}^{(1)}$ in Eq.~(\ref{eq:125a}) are equivalent in the parallel gauge or with the modified wavefunctions.
By contrast, for the evaluation of $F^{+(2)}$, the terms quadratic in 
$\psi_{||,i}^{(1)}$ in Eq.~(\ref{eq:125a})
are not left invariant. Instead of correcting them, it is better to stick with the formula for $F^{+(2)}$ in the parallel gauge.

In ABINIT, the following $\Theta$ function is implemented : 

\begin{align}
    \Theta(\occf_{i},\occf_{j}) = H ( |\occf_{i}| - |\occf_{j}| ) ,\label{eq:164} 
\end{align}
where $H(x)$ is the Heaviside step function, with value 1/2 at $x=0$:

\begin{align}
    H(x) = \Bigg\{
    \begin{matrix}
        1   & x > 0 \\
       1/2  & x = 0 \\
        0   & x < 0 \\
    \end{matrix}
    \label{eq:163}
\end{align}

The advantage of this formulation, beyond satisfying Eq.~(\ref{eq:155}) and Eq.~(\ref{eq:156}) trivially, comes from the fact that the sum $\sum_{j}^{{pocc}}$ in Eq.~(\ref{eq:154}) includes only the wavefunctions $| \psi_{j}^{(0)} \rangle$ with absolute occupation lower than the one of $| \psi_{i}^{(0)} \rangle$, that translates usually (when $f(\epsilon)$ is a monotonically decreasing function of $\epsilon$, bounded by 0 and 1) 
into energy $\epsilon_{j}^{(0)}$ higher than $\epsilon_{i}^{(0)}$.
This yields some CPU time saving, about a factor of two in that operation, instead of doing the sum $\sum_{j}^{{pocc}}$ on all states.

In practice, the parallel gauge first-order wavefunctions $| \psi_{||,i}^{(1)} \rangle$ are computed, at fixed $\hat{H}^{(1)}$, and then $| \psi_{\textrm{mod},i}^{(1)} \rangle$ is computed, that allows afterwards to compute $\rho^{(1)}$.
The computation of the second-order free energy can be done using the parallel gauge formula Eqs.~(\ref{eq:125a})-(\ref{eq:125f}), that is variational.


\section{Periodic systems}
\label{Sec:Periodic}


Although the occupation numbers and the density matrix have been explicitly treated, the DFT and DFPT formulas presented until now are valid for the case of finite systems, with a set of discretized levels where occupation number varies with temperature according to the Fermi-Dirac statistics. 
Systems are now treated with lattice periodicity, hence corresponding to the case of extended metals. The above theory is adapted to such case, with treatment of Brillouin Zone integral, and the appearance of a continuous band structure as a function of the wavevector.
Notations are obvious adaptations to the metallic case of those from Ref.~\onlinecite{Gonze1997}, Appendix A.
One focuses first on DFT then on DFPT.

\subsection{DFT for metallic periodic systems}
\label{sec:Metals-GS}

The DFT electronic free energy per unit cell writes
\begin{widetext}
\begin{eqnarray}
F[T;
\{u_{n
\textbf{k}}
\},\{\rho_{nm
\textbf{k}}
\}
]
&=& 
\frac{n_{\textrm{s}} \Omega_0}{(2\pi)^3}
\int_{\textrm{BZ}}
\sum_{nm}\rho_{nm\textbf{k}} \langle u_{m\textbf{k}} | \hat{K}_{\textbf{k}\textbf{k}} + \hat{v}_{\textrm{ext},\textbf{k}\textbf{k}} | u_{n\textbf{k}} \rangle
d\textbf{k}
+
E_{\textrm{Hxc}}[\rho] - TS[\{\rho_{nm\textbf{k}}\}].
\label{eq:F_metal}
\end{eqnarray}
\end{widetext}
This is a generalization of Eq.~(\ref{eq:free_energy})
to periodic solids. 
The matrix element of the kinetic operator and external potential operator is evaluated over the primitive cell with volume $\Omega_0$.
The Hartree and exchange-correlation energy $E_{\textrm{Hxc}}[\rho]$
is also evaluated for one primitive cell. Similarly for the entropy. 
The $u$ are periodic parts of Bloch wavefunctions.
The wavevector $\textbf{k}$ integral is performed over the Brillouin Zone, with volume $\frac{(2\pi)^3}{\Omega_0}$.
$n$ and $m$ are band indices.
The expression of the electronic density is 
\begin{eqnarray}
\rho(\textbf{r}) = 
\frac{n_{\textrm{s}}}{(2\pi)^3}
\int_{\textrm{BZ}}
\sum_{nm}\rho_{nm\textbf{k}}
u_{m\textbf{k}}^*(\textbf{r})
u_{n\textbf{k}}(\textbf{r})d\textbf{k}.
\label{eq:n_metal}
\end{eqnarray}
In Eq.~(\ref{eq:F_metal}) and~(\ref{eq:n_metal}), the wavefunctions are normalized as follows
\begin{equation}
\langle u_{m\textbf{k}} | u_{n\textbf{k}} \rangle=
\frac{1}{\Omega_0}
\int_{\Omega_0}
u_{m\textbf{k}}(\textbf{r})^*
u_{n\textbf{k}}(\textbf{r})
d\textbf{r}=\delta_{mn}.
\label{eq:ortho}
\end{equation}

The Hamiltonian and occupation matrix can be simultaneously diagonalized, as in the discrete situation, with
\begin{eqnarray}
\hat{H}_{\textbf{k}\textbf{k}} | u_{n\textbf{k}} \rangle &=&
\big( \hat{K}_{\textbf{k}\textbf{k}} + \hat{v}_{\textrm{ext},\textbf{k}\textbf{k}} 
+ \hat{v}_{\textrm{Hxc},\textbf{k}\textbf{k}}[\rho] 
\big)
| u_{n\textbf{k}} \rangle
\nonumber\\
&=& \epsilon_{n\textbf{k}} | u_{n\textbf{k}} \rangle.
\label{eq:KohnSham_metal}
\end{eqnarray}
Minimization of the free energy yields the same relationship between eigenenergy and occupation number than in the discrete case, Eq.~(\ref{eq:f_i_from_spm1}).

In this diagonal gauge, the Brillouin Zone integral entering the electronic density can be transformed to an energy integral, as follows.
The energy-resolved electronic density is defined as
\begin{eqnarray}
\rho(\textbf{r},\epsilon) = 
\frac{n_{\textrm{s}}}{(2\pi)^3}
\int_{\textrm{BZ}}
\sum_{n}\delta(\epsilon-\epsilon_{n\textbf{k}})
u_{n\textbf{k}}^*(\textbf{r})
u_{n\textbf{k}}(\textbf{r})d\textbf{k},
\nonumber\\
\label{n_e_metal}
\end{eqnarray}
such that
\begin{eqnarray}
\rho(\textbf{r}) = 
\int_{-\infty}^{+\infty}
f\Big( (\mu-\epsilon)/kT \Big)
\rho(\textbf{r},\epsilon)
d\epsilon . 
\label{eq:n_metal_int_e}
\end{eqnarray}

\subsection{DFPT for metallic periodic systems}
\label{sec:Metals-DFPT}

DFPT for periodic systems allows one to treat perturbations
that are characterized by a wavevector $\textbf{q}$: like
Bloch wavefunctions, they have a periodic part, and a phase.
In Sec.~IV of Ref.~\onlinecite{Gonze1997}, the strategy
to deal with such generic perturbations is explained,
and involves factorizing the phase in all DFPT equations.
We keep the same notations as in this reference, 
and proceed with the
systematic generalization of the quantities
developed in the DFPT for varying occupations, as obtained in the
previous sections, for the parallel gauge case. The generalization to other gauges proceeds in a similar way.

Starting with first-order quantities, one finds that
Eq.~(\ref{eq:124}) becomes
    \begin{align}
    \langle u_{m\textbf{k}+\textbf{q}}^{(0)} | u_{||,n\textbf{k},\textbf{q}}^{(1)} \rangle = 0  \quad [ m,n \in \textrm{S}_{pocc}], \label{eq:173}    
    \end{align}
    \indent that is similar to Eq.~(43) of 
    Ref.~\onlinecite{Gonze1997}.
    For Eq.~(\ref{eq:126}), one defines 
    \begin{align}
     \epsilon_{m\textbf{k}+\textbf{q},n\textbf{k}}^{(1)} = \langle u_{m\textbf{k}+\textbf{q}}^{(0)} | \hat{H}_{\textbf{k}+\textbf{q},\textbf{k}}^{(1)} | u_{n\textbf{k}}^{(0)} \rangle, \label{eq:174}
    \end{align}
    \indent then 
    \begin{align}
        \rho_{||,m\textbf{k}+\textbf{q},n\textbf{k}}^{(1)} = \frac{f_{m\textbf{k}+\textbf{q}}^{(0)}-f_{n\textbf{k}}^{(0)}}{\epsilon_{m\textbf{k}+\textbf{q}}^{(0)} - \epsilon_{n\textbf{k}}^{(0)}} \epsilon_{m\textbf{k}+\textbf{q},n\textbf{k}}^{(1)}.\label{eq:175}
    \end{align}
Eq.~(\ref{eq:57d})
becomes (see Eq.~(44) of Ref.~\onlinecite{Gonze1997}):
\begin{widetext}
    \begin{equation}
        \Bar{\rho}_{\textbf{q}}^{(1)}(\textbf{r})  
        = \frac{1}{(2 \pi)^{3}} \int_{\textrm{BZ}} n_\textrm{s}\Bigg[ \sum_{nm}^{{pocc}} \rho_{||,m\textbf{k}+\textbf{q},n\textbf{k}}^{(1)} \cdot u_{n\textbf{k}}^{*(0)}(\textbf{r}) u_{m\textbf{k}+\textbf{q}}^{(0)}(\textbf{r})
        + 2 \sum_{m}^{{pocc}} f_{m\textbf{k}}^{(0)} u_{m\textbf{k}}^{*(0)}(\textbf{r}) u_{m\textbf{k},\textbf{q}}^{(1)}(\textbf{r}) \Bigg] 
        d\textbf{k}
        \label{eq:176}
    \end{equation}

where $\Bar{\rho}_{\textbf{q}}^{(1)}(\textbf{r})$ is the periodic part
of the first-order density change.
The Sternheimer equation in the periodic case, coming 
from Eq.~(\ref{eq:118b}) is
\begin{equation}
\label{eq:118b_periodic}
\hat{P}_{\perp \textbf{k}+\textbf{q}} (\hat{H}_{\textbf{k}+\textbf{q},\textbf{k}+\textbf{q}}^{(0)} 
- \epsilon_{m\textbf{k}}^{(0)}) 
\hat{P}_{\perp \textbf{k}+\textbf{q}} | u_{||,m\textbf{k},\textbf{q}}^{(1)} \rangle = - \hat{P}_{\perp \textbf{k}+\textbf{q}} \hat{H}_{\textbf{k}+\textbf{q},\textbf{k}}^{(1)} | u_{m\textbf{k}}^{(0)} \rangle,
\end{equation}
where
\begin{equation}
\label{eq:H1_periodic}
\hat{H}_{\textbf{k}+\textbf{q},\textbf{k}}^{(1)}=
\hat{v}_{\textrm{ext},\textbf{k}+\textbf{q},\textbf{k}}^{(1)}
+
\int
K_{\textrm{Hxc}}[\rho](\textbf{r},\textbf{r}')
\bar{\rho}^{(1)}_\textbf{q}(\textbf{r}')
e^{-i\textbf{q}(\textbf{r}-\textbf{r}')}
d\textbf{r}'.
\end{equation}
\end{widetext}

Eqs. (\ref{eq:118b_periodic}) and (\ref{eq:H1_periodic}),
respectively, can be compared with Eqs.~(45) and (46)
of Ref.~\onlinecite{Gonze1997}, respectively. The two Sterneimer equations are
identical, while the definition of $\hat{H}_{\textbf{k}+\textbf{q},\textbf{k}}^{(1)}$ is similar, although in Ref.~\onlinecite{Gonze1997} an additional
term is also coming from a possible dependence of the Hartree and exchange-correlation potential on the perturbation, neglected in the present account for sake of simplicity, but
implemented in ABINIT.

Let us now examine the second-order free energy $F^{+(2)}$.
For a non-periodic perturbation, \textit{i.e.} $\textbf{q} \neq 0$, all diagonal elements of $\epsilon^{(1)}$ or $\rho_{||}^{(1)}$ vanish.
For such case, Eq.(\ref{eq:125f})
becomes

\begin{widetext}
\begin{align}
F_{el,-\textbf{q},\textbf{q}}^{+(2)} [T, \{ u_{||}^{(1)} \}, \{ \rho_{||}^{(1)} \} ] & = \frac{\Omega_{0}}{(2 \pi)^{3}} \int_{\textrm{BZ}} n_\textrm{s} \Bigg[ \sum_{m}^{{pocc}} f_{m\textbf{k}}^{(0)} F^{(2)}_{m\textbf{k}}[u_{||}^{(1)}]
+\frac{1}{2} \sum_{mn}^{{pocc}} \bigg( \rho_{||,n\textbf{k},m\textbf{k}+\textbf{q}}^{(1)} \langle u_{m,\textbf{k}+\textbf{q}}^{(0)} | \hat{v}_{\textrm{ext},\textbf{k}+\textbf{q},\textbf{k}}^{(1)} | u_{n\textbf{k}}^{(0)} \rangle + (\textrm{c.c.}) \bigg) \Bigg] d\textbf{k} \nonumber \\
& + \frac{1}{2} \int_{\Omega_{0}} \int K_\textrm{Hxc}(\textbf{r},\textbf{r'}) \Bar{\rho}_{\textbf{q}}^{*(1)}(\textbf{r}) \Bar{\rho}_{\textbf{q}}^{(1)}(\textbf{r'}) e^{-i\textbf{q}(\textbf{r}-\textbf{r'})}d\textbf{r} d\textbf{r'} \nonumber \\
& - \frac{n_\textrm{s}}{2} \frac{\Omega_{0}}{(2 \pi)^{3}} \int_{BZ} \sum_{mn}^{{pocc}} \frac{\epsilon_{m,\textbf{k}+\textbf{q}}^{(0)} - \epsilon_{n\textbf{k}}^{(0)}}{f_{m,\textbf{k}+\textbf{q}}^{(0)} - f_{n\textbf{k}}^{(0)}} |\rho_{||,m\textbf{k}+\textbf{q},n\textbf{k}}^{(1)}|^{2}
d\textbf{k}, \label{eq:177}
\end{align}
where
\begin{align}
    F^{(2)}_{m\textbf{k}}[u_{||}^{(1)}] = 
\langle u_{||,m\textbf{k},\textbf{q}}^{(1)} | \hat{H}_{\textbf{k}+\textbf{q},\textbf{k}+\textbf{q}}^{(0)} 
- \epsilon_{m\textbf{k}}^{(0)}
| u_{||,m\textbf{k},\textbf{q}}^{(1)} \rangle + \langle u_{m\textbf{k}}^{(0)} | \hat{v}_{\textrm{ext},\textbf{k},\textbf{k}}^{(2)} | u_{m\textbf{k}}^{(0)} \rangle
+ \bigg( \langle u_{||,m\textbf{k},\textbf{q}}^{(1)} | \hat{v}_{\textrm{ext},\textbf{k}+\textbf{q},\textbf{k}}^{(1)} | u_{mk}^{(0)} \rangle + (\textrm{c.c.}) \bigg).
\label{eq:F2mk}
\end{align}
\end{widetext}

Eqs.~(\ref{eq:177}) and (\ref{eq:F2mk}) can be compared with Eq.~(42) of Ref.~\onlinecite{Gonze1997}. In the latter, three additional
terms also come from a possible dependence of the Hartree and exchange-correlation potential on the perturbation, also not included in the present account, like in the equation for the 
first-order Hamiltonian. Also, the dependence of $F_{el,-\textbf{q},\textbf{q}}^{+(2)}$ on $\textbf{q}$ is not mentioned in the present Eq.~(\ref{eq:177}), for sake of simplicity. By the same token, the $\textbf{q}$
dependence is also not indicated for the
second-order $F^{(2)}_{m\textbf{k}}$. 
By contrast,
$\hat{v}_{\textrm{ext},\textbf{k},\textbf{k}}^{(2)}$ has no 
$\textbf{q}$ dependence, see 
the Eq.~(49) of Ref.~\onlinecite{Gonze1997}.

The commensurate perturbation case, that is, either $\textbf{q} = 0$, or $\textbf{q}$ is a vector of the reciprocal lattice,
is quite similar to the case of finite systems, so the explicit formula is obvious and will not be written down here.


\section{Applications}
\label{Sec:Applications}

As mentioned in the introduction, there have been many different applications of the formalism presented in the previous sections. 
However, in such studies, usually, results have been presented with little or no emphasis on understanding and characterizing the convergence characteristics with respect to the
temperature (or with respect to the smearing energy) jointly with
the sampling of the Brillouin Zone.
Interestingly, the target precision 
of the calculation, or its purpose, is seen to play an important role
for the definition of the convergence
regime.

In the following, the phonon frequencies
of copper, at the X point in the Brillouin Zone, for both transverse and longitudinal 
modes are taken as examples. 
The PBE (Perdew-Burke-Ernzerhof) exchange-correlation functional is used, 
with 
the optimized norm-conserving vanderbilt pseudopotential\cite{Hamann2013}
from the \verb|Pseudo-Dojo|\cite{VanSetten2018}, and
an energy cutoff of 46.0 Ha.
Bulk copper metal is FCC, with optimized lattice parameter 3.63 \AA{} for the
conventional cell edge.
Calculations have been done with ABINIT v9.8.3. 
 
Computing the phonon frequencies is often
done with a target of 1 cm$^{-1}$. 
This will be our reference target indeed for this property.
Such precision is not difficult to reach, 
and corresponds to a range of parameters
that might be called ``medium precision''. 
However, one might also be interested in the examination of the specific change of
phonon frequencies as a function of electronic temperature.
The changes are much smaller, 
and it is much more demanding to obtain reliably such temperature dependence.
This regime is called ``high precision''.

\begin{figure}
\begin{subfigure}{0.45\textwidth}
\centering
 \includegraphics[width=\linewidth]{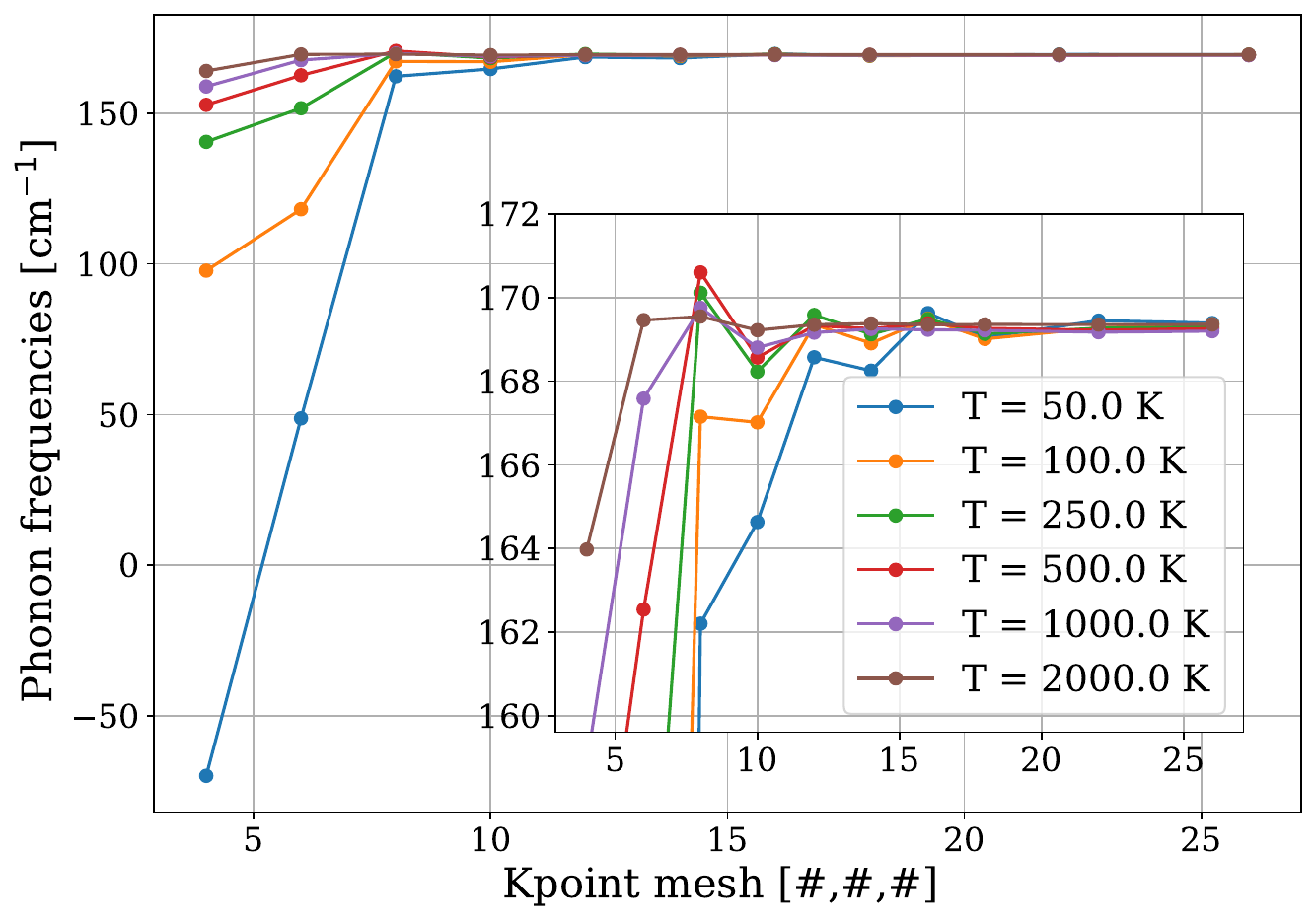}
\caption{Transverse phonon frequencies}
 \end{subfigure}
 \hfill
\begin{subfigure}{0.45\textwidth}
\centering
 \includegraphics[width=\linewidth]{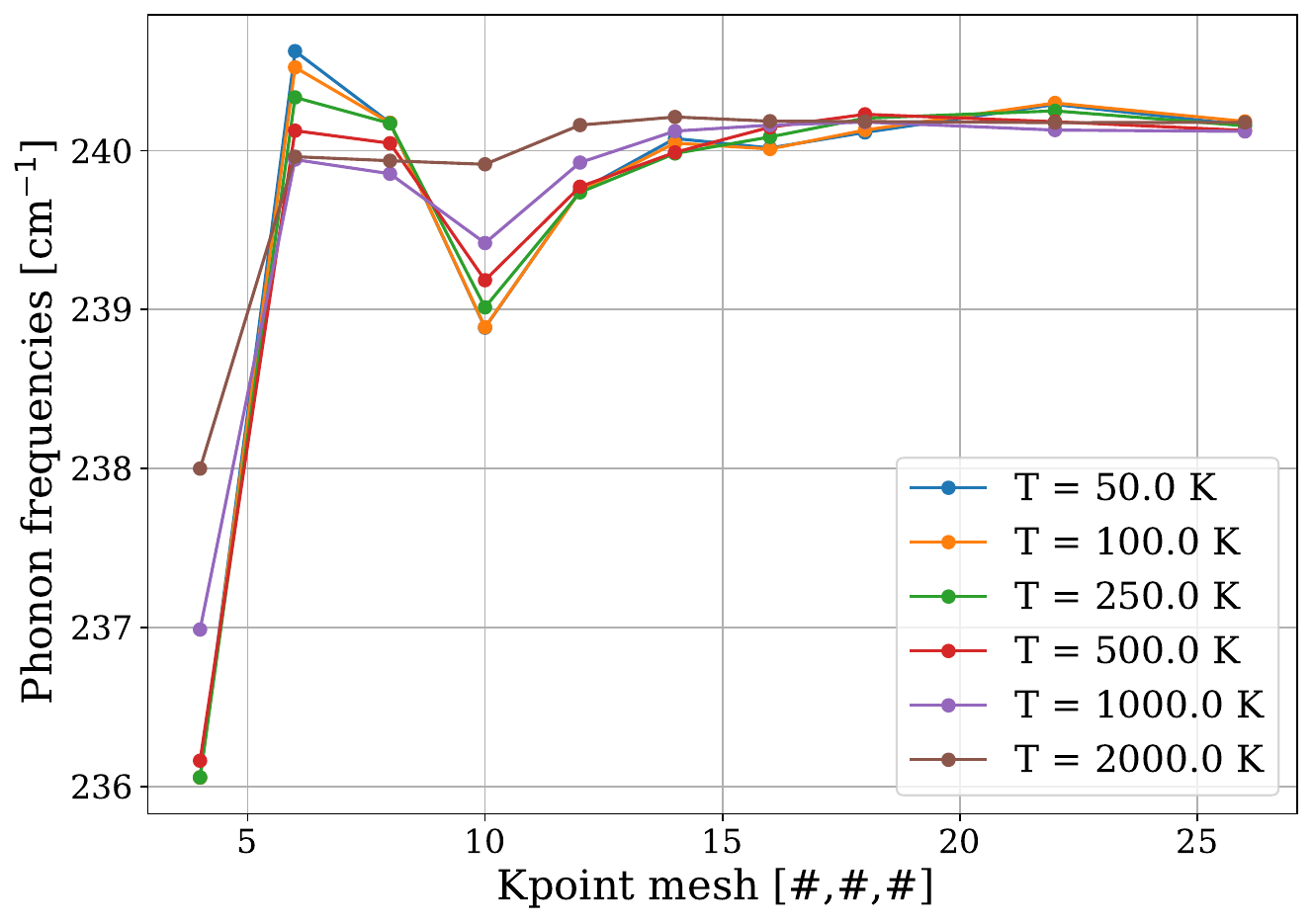}
 \caption{Longitudinal phonon frequencies }
 \end{subfigure}
 \caption{Phonon frequencies 
 as a function of the
 wavevector mesh 
 (linear discretization factor), using the Fermi-Dirac statistics, across a range of physical electronic temperatures \hbox{(50 K - 2000 K)}. The inset provides a closer view of the 160-172 cm$^{-1}$ range.}
\label{fig:phononfreq_FD}
\end{figure}

In Fig.~\ref{fig:phononfreq_FD}, the phonon frequencies are presented, as a function of the discreteness 
of the grid used to sample the Brillouin Zone. 
One targets the estimation of the
phonon frequencies at zero Kelvin.
The Fermi-Dirac broadening is used
here only for the purpose of 
alleviating some of the numerical burden.

The precision obtained for the longitudinal
and the transverse frequencies, for the
same parameters of the computation is quite different. 
With the coarse \hbox{4 $\times$ 4 $\times$ 4} mesh
at the lowest temperature (50 K), 
the longitudinal frequency (lower panel) is not so bad, and already close to the target precision of 1 cm$^{-1}$. At variance,
for such coarse grid,
the
transverse frequency is hardly significant.
Moreover, 
the computation of the low-temperature phonon frequencies actually does not need a low temperature: indeed, for the larger grid used in Fig.~\ref{fig:phononfreq_FD},
\hbox{26 $\times$ 26 $\times$ 26},
it is seen that the effect of the temperature is very small:
going from 50 K
to 2000 K modifies the phonon frequencies by much less than 1 cm$^{-1}$ for such grid.
Hence, the large smearing temperature
of 2000 K can be used 
for the coarser grids, the 
low-temperature phonon frequency is obtained
well within the target 
precision of 1 cm$^{-1}$

Being more quantitative, with a small broadening temperature of 50 K, one needs 
a \hbox{16 $\times$ 16 $\times$ 16} wavevector grid
to reach the target precision for the 
transverse mode frequency
(see the inset), while for a broadening
temperature of 2000 K, the same precision is obtained with a \hbox{8 $\times$ 8 $\times$ 8} wavevector grid. 
This amounts to a large
saving of computational resources.
Computing time and memory (or disk space) scale indeed
linearly with the number of wavevectors
in the Brillouin Zone. Hence the speed-up
obtained by using the coarser grid instead
 of the fine grid is about order of magnitude.

\begin{figure}
\begin{subfigure}{0.45\textwidth}
\centering
 \includegraphics[width=\linewidth]{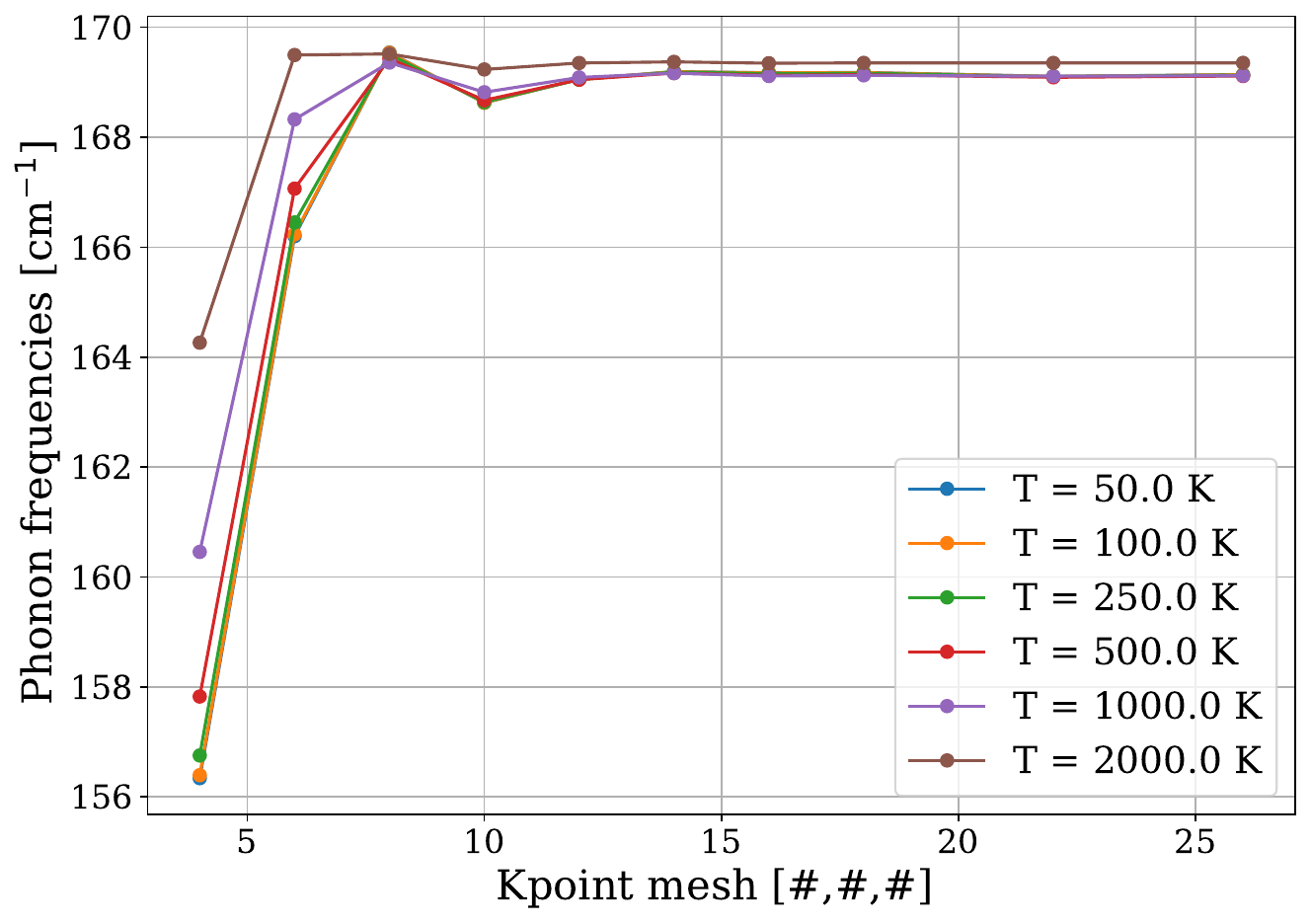}
\caption{Transverse phonon frequencies}
 \end{subfigure}
 \hfill
\begin{subfigure}{0.45\textwidth}
\centering
 \includegraphics[width=\linewidth]{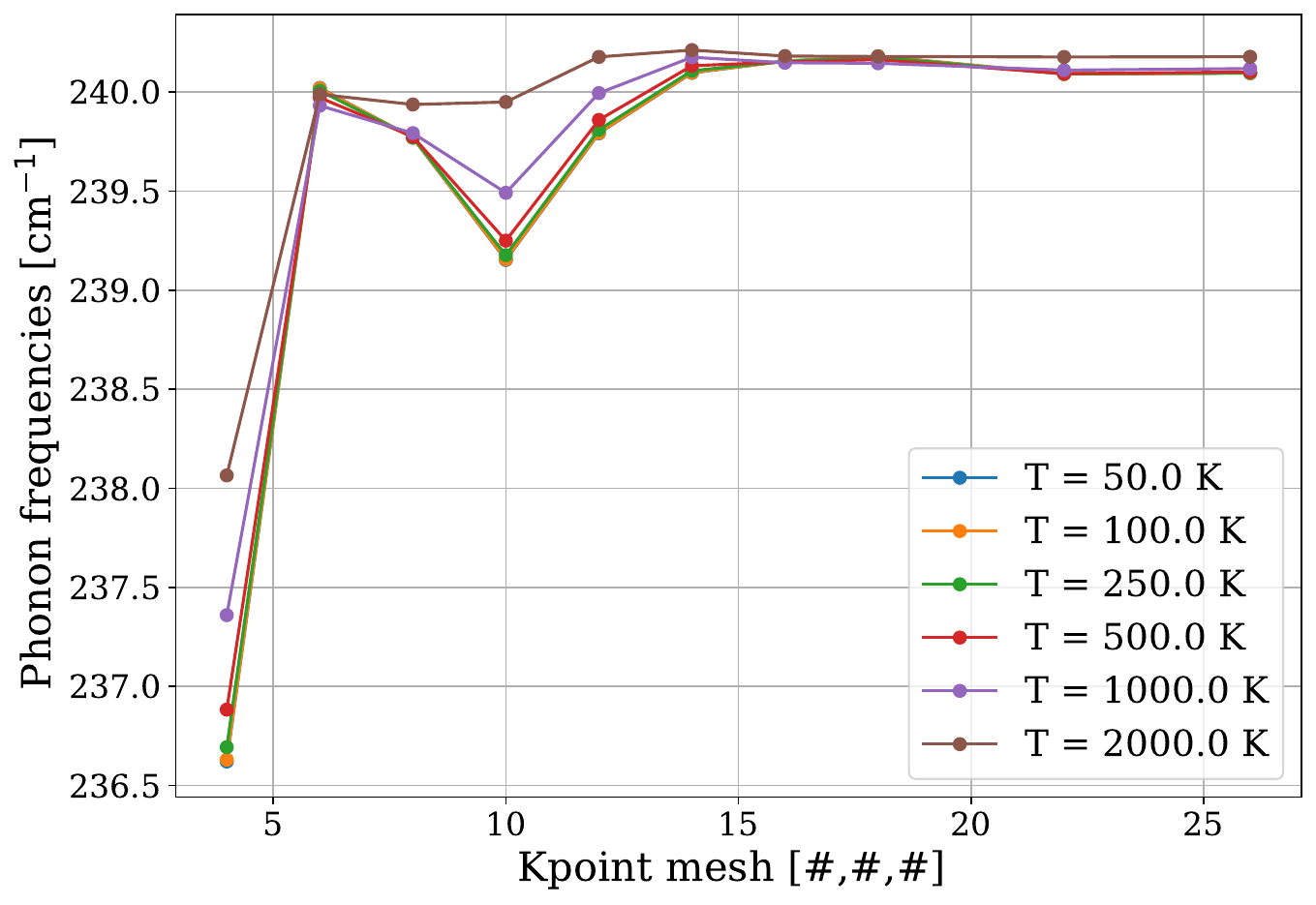}
 \caption{Longitudinal phonon frequencies}
 \end{subfigure}
 \caption{Phonon frequencies 
  as a function of the
 wavevector mesh (linear discretization factor) obtained with the resmearing scheme (broadening value 3000 K), across physical electronic temperatures from 50 K to 2000 K.}
\label{fig:phononfreq_resmearing}
\end{figure}

Fig.~\ref{fig:phononfreq_resmearing} presents
results obtained with the resmearing scheme (Fermi-Dirac statistics and Methfessel-Paxton smearing) where the MP smearing parameters corresponds to a temperature of 3000 K. 
One sees that irrespective of the
physical electronic temperature value,
the phonon frequencies are converged within the target value for 
a \hbox{8 $\times$ 8 $\times$ 8} wavevector grid. 

Let us now turn to the ``high-precision regime'', for which the target is to obtain the change of phonon frequencies as a function of the physical electronic temperature.
Using \hbox{4 $\times$ 4 $\times$ 4} wavevector grid,
\hbox{8 $\times$ 8 $\times$ 8} wavevector grid
or \hbox{16 $\times$ 16 $\times$ 16} wavevector grids
does not yield meaningful temperature
dependence of these phonon frequencies. 
Such a temperature dependence can be obtained with much finer grids,
\hbox{30 $\times$ 30 $\times$ 30}
or even \hbox{42 $\times$ 42 $\times$ 42} (for the latter, see the Supporting Information Sec.~S8).

\begin{figure}
\begin{subfigure}{0.45\textwidth}
\centering
 \includegraphics[width=\linewidth]{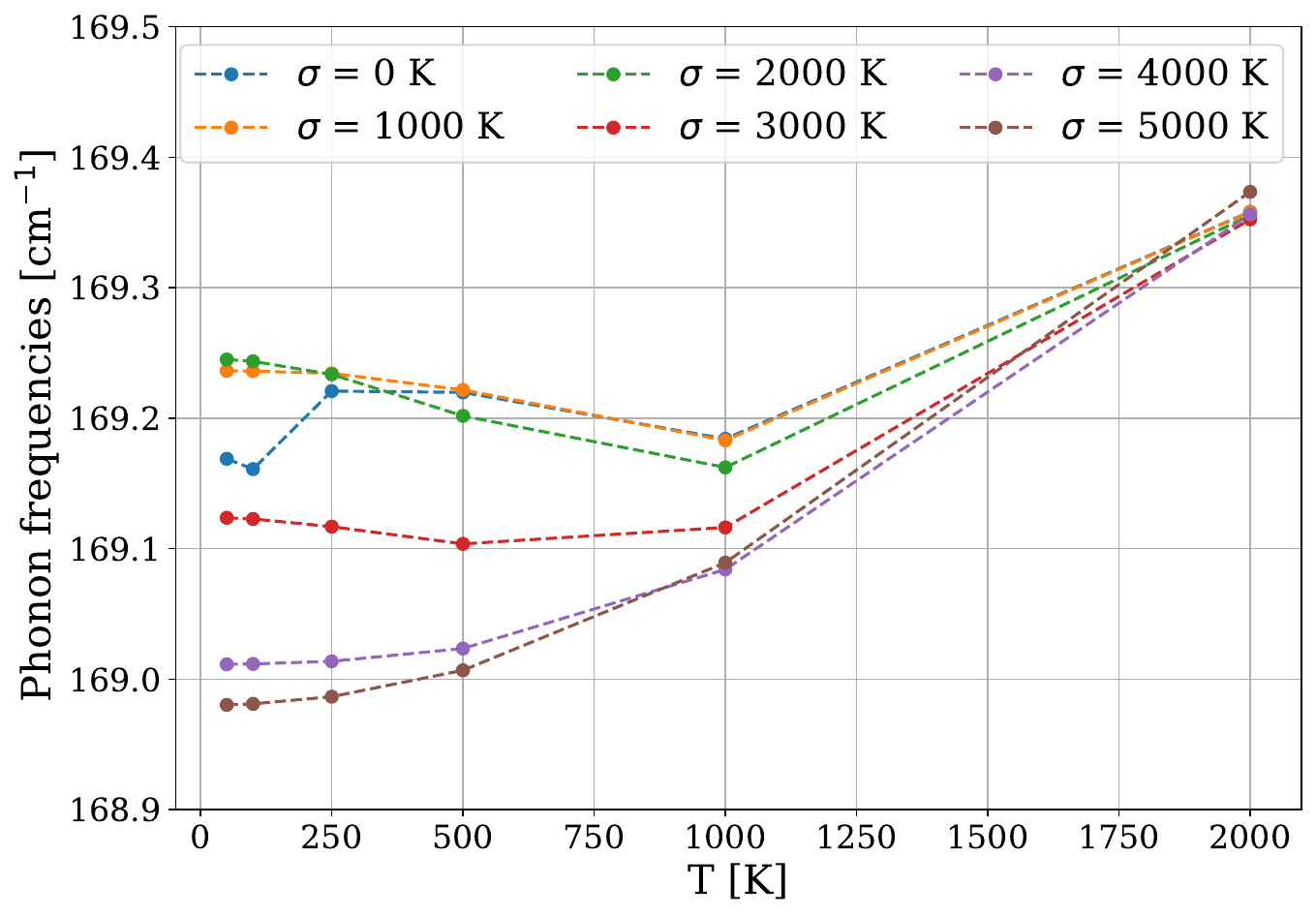}
\caption{Transverse phonon frequencies}
 \end{subfigure}
 \hfill
\begin{subfigure}{0.45\textwidth}
\centering
 \includegraphics[width=\linewidth]{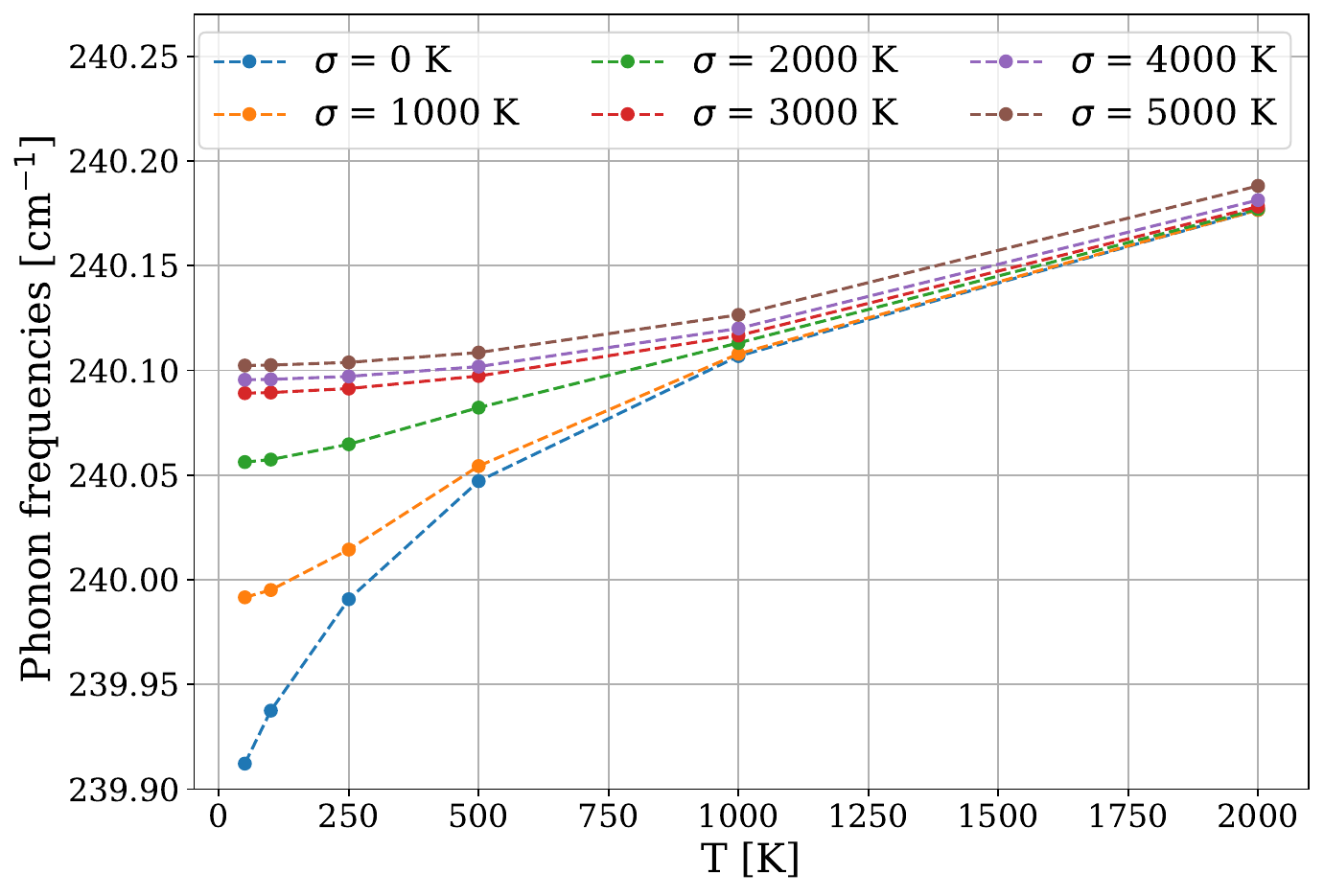}
 \caption{Longitudinal phonon frequencies}
 \end{subfigure}
 \caption{Phonon frequencies 
 obtained from various MP broadening values (0 - 5000 K), 
  as a function of physical electronic temperatures (50 K - 2000 K), obtained with a 30$\times$30$\times$30 wavevector grid.}
\label{fig:phononfreq_highprecision}
\end{figure}

Fig.~\ref{fig:phononfreq_highprecision} presents
the phonon frequencies as a function of the physical
temperature, for different values of the MP broadening
parameter, again for the transverse as well as longitudinal phonon modes. 
The scale of this figure
is quite different than the one of the previous
figures. 
Indeed, the change of phonon frequencies from a low temperature to the highest temperature (2000 K)
is on the order of 0.5 cm$^{-1}$ for the frequency
of the transverse
mode and even smaller for the frequency of the 
longitudinal mode, as seen previously.
Thus, the target precision must be much smaller
as well.
Having in mind the description of the global behavior, one sees that for MP broadening smaller than 3000-4000 K, at small physical electronic temperature,
there are considerable deviations from the expected
parabolic behavior, for this 
very fine \hbox{30 $\times$ 30 $\times$ 30} wavevector grid.
In the Sec.~S8 of the Supporting Information an even finer  
\hbox{42 $\times$ 42 $\times$ 42} grid is
used.
However, without MP broadening, the behavior
is not guaranteed to be even qualitatively correct.
We have not pushed beyond such
\hbox{42 $\times$ 42 $\times$ 42} grid.

\section{Underconverged ground-state wavefunctions}
\label{Sec:underconvergedGSwfs}

Until now, all the formulas in DFPT assume that the unperturbed wavefunctions $\psi_{i}^{(0)}$ are ``perfect" solutions to the unperturbed Schr\"{o}dinger equation.
In practice, while the occupied ones are usually excellent indeed, the unoccupied ones can be loosely converged, since they do not contribute to the ground-state unperturbed total energy
or to the density. Actually, they might be more difficult to converge than the lower lying ones, especially if there is a degeneracy between the highest state in the partially occupied space and the lowest state outside of it.

However, in DFPT, slightly incorrect $\psi_{i}^{(0)}$ in the partly occupied space (or even $\psi_{i}^{(0)}$ associated with vanishing occupations) will induce proportional errors in $F^{(2)}$.
This can be seen and quantified, as shown hereafter, analytically in a simple model, as well as numerically.

Let us first examine a three-state model, in the non-interacting case.
The three exact eigenstates are denoted 
$|\psi_{1}^{(0)} \rangle$, $|\psi_{2}^{(0)} \rangle$ and $|\psi_{3}^{(0)} \rangle$ with exact eigenvalues $\epsilon_{1}^{(0)}$, $\epsilon_{2}^{(0)}$ and $\epsilon_{3}^{(0)}$.

The first state occupation number is $1-\delta f$ where $\delta f$ is not very large, still finite, while the second state occupation number is $\delta f$, and the third state is unoccupied.
This is the S$_{pocc}$ space of the problem.

The ground-state total energy of this independent-particle system, taking into account the spin degeneracy, as done in the previous sections, is 

\begin{align}
    E^{(0)} = n_\textrm{s} \big[ (1 - \delta f) \epsilon_{1}^{(0)} + \delta f \epsilon_{2}^{(0)}  \big]. \label{eq:191}
\end{align}

The perturbation couples the different states, with matrix elements
denoted

\begin{align}
    H_{ij} = \langle \psi_{i}^{(0)} | \hat{H}^{(1)} | \psi_{j}^{(0)} \rangle. \label{eq:190} 
\end{align}

The computation of $E^{(2)}$ gives
\begin{widetext}
\begin{align}
  E^{(2)} & = \frac{n_\textrm{s}}{2} \sum_{i \neq j} \frac{f_{i}^{(0)} - f_{j}^{(0)}}{\epsilon_{i}^{(0)} - \epsilon_{j}^{(0)}} |\langle \psi_{i}^{(0)}| \hat{H}^{(1)} | \psi_{j}^{(0)} \rangle |^{2} \nonumber \\
  & = n_\textrm{s} \Bigg[ \frac{2\delta f - 1}{\epsilon_{2}^{(0)} - \epsilon_{1}^{(0)}} |H_{12}|^{2}
  - \frac{1 - \delta f}{\epsilon_{3}^{(0)} - \epsilon_{1}^{(0)}} |H_{13}|^{2} 
  - \frac{\delta f}{\epsilon_{3}^{(0)} - \epsilon_{2}^{(0)}} |H_{23}|^{2} \Bigg], \label{eq:192}
\end{align}
\end{widetext}
with the hypothesis that the occupation numbers are frozen (this hypothesis might be removed, and does not affect the final proportionality relation).

Now let us suppose that the ground-state Schr\"{o}dinger equation has not been solved exactly, but approximately, so that there is a small contamination of the second eigenvector $| \psi_{2}^{(0)} \rangle $ by the third eigenvector $| \psi_{3}^{(0)} \rangle $ and vice-versa. 
The ``contaminated'' quantities are denoted with a tilde. This 
contamination is determined by the admixture angle $\alpha$, that should be small:

\begin{align}
| \tilde\psi_{2}^{(0)} \rangle = \cos{\alpha} |\psi_{2}^{(0)} \rangle + \sin{\alpha} | \psi_{3}^{(0)} \rangle. \label{eq:193}
\end{align}
Similarly the third eigenvector is contaminated by
$|\psi_{2}^{(0)} \rangle $, and both contaminated vectors are kept orthogonal:

\begin{align}
| \tilde\psi_{3}^{(0)} \rangle = -\sin{\alpha} |\psi_{2}^{(0)} \rangle + \cos{\alpha} | \psi_{3}^{(0)} \rangle. \label{eq:194}
\end{align}

The error in the second eigenvector is quantified in term of its ``residual" $| R_{2} \rangle$ defined as
\begin{align}
| R_{2} \rangle & = (\hat{H}^{(0)} - \tilde\epsilon_2^{(0)}) | \tilde\psi_2^{(0)} \rangle, \label{eq:196}
\end{align}
where
\begin{align}
\tilde\epsilon_2^{(0)} & = \langle \tilde\psi_2^{(0)} | \hat{H}^{(0)} | \tilde\psi_2^{(0)} \rangle.
\label{eq:195}
\end{align}
The norm of the residual vector (or its square) of an approximate 
eigenvector is a common measure of the convergence of a solution of the Schr\"odinger
equation.

After some intermediate calculation (see the Sec.~S9 of the Supporting Information), the squared norm $R^2$ of the residual of the second eigenvector is obtained :
\begin{align}
R^2_2=\langle R_{2} | R_{2} \rangle = \cos^{2}{\alpha} \sin^{2}{\alpha} \cdot (\epsilon_{3}^{(0)} - \epsilon_{2}^{(0)})^{2}. \label{eq:197}
\end{align}
It is proportional to $\sin^{2}{\alpha}$, hence the square of the admixture angle when the latter is small.

Then, the ``contaminated", approximate, second-order derivative of the energy is computed, starting from
\begin{align}
    \tilde{E}^{(2)} = \frac{n_\textrm{s}}{2} \sum_{i \neq j}  \frac{f_{i}^{(0)} - f_{j}^{(0)}}{\tilde\epsilon_i^{(0)} - \tilde\epsilon_j^{(0)}}  |\langle \tilde\psi_i^{(0)}| \hat{H}^{(1)} | \tilde\psi_j^{(0)} \rangle |^{2}. \label{eq:198}
\end{align}

\indent For the three-state model, $\tilde\epsilon_1^{(0)} = \epsilon_{1}^{(0)}$ (no contamination), $\tilde\epsilon_2^{(0)}$ is given by
Eq.~(\ref{eq:195}) and a similar formula holds for $\tilde\epsilon_3^{(0)}$.
This expression is worked out, see Sec.~S9 of the Supporting Information, and a Taylor expansion of $\tilde{E}^{(2)}$ in term of the small admixture angle is performed, where quadratic contributions are discarded (e.g. $\cos^{2}{\alpha} \simeq 1$).

After such computation, the difference between the approximate $\tilde{E}^{(2)}$ and $E^{(2)}$ is found

\begin{align}
 \tilde{E}^{(2)} - E^{(2)} \cong n_\textrm{s} \cdot 2 \sin{\alpha} \cdot \Re{e}(H_{12}^{*} H_{13}) \nonumber \\
 \Bigg[ 
 \frac{-1 + 2 \delta f}{\epsilon_{2}^{(0)} -\epsilon_{1}^{(0)}}
 + \frac{1 - \delta f}{\epsilon_{3}^{(0)} -\epsilon_{1}^{(0)}}
 \Bigg] + O(\sin^{2}{\alpha}). \label{eq:202}
\end{align}

Thus, the error $\tilde{E}^{(2)}-E^{(2)}$ is proportional to the admixture angle, not to its square. This is at variance with the error of $E^{(2)}$
with respect to an error in $\psi^{(1)}$, since $E^{(2)}$ is variational
with respect to $\psi^{(1)}$.
It emphasizes that the determination of eigenvectors in the potentially occupied space must be rather accurate in order for
$E^{(2)}$ to be accurate.
Ignoring prefactors, improving $R^{2}$ by about $10^{-6}$ brings only $10^{-3}$ decrease of the difference $\tilde{E}^{(2)}-E^{(2)}$.

Also, Eq.~(\ref{eq:202}) reveals that, in this three-band model, an error is present even if $\delta f = 0$, provided $\epsilon_{2}^{(0)} \neq \epsilon_{3}^{(0)}$. 
For the more general many-band case, there will always be unoccupied states with different energies, so that the outcome of this analysis is that, whatever occupation, metallic or insulating materials, if some states in the space of explicitly treated, unperturbed, wavefunctions are approximate, there will be non-negligible errors.
Finally, the $\Re(H_{12}^{*}H_{13})$ factors indicates that an interference effect between the transition from state 1 to state 2 and the transition from state 1 to state 3 is at the origin of the dominant error.

In order to substantiate these statements, numerical tests have been made, in which the convergence of the explicitly treated highest-lying states was not perfect, and the second-order derivative of the total energy error was monitored.

\begin{figure}[h]
	\includegraphics[width=0.45\textwidth]{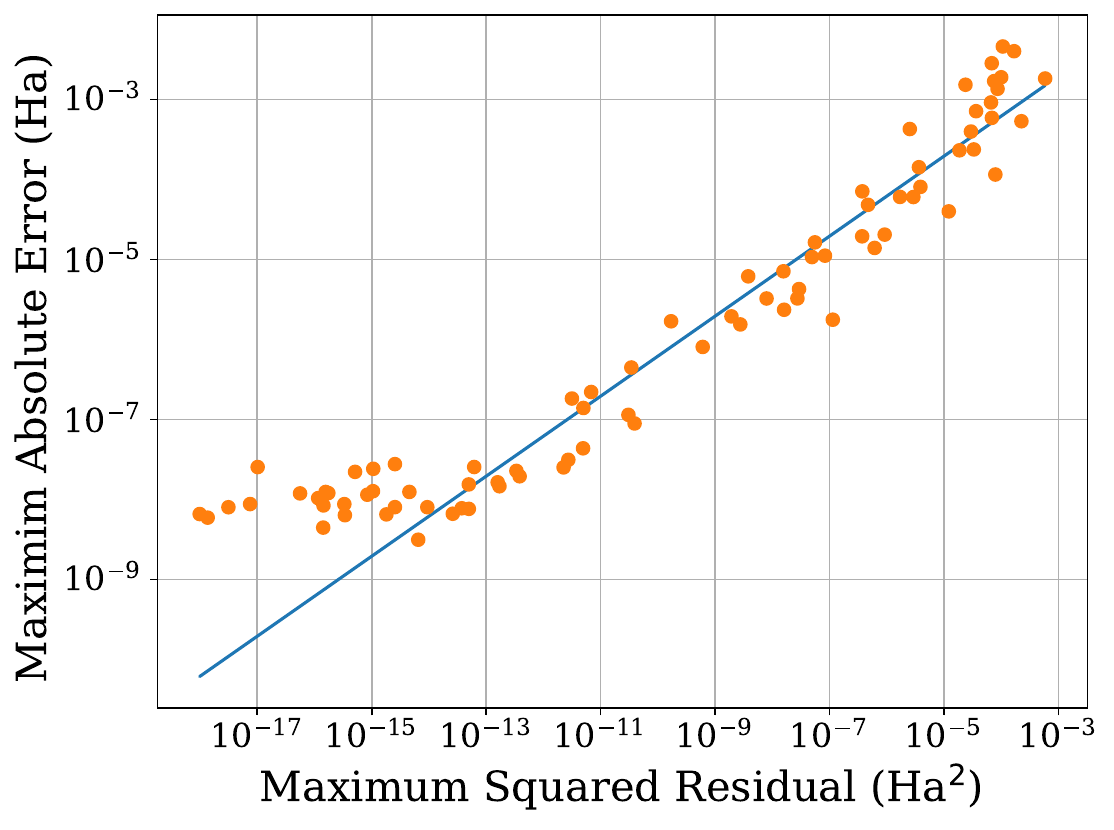}
	\caption{Relationship between the resulting errors in the second-order derivative of the free energy and the square of the wavefunction residual. The scatter plot illustrates the observed errors, while the solid line corresponds to a square root behavior.
		 }
	\label{fig:underconverged}
 \end{figure}
 
As for the previous section, calculations were done for phonon frequencies in copper.
The planewave kinetic energy cutoff was 50 Ha and a 12 \hbox{$\times$ 12 $\times$ 12} wavevector grid was used for the Brillouin Zone sampling. Phonon calculations were done on a \hbox{6 $\times$ 6 $\times$ 6} phonon wavevector sampling grid, such as to accumulate statistics.
In the ground-state calculations, the number of explicitly treated bands was varied between 10 and 22 .
 A stringent convergence criterion was set for the potential residual at $10^{-20}$, ensuring precise results for the potential and density. Still, the higher-lying
 bands, that do not contribute to the density,
 were not fully converged.
 We systematically varied the number of line steps for the conjugate gradient minimization
 in the Sternheimer equation from 4 to 18 and monitored the maximum of the square of the wavefunction residuals. 
Fig.~\ref{fig:underconverged} collects the resulting errors in the second-order derivative of the free energy for a whole set of elements of the dynamical matrices, as a function
of the maximum squared wavefunction residual.
The global trend is in line with the expectations, namely, the maximum absolute error is roughly proportional to square root of the the maximum squared wavefunction residual (or equivalently proportional to the wavefunction residual).
When the maximum squared residual is lower than
about 10$^{-13}$, the maximum absolute error
saturates at about 10$^{-8}$ Ha. Withouth having pursued further this matter, it seems plausible that sources of errors independent of the wavefunction residual exist at that numerical level, and start to dominate.


\section{Conclusion}
\label{Sec:Conclusion}

In the present work, a variational formulation of density-functional perturbation theory for metals has been described, covering in detail : the consequences 
of the presence of an entropy contribution; different smearing schemes, including  a resmearing scheme to deal with finite temperatures; the treatment of the space of potentially occupied wavefunctions; the different possible gauges, their advantages and drawbacks; specificities of the treatment of periodic systems.
In line with the well-established generic theorems in DFPT,
the
second-order derivative of the free energy is formulated
as a variational functional of trial first-order wavefunctions and trial first-order density matrix. 
A contribution from the second-order entropy is present in the second-order free energy. The
changes of the occupation numbers are explicitly 
taken into account. 

Concerning applications, this formalism has been available for some time in ABINIT, and 
has already yielded many publications. Nevertheless, two advanced application-related topics have been covered. For the first topic, the study of the convergence of phonon frequencies with
respect to wavevector sampling, two regimes, ``medium precision'' and ``high precision'', naturally emerge,
corresponding to whether the phonon frequencies
as such are the target property, or whether their
temperature dependence is the target property. 
The second topic relates to the impact of the
preliminary unperturbed calculation on the
subsequent DFPT calculation if the unoccupied wavefunctions have not been sufficiently accurately computed. 


\begin{acknowledgments}
This work has been supported by the Fonds de la Recherche
Scientifique (FRS-FNRS Belgium) through the PdR Grant No.
T.0103.19 – ALPS. 
It is an outcome of the Shapeable
2D magnetoelectronics by design project (SHAPEme, EOS
Project No. 560400077525) that has received funding from
the FWO and FRS-FNRS under the Belgian Excellence of
Science (EOS) program.
\\
Ch.T. acknowledges support from the Research Council of Norway through
its Centres of Excellence scheme (262695), through the FRIPRO grant ReMRChem
(324590),  and from NOTUR -- The Norwegian Metacenter for Computational Science
through grant of computer time (nn14654k).
\end{acknowledgments}



\bibliography{main} 

\end{document}


\renewcommand{\thesection}{S\arabic{section}}
\renewcommand{\thepage}{S\arabic{page}}
\renewcommand{\thefigure}{S\arabic{figure}}
\renewcommand{\theequation}{S\arabic{equation}}

\title{Supplemental Material \\
of \\
Variational Density Functional Perturbation Theory for Metals}

\author{Xavier Gonze*}
\affiliation{European Theoretical Spectroscopy Facility, Institute of Condensed Matter and Nanosciences, Universit\'{e} catholique de Louvain, Chemin des \'{e}toiles 8, bte L07.03.01, B-1348 Louvain-la-Neuve, Belgium}

\author{Samare Rostami}
\affiliation{European Theoretical Spectroscopy Facility, Institute of Condensed Matter and Nanosciences, Universit\'{e} catholique de Louvain, Chemin des \'{e}toiles 8, bte L07.03.01, B-1348 Louvain-la-Neuve, Belgium}

\author{Christian Tantardini*}
\affiliation{Hylleraas center, Department of Chemistry, UiT The Arctic University of Norway, PO Box 6050 Langnes, N-9037 Troms\o, Norway.}
\affiliation{Department of Materials Science and 
Nanoengineering, Rice University, Houston, Texas 77005, USA.}

\email{xavier.gonze@uclouvain.be; \\
christiantantardini@ymail.com}

\date{\today}

\maketitle

\section{Smearing scheme for finite-temperature electronic-structure calculations}

A couple of equations in Ref.~\onlinecite{Verstraete2001} (denoted V01 in what follows, for convenience) appear to be incorrect, or to rely on fuzzy (ambiguous) notations.
This section fixes such problem, and makes sure that the results presented in the main text are in line with those of Ref.~\onlinecite{DosSantos2023}.
Also, additional comments are made.
In this section S1 of the supplementary information, in order to ease the comparison with V01, we stick to V01 notations in the few cases where they differ from those used in the body of the present paper, and also the Boltzmann constant $k$ is set to 1, like in V01 (to recover the equations in the body of the paper, simply replace $T$ by $kT$).
In particular, the ``distribution function'' $\delta$ considered here
is not a function of the occupation numbers, but is a function of the energy (or the rescaled energy). This function is denoted $\tilde\delta$ in the main part of this paper.

Such ``$\delta$'' function is supposed to be normalized with respect to its argument. 
Checking that this is indeed the case brings us straight
to the ambiguity of notation in V01.
$\delta_{1}(x)$ is indeed normalized, in Eq.~(4) of V01 as

\begin{align}
    \int_{- \infty}^{+ \infty} \delta_{1}(x) dx =  1. \label{eq:1}
\end{align}

This is also true of $\Tilde{\delta}_{T}(\epsilon)$ in Eq. (3) of V01 from 

\begin{equation}
    \int_{- \infty}^{+ \infty} \delta_{T}(\epsilon) d\epsilon =  \int_{- \infty}^{+ \infty} \delta_{1} \bigg( \frac{\epsilon}{T} \bigg) \frac{d\epsilon}{T}
 \int_{- \infty}^{+ \infty} \delta_{1}(x) dx =  1 .\label{eq:2}
\end{equation}

However this is not clear to be valid for the non-numbered original equations at the end of page 035111-2\cite{Verstraete2001}. Indeed, the strict interpretation of these equations, that corresponds to the following ``original'' function

\begin{align}
    \delta_{\textrm{tot}}^{\textrm{ori}} \bigg( x, \frac{\sigma}{T} \bigg) = \frac{1}{T} \int \, \delta_{2}(z) \, \delta_{1} \bigg( x -\frac{\sigma}{T}z  \bigg) dz, \label{eq:3}
\end{align}
is not normalized:

\begin{align}
    \int \delta_{\textrm{tot}}^\textrm{ori} \bigg( x, \frac{\sigma}{T} \bigg) dx = \frac{1}{T} \int \, \delta_{2}(z) \Bigg[ \int \delta_{1}  \bigg( x -\frac{\sigma}{T}z  \bigg) \, dx \Bigg] dz
    \nonumber = \frac{1}{T}. \label{eq:5}
\end{align}

For such $\delta_{\textrm{tot}}^\textrm{ori}$ the usual normalization is obtained by considering it as a function of $\epsilon$ with $x(\epsilon) = \frac{\mu - \epsilon}{kT}$, and not as a function of $x$. 

\begin{equation}
    \int \delta_{\textrm{tot}}^\textrm{ori} \bigg( x(\epsilon), \frac{\sigma}{T} \bigg) d\epsilon = 1. \label{eq:5+}
\end{equation}

However, this is only loosely implied by the notation

\begin{align}
    \delta_{\textrm{tot}}\bigg( x = \frac{\mu - \epsilon}{kT}, ...  \bigg) 
\end{align}
used in V01.

Having identified the ambiguity, the present
supplementary information provide rigorous mathematical formulas for the
resmearing scheme presented in V01. 
The numerical results of V01 are not modified.

So, more rigorously, one defines

\begin{align}
    \delta_{\textrm{tot}}(x,T,\sigma) = \int \, \Tilde{\delta}_{T1}(z) \, \Tilde{\delta}_{\sigma 2} (x-z) \, dz, \label{eq:6}
\end{align}

\indent where

\begin{align}
    \Tilde{\delta}_{T1}(z) = \frac{1}{T} \delta_{1} \bigg( \frac{z}{T} \bigg),  \label{eq:205new} \\
    \Tilde{\delta}_{\sigma 2}(z) = \frac{1}{\sigma} \delta_{2} \bigg( \frac{z}{\sigma} \bigg).
\end{align}
In these formulas, the $z$ variable has the dimension of an energy (or a temperature, as $k$=1 in this SI).
The arguments of $\delta_1$ and $\delta_2$ are adimensional, while the argument of the $\Tilde{\delta}$
functions are energies. Likewise, all the arguments of 
$\delta_{tot}$, including $x$, have the dimension of an energy. Note that these conventions, coherent with V01, are not the same as in the main text.

$\delta_{\textrm{tot}}(x,T,\sigma)$ is correctly normalized,

\begin{align}
    \int \delta_{\textrm{tot}}(x,T,\sigma) dx & = \int \Tilde{\delta}_{T1}(z) \Bigg[ \int  \Tilde{\delta}_{\sigma 2}(x-z) \, \Bigg] dz = \int \Tilde{\delta}_{T1}(z) dz = 1. 
\end{align}

By the variable change $z'=x-z$ one obtains the symmetric formula 

\begin{align}
    \delta_{\textrm{tot}}(x,T,\sigma) = \int \Tilde{\delta}_{T1}(x-z')  \Tilde{\delta}_{\sigma 2}(z') dz'. \label{eq:7}
\end{align}

In terms of $\delta_{1}$ and $\delta_{2}$,

\begin{align}
    \delta_{\textrm{tot}}(x,T,\sigma) & = \frac{1}{T \sigma} \int \delta_{1} \bigg( \frac{z}{T} \bigg) \delta_{2} \bigg( \frac{x-z}{\sigma} \bigg) dz = \frac{1}{T \sigma} \int \delta_{1} \bigg( \frac{x-z'}{T} \bigg) \delta_{2} \bigg( \frac{z'}{\sigma} \bigg) dz'. \label{eq:8b} 
\end{align}

From Eq.~(\ref{eq:8b}), a change of variable $w = z'/ \sigma$ ($w$ is adimensional) brings

\begin{equation}
    \delta_{\textrm{tot}}(x,T,\sigma) = \frac{1}{T} \int \ \delta_{1} \bigg( \frac{x}{T} - \frac{\sigma}{T}w \bigg) \delta_{2}(w) dw. \label{eq:9b}
\end{equation}

Now one sets $x=Tv$ ($v$ is adimensional) and exchanges the two delta functions,

\begin{align}
    \delta_{\textrm{tot}}(Tv,T,\sigma) = \frac{1}{T} \int \delta_{2}(w) \delta_{1} \bigg( v - \frac{\sigma}{T}w \bigg) dw . \label{eq:10}
\end{align}

The right-hand side is the right-hand side of Eq.~(\ref{eq:3}),
with $z$ replaced by $w$, while the argument of $\delta_{\textrm{tot}}$ is actually $Tv$. This gives
another correct mathematical formula that might have been used in V01.

%

The definition in Eq.(\ref{eq:6}), or equivalently Eqs.~(\ref{eq:7}) or ~(\ref{eq:8b}), allows one to recover Eq.~(7) of V01. 
Indeed, one defines
\begin{align}
    \Tilde{n}(\epsilon) = \int  n(\epsilon_{2}) \delta_{\textrm{tot}}(\epsilon-\epsilon_{2}, T, \sigma) d \epsilon_{2},\label{eq:12appendix}
\end{align}
(in this equation, we use the notations of Ref.~\onlinecite{DosSantos2023}, namely $\Tilde{n}$ and $n$, instead of $n$ and $n_{0}$ of V01), that is, $\tilde n$ is 
 $n$ smeared with $\delta_{\textrm{tot}}$. Using Eq.~(\ref{eq:8b}), it follows

\begin{align}
   \Tilde{n}(\epsilon) = \int n(\epsilon_{2}) 
   \Bigg[ \frac{1}{T \sigma} \int  \delta_{1} \bigg( \frac{\epsilon - \epsilon_{2} - \epsilon_{3}}{T} \bigg) \delta_{2} \bigg( \frac{\epsilon_{3}}{\sigma} \bigg) 
   d\epsilon_{3}\Bigg]
   d\epsilon_{2},\label{eq:13}
\end{align}
that is indeed Eq.~(7) of V01, except for the typo 
present in Eq.~(7) of V01, the argument of $\delta_{2}$ being typed $\frac{\epsilon_{2}}{\sigma}$ instead of the above correct $\frac{\epsilon_{3}}{\sigma}$.

Now one focuses on occupation functions, also presented in connection to those in V01, but here with mathematical rigor. 
Following the usual definition,

\begin{align}
    f_{\textrm{tot}}(y,T,\sigma) & = \int_{- \infty}^{y} \delta_{\textrm{tot}}(x,T,\sigma) dx \label{eq:17} \\
    & = \int_{- \infty}^{y} \Bigg[ \int \Tilde{\delta}_{T1}(z) \Tilde{\delta}_{\sigma 2}(x -z) dz \Bigg] dx
    = \int \Tilde{\delta}_{T1}(z) f_{\sigma 2}(y -z) dz 
    = \int \frac{1}{T} \delta_{1} \bigg( \frac{z}{T} \bigg) f_{\sigma 2}(y-z) dz, \label{eq:19x}
\end{align}

\indent with the definition

\begin{align}
    f_{\sigma 2}(x) = \int_{-\infty}^{x}  \Tilde{\delta}_{\sigma 2} (x') dx' = \frac{1}{\sigma} \int_{- \infty}^{x} \delta_{2} \bigg( \frac{x'}{\sigma} \bigg) dx'. \label{eq:18}
\end{align}

Also, following the usual definition,

\begin{align}
    f_{2}(u) = \int_{- \infty}^{u} \delta_{2} (w) dw,  \label{eq:20}
\end{align}

one gets

\begin{align}
    f_{\sigma 2}(x) & = \frac{1}{\sigma} \int_{- \infty}^{x} \delta_{2} \bigg( \frac{x'}{\sigma} \bigg) dx' = \int_{- \infty}^{x/\sigma} \delta_{2}(w) dw  = f_{2} \bigg( \frac{x}{\sigma} \bigg). \label{eq:21}
\end{align}

Thus 

\begin{align}
    f_{\textrm{tot}}(y,T,\sigma) = \int \frac{1}{T} \delta_{1} \bigg( \frac{z}{T} \bigg) f_{2} \bigg( \frac{y -z}{\sigma} \bigg) dz. \label{eq:22}
\end{align}

\indent After rescaling with $w = z/T$,

\begin{align}
    f_{\textrm{tot}}(y,T,\sigma) = \int  \delta_{1} (w) f_{2} \bigg( \frac{y}{\sigma} - \frac{T}{\sigma} w \bigg) dw. \label{eq:23}
\end{align}

Thus, the two non-numbered equations, present after Eq.(8) of V01, should be written
\begin{align}
    f_{\textrm{tot}}(Tu,T,\sigma) = \int  \delta_{2}(w) f_{1} \bigg( u-\frac{\sigma}{T}w \bigg) dw, \label{eq:24}
\end{align}

\indent and

\begin{align}
    f_{\textrm{tot}}(\sigma u,T,\sigma) = \int \delta_{1}(w) f_{2} \bigg( u-\frac{T}{\sigma} w \bigg) dw. \label{eq:25}
\end{align}

Let us now turn to the formulas for the entropy, formulated rigorously mathematically. Following the usual definition,

\begin{align}
    s_{\textrm{tot}}(y,T,\sigma) & = - \int_{- \infty}^{y} \quad x \delta_{\textrm{tot}}(x,T,\sigma) dx \label{eq:26} \\
    & = - \int  \Tilde{\delta}_{T1}(z)
     \int_{- \infty}^{y} x \Tilde{\delta}_{\sigma 2}(x - z) \, dx  dz 
    = - \int \Tilde{\delta}_{T1}(z) \int_{- \infty}^{(y-z)}  (x'+z) \Tilde{\delta}_{\sigma 2}(x') dx' 
    dz \\
    & = \int \Tilde{\delta}_{T1}(z) s_{\sigma 2}(y-z) dz 
    - \int  \Tilde{\delta}_{T1}(z) \cdot z f_{\sigma 2}(y - z) dz, \label{eq:27b}
\end{align}

\indent with the definition

\begin{align}
   s_{\sigma 2}(x) & = - \int_{- \infty}^{x} x' \Tilde{\delta}_{\sigma 2}(x') dx' = - \frac{1}{\sigma} \int_{-\infty}^{x} x' \delta_{2} \bigg( \frac{x'}{\sigma} \bigg) dx'.\label{eq:28}
\end{align}

Let us define

\begin{align}
    s_{2}(x) = - \int_{- \infty}^{x}  w 
    \delta_{2}(w) dw, \label{eq:29}
\end{align}
where $w$ is adimensional.

Hence,

\begin{align}
    s_{\sigma 2}(x) & = - \sigma \int_{- \infty}^{x/\sigma} w \delta_{2}(w) dw = \sigma \cdot s_{2} \bigg( \frac{x}{\sigma} \bigg). \label{eq:31}
\end{align}

After further manipulations of the second term in the total entropy, Eq.~(\ref{eq:27b}), that involve using Eq.~(\ref{eq:205new}) and expressing differently the obtained two-dimensional domain, Eq.~(\ref{eq:26}) becomes

\begin{align}
    s_{\textrm{tot}}(y,T,\sigma) & = \int_{-\infty}^{\infty} \Tilde{\delta}_{T1}(z) s_{\sigma 2}(y -z) dz  + \int_{-\infty}^{\infty} \Tilde{\delta}_{\sigma 2}(z) s_{T 1}(y -z) dz, \label{eq:37b}
\end{align}

\indent that is, the convolution of $s_{\sigma 2}$ by $\Tilde{\delta}_{T1}$ plus the convolution of $s_{T 1}$ by $\Tilde{\delta}_{\sigma 2}$, as announced in V01 paper, now with the appropriate smearing functions.
Equivalently,

\begin{align}
    s_{\textrm{tot}}(y,T,\sigma) & = T \int_{-\infty}^{\infty} \frac{1}{\sigma} \delta_{2} \bigg( \frac{z}{\sigma} \bigg) s_{1}\bigg( \frac{y-z}{T} \bigg) dz + \sigma \int_{-\infty}^{\infty} \frac{1}{T} \delta_{1} \bigg( \frac{z}{T} \bigg) s_{2}\bigg( \frac{y-z}{\sigma} \bigg) dz, \label{eq:38b}
\end{align}

\indent a formulation that is closer to the equations in V01.

Finally, in the resmearing scheme, the free energy becomes

\begin{align}
    \Tilde{E}_{free} & = \int_{-\infty}^{\mu}\epsilon\Tilde{n}(\epsilon) d\epsilon
    \label{eq:14a} \\
    & = \int \epsilon_{2} n(\epsilon_{2}) f_{\textrm{tot}} (\mu-\epsilon_{2}, T, \sigma)  d \epsilon_{2} \nonumber \\
    & - T \int n(\epsilon_{2}) \Bigg[ \int \frac{1}{\sigma} \delta_{2} \bigg( \frac{z}{\sigma} \bigg) s_{1} \bigg( \frac{\mu - \epsilon_{2} - z}{T} \bigg) 
    dz \Bigg] d \epsilon_{2}
    - \sigma \int n(\epsilon_{2}) \Bigg[ \int \frac{1}{T} \delta_{2} \bigg( \frac{z}{T} \bigg) s_{2} \bigg( \frac{\mu - \epsilon_{2} - z}{\sigma} \bigg)
    dz \Bigg] d \epsilon_{2}, \label{eq:39c}
\end{align}

\indent that is in line with Eq.~(8) of V01, except that $f_{\textrm{tot}} \bigg( \epsilon_{2}, \frac{\sigma}{T} \bigg)$ in Eq.~(8) of V01 should be replaced by 

\begin{align}
    f_{\textrm{tot}} \bigg( x=\frac{\mu - \epsilon_{2}}{T}, \frac{\sigma}{T} \bigg) \textnormal{in the fuzzy notations of V01.} \nonumber 
\end{align}

Thus, the equations of V01 can be fixed, albeit by a meticulous approach, and notwithstanding
some typos (in addition to thos already signaled, note sign errors in Eqs.~(5) and (6), and a typo in Eq.~(8)).

$\delta_{\textrm{tot}}$ can be rewritten, in order to examine how the original Fermi-Dirac $\delta_{\textrm{FD}} = \delta_{1}$ with fixed \textit{T} is broadened by $\delta_{2}$, where the ratio $R=\frac{\sigma}{T}$ appears explicitly.

One starts from Eq.~(\ref{eq:8b}):

\begin{align}
    \delta_{\textrm{tot}}(x,T,\sigma) & = \frac{1}{T \sigma} \int \delta_{1} \bigg( \frac{x-z'}{T} \bigg) \delta_{2} \bigg( \frac{z'}{\sigma} \bigg) dz' = \frac{1}{T} \int \quad \delta_{1} \bigg( \frac{x}{T} - \frac{\sigma}{T}w \bigg) \cdot \delta_{2}(w) dw.\label{eq:40}
\end{align}

Then, the ``resmeared'' delta function $\delta_{\textrm{rsm}}$ is defined, as
\begin{align}
    \delta_{\textrm{rsm}}(u,R) = \int \delta_{1}(u-Rw) \delta_{2}(w) dw , \label{eq:41SI}
\end{align}
where all arguments are adimensional, and 
\indent in term of which

\begin{align}
    \delta_{\textrm{tot}}(x,T,\sigma) = \frac{1}{T} \delta_{\textrm{rsm}} \bigg( \frac{x}{T}, \frac{\sigma}{T} \bigg). \label{eq:42appendixSI}
\end{align}

So, for fixed $T$, the ratio $R=\frac{\sigma}{T}$ can be scanned, and the behavior of $\delta_{\textrm{tot}}(x,T,\sigma)$ can be plotted for different smearings $\sigma$.
\newline
\\

Let us now examine the 
asymptotic behavior of $\delta_{\textrm{rsm}}$,
for different values of $R$. It will be shown
that, at the leading order, this function has an exponentially decreasing tail, with either positive or negative coefficient, depending on the value of $R$, and changing sign at $R=2$.
So, at $R=2$, this leading asymptotic order vanishes, and numerical evidence shows that 
the function is positive everywhere, see Fig. 1
 in the main text.

The analysis starts by expanding $\delta_{1}(x)$ in a serie for $x \rightarrow + \infty $,

\begin{align}
\delta_{1}(x \rightarrow + \infty) & = \frac{e^{-x}}{(1 + e^{-x})^{2}} \label{eq:8_rsm} \\
& \simeq e^{-x} (1 - 2e^{-x} + 3e^{-2x} - 4e^{-3x} \cdots) 
 = e^{-x} - 2e^{-2x} + 3e^{-3x} - 4e^{-4x} + \cdots 
 = \sum_{n=1}^{\infty} n e^{-nx}. \label{eq:9_rsm}
\end{align}

The series expansion converges for $|e^{-x}| < 1$.
At the lowest order,
\begin{align}
\delta_{1}(x) \simeq e^{-x} + \mathcal{O}(e^{-2x}). \label{eq:10_rsm}
\end{align}
Focusing on this first term,
\begin{align}
    \delta_{\textrm{rsm}}(y,R) & \cong \int_{-\infty}^{+ \infty} \frac{1}{\sqrt{\pi}} \bigg( \frac{3}{2} - z^{2} \bigg) e^{-z^{2}} e^{-(y-Rz)} dz  + \mathcal{O}\big(e^{-2(y-Rz)}\big) \nonumber \\
    & \cong \frac{1}{\sqrt{\pi}} e^{-y} \int_{-\infty}^{+ \infty} \bigg( \frac{3}{2} - \frac{\partial^{2}}{\partial R^{2}} \bigg) e^{-z^{2}+Rz} dz \nonumber \\
    & \cong \frac{1}{\sqrt{\pi}} e^{-y} \bigg( \frac{3}{2} - \frac{\partial^{2}}{\partial R^{2}} \bigg) \int_{-\infty}^{+ \infty} e^{\big( -z^{2} + Rx - \frac{R^{2}}{4} \big)} e^{\frac{R^{2}}{4}} dz \nonumber \\
    & \cong e^{-y} \bigg( \frac{3}{2} - \frac{\partial^{2}}{\partial R^{2}} \bigg) e^{\frac{R^{2}}{4}} \frac{1}{\sqrt{\pi}} \int_{-\infty}^{+ \infty}  e^{- \big( z - \frac{R}{2}\big)^{2}} dz \nonumber \\
    & \cong e^{-y} \bigg( \frac{3}{2} e^{\frac{R^{2}}{4}} - \frac{\partial}{\partial R} \bigg( \frac{2R}{4} e^{\frac{R^{2}}{4}} \bigg) \bigg) \nonumber \\
    & \cong e^{-y} \bigg( \frac{3}{2} e^{\frac{R^{2}}{4}} - \frac{1}{2} e^{\frac{R^{2}}{4}} - \bigg(\frac{R}{2}\bigg)^{2} e^{\frac{R^{2}}{4}} \bigg) \nonumber \\
    & \cong e^{-y} \bigg( 1 - \bigg( \frac{R}{2} \bigg)^{2} \bigg) e^{\frac{R^{2}}{4}} + \mathcal{O}\big(e^{-2(y-Rz)}\big). \label{eq:12_rsm}  
\end{align}

As expected, for $R=2$, the leading contribution vanishes. For $R$ smaller than 2 the function is asymptotically exponentially decreasing, with a positive prefactor, while for $R$ larger than 2 , it is also exponentially decreasing, but with a negative prefactor.

\section{Derivation of the expression for the second-order free energy}
\label{sec:derivationF2}

In this section, the details of the derivation of the
variational expression for the second-order free energy, Eq.~(34) in the main text, are explained.
One starts from
\begin{align}
    F^{+(2)} = \min_{\{ \psi_{i}^{(1)},\rho_{ij}^{(1)}\}} F^{+(2)} [ T, \{ \psi_{i}^{(1)} \}, \{ \rho_{ij}^{(1)} \} ], \label{eq:43}
\end{align}

\indent where $F^{+(2)}$ is obtained from Eq.~(20) in the main text
by the application of the methodology explained in Ref.~\onlinecite{Gonze1995},

\begin{align}
    F^{+(2)} & [ T, \{ \psi_{i}^{(1)} \}, \{ \rho_{ij}^{(1)} \} ] = \Big( F^{+} [ T, \{ \psi_{i}^{(0)} + \lambda \psi_{i}^{(1)} \}, 
    \{ \rho_{ij}^{(0)} + \lambda \rho_{ij}^{(1)} \} ] \Big)^{(2)}. \label{eq:44}
\end{align}

\indent The zeroth order quantities have been obtained previously in the diagonal gauge, self-consistently, see Sec.~II in the main text.

\begin{widetext}
The five terms of Eq.~(20) in the main text 
are treated in turn,
\begin{align}
    F^{+(2)} = F_{1}^{+(2)} + F_{2}^{+(2)} + F_{3}^{+(2)} + F_{4}^{+(2)} + F_{5}^{+(2)} \label{eq:85}.
\end{align}

\underline{Contribution 1}

$n_\textrm{s} \sum_{ij} \rho_{ij} h_{ij}$ becomes      \begin{align}
        F_{1}^{+(2)} & = n_\textrm{s} \sum_{ij}^{{pocc}} ( \rho_{ij}^{(0)} + \lambda \rho_{ij}^{(1)}) 
         \langle \psi_{i}^{(0)} + \lambda \psi_{i}^{(1)} | \hat{K} + \lambda \hat{v}_{\textrm{ext}}^{(0)} + \lambda \hat{v}_{\textrm{ext}}^{(1)} + \lambda^{2} \hat{v}_{\textrm{ext}}^{(2)} | \psi_{j}^{(0)} + \lambda \psi_{j}^{(1)} \rangle \label{eq:52} \\
        & = n_\textrm{s} \sum_{i}^{{pocc}} f_{i}^{(0)} \bigg( \langle \psi_{i}^{(1)} | \hat{K} + \hat{v}_{\textrm{ext}}^{(0)} | \psi_{i}^{(1)} \rangle + \langle \psi_{i}^{(0)} | \hat{v}_{\textrm{ext}}^{(2)} | \psi_{i}^{(0)} \rangle + \big( \langle \psi_{i}^{(1)} | \hat{v}_{\textrm{ext}}^{(1)} | \psi_{i}^{(0)} \rangle + (\textrm{c.c.}) \big) \bigg) \nonumber \\
        & + n_\textrm{s} \sum_{ij}^{{pocc}} \rho_{ji}^{(1)} \bigg( \langle \psi_{i}^{(0)} | \hat{v}_{\textrm{ext}}^{(1)} | \psi_{j}^{(0)} \rangle + \langle \psi_{i}^{(1)} | \hat{K} + \hat{v}_{\textrm{ext}}^{(0)} | \psi_{j}^{(0)} \rangle + \langle \psi_{i}^{(0)} | \hat{K} + \hat{v}_{\textrm{ext}}^{(0)} | \psi_{j}^{(1)} \rangle \bigg). \label{eq:53b}
    \end{align}
        
    \underline{Contribution 2}

         $E_\textrm{Hxc}[\rho]$ becomes
        \begin{align}
        F_{2}^{+(2)} & = \bigg( E_\textrm{Hxc} \big[ \{ \psi_{i}^{(0)} + \lambda \psi_{i}^{(1)} \}, \{ \rho_{ji}^{(0)} + \lambda \rho_{ji}^{(1)} \}    \big] \bigg)^{(2)} \label{eq:54} \\
        & = E_\textrm{Hxc} \bigg[ \rho^{(0)} + \lambda \rho^{(1)} + \lambda^{2} \bigg( n_\textrm{s} \bigg\{ \sum_{ij}^{{pocc}} \rho_{ji}^{(0)} \psi_{i}^{*(1)}(\textbf{r}) \psi_{j}^{(1)} + \sum_{ij}^{{pocc}} \rho_{ji}^{(1)} \bigg( \psi_{i}^{(0)*}(\textbf{r}) \psi_{j}^{(1)}(\textbf{r}) + \psi_{i}^{(1)*}(\textbf{r}) \psi_{j}^{(0)}(\textbf{r}) \bigg) \bigg\} \bigg) \bigg]^{(2)} \label{eq:55} \\
        & = \frac{1}{2} \int \int K_\textrm{Hxc}(\textbf{r},\textbf{r'}) \rho^{(1)}(\textbf{r}) \rho^{(1)}(\textbf{r'}) d\textbf{r} d\textbf{r'} 
        \nonumber\\
        & + \int V_\textrm{Hxc}^{(0)}(\textbf{r}) \cdot \bigg\{ \sum_{i}^{{pocc}} f_{i}^{(0)} \psi^{(1)*}_{i} (\textbf{r}) \psi_{i}^{(1)} (\textbf{r'})  + \sum_{ij}^{{pocc}} \rho_{ji}^{(1)} \bigg( \psi_{i}^{(0)*}(\textbf{r})\psi_{j}^{(1)}(\textbf{r}) + \psi_{i}^{(1)*}(\textbf{r})\psi_{j}^{(0)}(\textbf{r}) \bigg) \bigg\} \label{eq:56}\\
        & = \frac{1}{2} \int \int K_\textrm{Hxc}(\textbf{r},\textbf{r'}) \rho^{(1)}(\textbf{r}) \rho^{(1)}(\textbf{r'}) d\textbf{r} d\textbf{r'} 
        \nonumber \\
        & + n_\textrm{s} \sum_{i}^{{pocc}} f_{i}^{(0)} \langle \psi_{i}^{(1)} | \hat{v}_\textrm{Hxc}^{(0)} | \psi_{i}^{(1)} \rangle + n_\textrm{s} \sum_{ij}^{{pocc}} \rho_{ji}^{(1)} \bigg( \langle \psi_{i}^{(0)} | \hat{v}_\textrm{Hxc}^{(0)} | \psi_{j}^{(1)} \rangle + \langle \psi_{i}^{(1)} | \hat{v}_\textrm{Hxc}^{(0)} | \psi_{j}^{(0)} \rangle \bigg), \label{eq:57c}
        \end{align}
        \indent where $\rho^{(1)}(\textbf{r})$ is obtained from Eq.(38) in the main text.
\newline\\

\underline{Contribution 3}

This contribution comes from the term $-TS[ \{\rho_{ij}\}]$.
One needs to understand the second-order expansion of the entropy.
One starts from 
\begin{align}
S[\hat{\rho}]  = n_{\textrm{s}} \sum_{\gamma}^{pocc} ks\big( \occf_{\gamma}^{(0)} + \Delta 
\occf_{\gamma} \big), \label{eq:58}
\end{align}
\indent where 
\begin{align}
   \Delta \occf_{\gamma} = \lambda \occf_{\gamma}^{(1)} + \lambda^{2} \occf_{\gamma}^{(2)} + ... \label{eq:59}  
\end{align}
A Taylor series is used,
\begin{align}
    & S[\hat{\rho}] = n_\textrm{s} 
     \sum_{\gamma}^{{pocc}} 
     k\bigg( s(\occf_{\gamma}^{(0)}) + s'(\occf_{\gamma}^{(0)}) \Delta \occf_{\gamma} + \frac{1}{2} s''(\occf_{\gamma}^{(0)})(\Delta \occf_{\gamma})^{2} + ...\bigg), \label{eq:60}
\end{align}
from which
\begin{align}
    & S^{(2)}\big[\hat{\rho}^{(0)} + \lambda \hat{\rho}^{(1)}\big] = n_\textrm{s}  
    \sum_{\gamma}^{{pocc}} k\bigg( s'(\occf_{\gamma}^{(0)}) \cdot \occf_{\gamma}^{(2)} + \frac{1}{2}s''(\occf_{\gamma}^{(0)}) \cdot (\occf_{\gamma}^{(1)})^{2}\bigg), \label{eq:61}
\end{align}

where, in the last equation, no contribution from
$\hat{\rho}^{(2)}$ is included. Using Eq.~(10) in the main text, one obtains

\begin{equation}
F_{3}^{+(2)} = -TS^{(2)}\big[\hat{\rho}^{(0)} + \lambda \rho^{(1)}\big] = - n_\textrm{s} \sum_{\gamma}^{{pocc}} \Bigg( (\epsilon_{\gamma}^{(0)} - \mu^{(0)}) \cdot \occf_{\gamma}^{(2)} + \frac{1}{2} \frac{\partial \epsilon}{\partial f} \bigg|_{\occf_{\gamma}^{(0)}} \cdot (\occf_{\gamma}^{(1)})^{2} \Bigg) \label{eq:64ter},
\end{equation}

where also $s''(\occf_{\gamma}^{(0)}) = \frac{1}{kT}\frac{\partial \epsilon}{\partial f} \bigg|_{\occf_{\gamma}^{(0)}}$, see Eq.(32) in the main text, has been used. 

Now, one has to compute $\occf_{\gamma}^{(1)}$
and $\occf_{\gamma}^{(2)}$.
Standard perturbation theory treatment of the eigenvalues and eigenvectors of the (hermitian) one-particle density matrix 
$\hat{\rho}$, at first order, delivers
\begin{equation}
 \occf_{\gamma}^{(1)} = \rho_{\gamma \gamma}^{(1)}, \label{eq:73}
\end{equation}
that is, the Hellmann-Feynman theorem.
For the second-order $\occf_{\gamma}^{(2)}$, one finds
\begin{equation}
\occf_{\gamma}^{(2)} = \rho_{\gamma \gamma}^{(2)} - \frac{1}{2} \sum_{i (\neq \gamma)}^{{pocc}} \frac{|\rho_{i \gamma}^{(1)}|^{2}}{\occf_{i}^{(0)} - \occf_{\gamma}^{(0)}}, \label{eq:78}
\end{equation}
where $\rho_{\gamma \gamma}^{(2)}$ has to be set to zero,
see Eq.~(\ref{eq:61}) and the comment afterwards.
Then, Eq.~(\ref{eq:64ter}) becomes 

\begin{equation}
F_{3}^{+(2)} = n_\textrm{s} \sum_{\gamma}^{{pocc}} \Bigg[ (\epsilon_{\gamma}^{(0)} - \mu^{(0)})  \sum_{i (\neq \gamma)}^{{pocc}} \frac{|\rho_{i \gamma}^{(1)}|^{2}}{\occf_{i}^{(0)} - \occf_{\gamma}^{(0)}} - \frac{1}{2}  (\occf_{\gamma}^{(1)})^{2} \frac{\partial \epsilon}{\partial f} \Bigg|_{\occf_{\gamma}^{(0)}} \Bigg].  \label{eq:79bisSI} 
\end{equation}

This contribution can be reformulated switching $\gamma$ and $i$ for half the contribution, and also using the hermiticity of
$\rho^{(1)}$ to obtain

\begin{equation}
F_{3}^{+(2)} = -\frac{n_\textrm{s}}{2} \sum_{\gamma}^{{pocc}} \Bigg[ \sum_{i (\neq \gamma)}^{{pocc}}  \frac{\epsilon_{i}^{(0)} - \epsilon_{\gamma}^{(0)}}{\occf_{i}^{(0)} - \occf_{\gamma}^{(0)}} |\rho_{i \gamma}^{(1)}|^{2}
+  (\occf_{\gamma}^{(1)})^{2} \frac{\partial \epsilon}{\partial f} \Bigg|_{\occf_{\gamma}^{(0)}} \Bigg] . \label{eq:79bis} 
\end{equation}

Note that if $f(\epsilon)$ is monotonically decreasing as a function of $\epsilon$, the term between parentheses is negative, as well as the derivative $\frac{\partial \epsilon}{\partial f}$, hence this contribution $F_{3}^{+(2)}$ is definite positive. 
This is not the case for advanced smearing schemes.
\newline
\\

\underline{Contribution 4}

$-n_\textrm{s} \sum_{ij}^{{pocc}} \Lambda_{ji} \bigg( \langle \psi_{i} | \psi_{j} \rangle - \delta_{ij} \bigg)$ becomes 

\begin{equation}
F_{4}^{+(2)}  = -n_\textrm{s} \Bigg[ \sum_{ij}^{{pocc}} \Lambda_{ji}^{(2)} \bigg( \langle \psi_{i}^{(0)} | \psi_{j}^{(0)} \rangle - \delta_{ij} \bigg) + \Lambda_{ji}^{(1)} \bigg( \langle \psi_{i}^{(1)} | \psi_{j}^{(0)} \rangle + \langle \psi_{i}^{(0)} | \psi_{j}^{(1)} \rangle \bigg) 
 + \Lambda_{ji}^{(0)} \langle \psi_{i}^{(1)} | \psi_{j}^{(1)} \rangle  \Bigg]. \label{eq:80}
\end{equation}

The expansion of the constraints (for $i,j \in S_{pocc}$) $\langle \psi_{i}^{(0)} | \psi_{j}^{(0)} \rangle - \delta_{ij} = 0$, at first order, delivers the first-order constraints shown in Eq.~(40) in the main text. This can be used to obtain

\begin{equation}
    F_{4}^{+(2)}  = -n_\textrm{s} \Bigg[ \sum_{i}^{{pocc}} \occf_{i}^{(0)} \epsilon_{i}^{(0)} \langle \psi_{i}^{(1)} | \psi_{i}^{(1)} \rangle
    + \sum_{ij}^{{pocc}} \Lambda_{ji}^{(1)} \bigg( \langle \psi_{i}^{(1)} | \psi_{j}^{(0)} \rangle + \langle \psi_{i}^{(0)} | \psi_{j}^{(1)} \rangle \bigg) \Bigg]. \label{eq:82}
\end{equation}

\underline{Contribution 5}

$- \mu \big( n_\textrm{s} \sum_{i}^{{pocc}}\rho_{ii} - N \big) $ becomes

\begin{equation}
    F_{5}^{+(2)} = - \mu^{(2)} \big( n_\textrm{s} \sum_{i}^{{pocc}} \rho_{ii}^{(0)} - N \bigg) 
    - \mu^{(1)} n_\textrm{s} \sum_{i}^{{pocc}} \rho_{ii}^{(1)} 
    = - \mu^{(1)} n_\textrm{s} \sum_{i}^{{pocc}} \rho_{ii}^{(1)} , \label{eq:83}
\end{equation}

with the first-order constraint 

\begin{align}
    \sum_{i}^{{pocc}} \rho_{ii}^{(1)} = 0. \label{eq:84} 
\end{align}

Thus, gathering the terms, and with several rearrangements, one finds
the expressions in Sec.~III~A.

There might be a slight rewriting of such expressions. 
For example, the contribution Eq.~(36) in the main text coming from $\rho_{ij}^{(1)}[\psi_i^{(1)}]$
might be modified, thanks to

\begin{align}
    n_\textrm{s} \sum_{ij}^{{pocc}} & \rho_{ji}^{(1)} \bigg( \langle \psi_{i}^{(0)} | \hat{v}_{\textrm{ext}}^{(1)} | \psi_{j}^{(0)} \rangle + \epsilon_{j}^{(0)} \langle \psi_{i}^{(1)} | \psi_{j}^{(0)} \rangle + \epsilon_{i}^{(0)} \langle \psi_{i}^{(0)} | \psi_{j}^{(1)} \rangle \bigg) \nonumber \\
    & = n_\textrm{s} \sum_{ij}^{{pocc}} \rho_{ji}^{(1)} \bigg( \langle \psi_{i}^{(0)} | \hat{v}_{\textrm{ext}}^{(1)} | \psi_{j}^{(0)} 
    + \bigg( \frac{\epsilon_{j}^{(0)} + \epsilon_{i}^{(0)}}{2}  + \frac{\epsilon_{j}^{(0)} - \epsilon_{i}^{(0)}}{2} \bigg) \langle \psi_{i}^{(1)} | \psi_{j}^{(0)} \rangle 
    + \bigg( \frac{\epsilon_{j}^{(0)} + \epsilon_{i}^{(0)}}{2}  - \frac{\epsilon_{j}^{(0)} - \epsilon_{i}^{(0)}}{2} \bigg) \langle \psi_{i}^{(0)} | \psi_{j}^{(1)} \rangle \bigg) \nonumber \\
    & = n_\textrm{s} \sum_{ij}^{{pocc}} \rho_{ji}^{(1)} \bigg( \langle \psi_{i}^{(0)} | \hat{v}_{\textrm{ext}}^{(1)} | \psi_{j}^{(0)} \rangle + \big( \epsilon_{j}^{(0)} + \epsilon_{i}^{(0)} \big) \Re\textrm{e} \langle \psi_{i}^{(1)} | \psi_{j}^{(0)} \rangle   + \big( \epsilon_{j}^{(0)} - \epsilon_{i}^{(0)} \big) \Im\textrm{m} \langle \psi_{i}^{(1)} | \psi_{j}^{(0)} \rangle  \bigg). \label{eq:87cprime}
\end{align}

\section{Non-hermiticity freedom for the first-order off-diagonal occupation matrix elements}
\label{sec:non-hermiticity}

Until now, one has supposed that the $\rho^{(1)}$ matrix is hermitian.
However, this condition can be relaxed at this stage.
Indeed, let us start from an unrestricted $\tilde{\rho}^{(1)}$ matrix.
One can generate a hermitian matrix $f$ from 

\begin{equation}
\label{eq:88}
f^{(1)} = \frac{1}{2} \bigg( \tilde{f}^{(1)} + \tilde{f}^{+(1)} \bigg).
\end{equation}

The first-order density Eq.~(38) in the main text becomes 

\begin{equation}
\label{eq:89}
    \rho^{(1)}(\textbf{r}) =  
    n_\textrm{s} \bigg[ \sum_{ij}^{{pocc}} \frac{1}{2} \bigg( \tilde{\rho}_{ij}^{(1)} + \tilde{\rho}_{ji}^{*(1)} \bigg) \psi_{i}^{*(0)}(\textbf{r})\psi_{j}^{(0)}(\textbf{r})
 + \sum_{i}^{{pocc}} \occf_{i}^{(0)} \big( \psi_{i}^{*(1)}(\textbf{r})\psi_{i}^{(0)}(\textbf{r}) + \psi_{i}^{*(0)}(\textbf{r})\psi_{i}^{(1)}(\textbf{r}) \big) \bigg]
\end{equation}

The contribution of $\rho_{ij}^{(1)}[\psi_i^{(1)}]$
to Eq.~(34) in the main text becomes 
\begin{equation}
\label{eq:90}
n_\textrm{s} \sum_{ij}^{{pocc}} \frac{1}{2} \bigg( \tilde{\rho}_{ij}^{(1)} + \tilde{\rho}_{ji}^{*(1)} \bigg)
\big( \langle \psi_{i}^{(0)} | v_{\textrm{ext}}^{(1)} | \psi_{j}^{(0)} \rangle
+ \langle \psi_{i}^{(1)} | H^{(0)} | \psi_{j}^{(0)} \rangle 
+ \langle \psi_{i}^{(0)} | H^{(0)} | \psi_{j}^{(1)} \rangle \big)  \\
\end{equation}

The contribution to Eq.~(\ref{eq:79bis}) that is quadratic in $\tilde{\rho}_{ij}^{(1)}$ becomes 
\begin{equation}
\label{eq:91}
- \frac{n_\textrm{s}}{2} \sum_{i}^{{pocc}} 
\sum_{j (\neq i)}^{{pocc}} \frac{\epsilon_{i}^{(0)}-\epsilon_{j}^{(0)}}{\occf_{i}^{(0)} - \occf_{j}^{(0)}}  \frac{\tilde{\rho}_{ij}^{(1)} + \tilde{\rho}_{ji}^{*(1)}}{2} \frac{\tilde{\rho}_{ij}^{*(1)} + \tilde{\rho}_{ji}^{(1)}}{2} 
\end{equation}

In this way, $\tilde{\rho}_{ij}^{(1)}$ and $\tilde{\rho}_{ji}^{(1)}$ can be varied independently of each other.
Note that some terms in Eq.~(34) in the main text, like the 
last one, 
involve only diagonal elements of $\rho$, that must be real.
Full freedom of $\tilde{\rho}$, implies the possibility to have complex diagonal elements, that will nevertheless induce real diagonal elements of $\rho$ through Eq.~(\ref{eq:88}).

\section{Gradients of the second-order free energy}
\label{sec:gradients}

The expression to be minimized is $F^{+(2)} [ T, \{ \psi_{i}^{(1)} \}, \{ \rho_{ij}^{(1)} \} ]$, Eq.(34) from Sec.~III~A.
In the current section, the gradients of $F^{+(2)} [ T, \{ \psi_{i}^{(1)} \}, \{ \rho_{ij}^{(1)} \} ]$
with respect to its arguments are listed, taking into account
the freedom discussed in the previous section.
The equalities obtained at the minimum, i.e., when the gradient vanishes, are also deduced, as listed in Sec.~III~B.

Let us start with the gradients with respect to occupation matrix elements:
\begin{align}
    \frac{\partial F^{+(2)}}{\partial (\Tilde{\rho}_{ij}^{(1)*})} & = \frac{n_\textrm{s}}{2} \Bigg( \langle \psi_{i}^{(0)} | \hat{v}_{\textrm{ext}}^{(1)} | \psi_{j}^{(0)} \rangle + \langle \psi_{i}^{(1)} | \hat{H}^{(0)} | \psi_{j}^{(0)} \rangle + \langle \psi_{i}^{(0)} |\hat{H}^{(0)} | \psi_{j}^{(1)} \rangle  \bigg) \nonumber \\
    & + \frac{1}{2} \int \int K_\textrm{Hxc}(\textbf{r},\textbf{r'}) \rho^{(1)}(\textbf{r})  \cdot n_\textrm{s} \psi_{i}^{(0)*}(\textbf{r}) \psi_{j}^{(0)}(\textbf{r}) 
    d\textbf{r} d\textbf{r}'
    \nonumber\\
    & - (1 - \delta_{ij}) \frac{n_\textrm{s}}{2}  \frac{\epsilon_{i}^{(0)} - \epsilon_{j}^{(0)}}{\occf_{i}^{(0)} - \occf_{j}^{(0)}}  \frac{\Tilde{\rho}_{ij}^{(1)} + \Tilde{\rho}_{ji}^{(1)}}{2} - \delta_{ij} \frac{1}{2} n_\textrm{s} \rho_{ii}^{(1)} \frac{\partial \epsilon}{\partial f} \Bigg|_{\occf_{i}^{(0)}}  - \frac{1}{2} n_\textrm{s} \mu^{(1)} \delta_{ij}. \label{eq:106e}
\end{align}

\indent Also,

\begin{align}
    \frac{\partial F^{+(2)}}{\partial (\Tilde{\rho}_{ij}^{(1)*})} = \frac{\partial F^{+(2)}}{\partial (\Tilde{\rho}_{ji}^{(1)})}.\label{eq:108}
\end{align}

When considering only the gradients with respect to diagonal occupations, Eq.~(\ref{eq:106e}) simplifies to

\begin{align}
    \frac{\partial F^{+(2)}}{\partial (\Tilde{\rho}_{ii}^{(1)*})} = \frac{\partial F^{+(2)}}{\partial (\Tilde{\rho}_{ii}^{(1)})} = \frac{n_\textrm{s}}{2} \bigg( \langle \psi_{i}^{(0)} | \hat{H}^{(1)} | \psi_{i}^{(0)} \rangle + \langle \psi_{i}^{(1)} | \hat{H}^{(0)} | \psi_{i}^{(0)} \rangle + \langle \psi_{i}^{(0)} |\hat{H}^{(0)} | \psi_{i}^{(1)} \rangle - \rho_{ii}^{(1)} \frac{\partial \epsilon}{\partial f} 
    \Bigg|_{f_{i}^{(0)}}  - \mu^{(1)} \bigg). \label{eq:109b}
\end{align}

Imposing now the constraint Eq.~(40) in the main text on the diagonal occupation gradient, these gradients become

\begin{align}
    \frac{\partial F^{+(2)}}{\partial (\Tilde{\rho}_{ii}^{(1)*})} & = \frac{\partial F^{+(2)}}{\partial (\Tilde{\rho}_{ii}^{(1)})} = \frac{n_\textrm{s}}{2} \bigg(  \epsilon_{i}^{(1)} -   \rho_{ii}^{(1)} \frac{\partial \epsilon}{\partial f} 
    \Bigg|_{f_{i}^{(0)}}  - \mu^{(1)}  \bigg), \label{eq:110}
\end{align}

\indent with 

\begin{align}
    \epsilon_{i}^{(1)} = \langle \psi_{i}^{(0)} | \hat{H}^{(1)} | \psi_{i}^{(0)} \rangle. \label{eq:110bis}
\end{align}

At the minimum of the variational expression, the gradient vanishes, and so one obtains

\begin{align}
    \rho_{ii}^{(1)} = \frac{\partial f}{\partial \epsilon} \Bigg|_{\epsilon_{i}^{(0)} - \mu^{(0)}} \big( \epsilon_{i}^{(1)} - \mu^{(1)} \big) \label{eq:111-appendix}.
\end{align}

The $\mu^{(1)}$ can be determined from the condition Eq.~(\ref{eq:84}):

\begin{align}
    \mu^{(1)} = \Bigg( \sum_{i} \frac{\partial f}{\partial \epsilon} \Bigg|_{\epsilon_{i}^{(0)} - \mu^{(0)}} \epsilon_{i}^{(1)} \Bigg) / \Bigg( \sum_{i} \frac{\partial f}{\partial \epsilon} \Bigg|_{\epsilon_{i}^{(0)} - \mu^{(0)}} \Bigg). \label{eq:111bis-appendix}
\end{align}

Considering now the off-diagonal occupation gradients (with $i \neq j$), Eq.~(\ref{eq:106e}) simplifies to

\begin{align}
    \frac{\partial F^{+(2)}}{\partial (\Tilde{\rho}_{ij}^{(1)*})} = \frac{n_\textrm{s}}{2} \bigg( \langle \psi_{i}^{(0)} | \hat{H}^{(1)} | \psi_{j}^{(0)} \rangle
     + \epsilon_{j}^{(0)} \langle \psi_{i}^{(1)} | \psi_{j}^{(0)} \rangle + \epsilon_{i}^{(0)} \langle \psi_{i}^{(0)} | \psi_{j}^{(1)} \rangle  -  \frac{\epsilon_{i}^{(0)} - \epsilon_{j}^{(0)}}{\occf_{i}^{(0)} - \occf_{j}^{(0)}} \rho_{ij}^{(1)} \bigg).\label{eq:112b}
\end{align}

Imposing the constraint Eq.~(40) in the main text on the non-diagonal occupation gradient gives

\begin{align}
    \frac{\partial F^{+(2)}}{\partial (\Tilde{\rho}_{ij}^{(1)*})} = \frac{n_\textrm{s}}{2} \bigg( \langle \psi_{i}^{(0)} | \hat{H}^{(1)} | \psi_{j}^{(0)} \rangle + \frac{\epsilon_{j}^{(0)} - \epsilon_{i}^{(0)}}{2}  \bigg( \langle \psi_{i}^{(1)} | \psi_{j}^{(0)} \rangle - \langle \psi_{i}^{(0)} | \psi_{j}^{(1)} \rangle \bigg)  - \frac{\epsilon_{i}^{(0)} - \epsilon_{j}^{(0)}}{\occf_{i}^{(0)} - \occf_{j}^{(0)}} \rho_{ij}^{(1)}  \bigg).\label{eq113b}
\end{align}

At the minimum, the gradient vanishes, and 

\begin{align}
    \rho_{ij}^{(1)} = (\occf_{i}^{(0)} - \occf_{j}^{(0)}) \Bigg( \frac{\langle \psi_{i}^{(0)} | \hat{H}^{(1)} | \psi_{j}^{(0)} \rangle}{\epsilon_{i}^{(0)} - \epsilon_{j}^{(0)}} - \bigg( \langle \psi_{i}^{(1)} | \psi_{j}^{(0)} \rangle - \langle \psi_{i}^{(0)} | \psi_{j}^{(1)} \rangle \bigg) \Bigg). \label{eq:114-appendix}
\end{align}

Note that Eq.~(\ref{eq:114-appendix}) shows that $\rho_{ij}^{(1)}$ is hermitian.

Let us focus now on the
gradients with respect to the $1^{st}$-order wavefunction $\langle \psi^{(1)}_i|$,

\begin{align}
    \frac{\partial F^{+(2)}}{\partial \langle \psi^{(1)}_i|} = n_\textrm{s} \occf_{i}^{(0)} \bigg[ (\hat{H}^{(0)} - \epsilon_{i}^{(0)}) | \psi_{i}^{(1)} \rangle + \hat{H}^{(1)} | \psi^{(0)}_i \rangle \bigg] + n_\textrm{s} \sum_{j}^{{pocc}} \rho_{ji}^{(1)} \epsilon_{j}^{(0)} | \psi_{j}^{(0)} \rangle -n_\textrm{s} \sum_{j}^{{pocc}} \Lambda_{ji}^{(1)} |\psi_{j}^{(0)} \rangle. \label{eq:115c}
\end{align}

At the minimum, the projected gradient on $\langle \psi_{k}^{(0)}|$ in the S$_{pocc}$ space must vanish:

\begin{align}
0 = \Bigg\langle \psi_{k}^{(0)} \Bigg| \frac{\partial F^{+(2)}}{\partial \langle \psi^{(1)}_i|} \Bigg\rangle = n_\textrm{s} \occf_{i}^{(0)} 
\bigg[ (\epsilon_{k}^{(0)} - \epsilon_{i}^{(0)}) 
 \langle \psi_{k}^{(0)} | \psi_{i}^{(1)} \rangle + \langle \psi_{k}^{(0)} | \hat{H}^{(1)} | \psi_{i}^{(0)} \rangle \bigg] + n_\textrm{s} f_{ki}^{(1)} \epsilon_{k}^{(0)} -n_\textrm{s} \Lambda_{ki}^{(1)}. \label{eq:116c}
\end{align}

For the diagonal case, $i$=$k$, one gets

\begin{align}
    0 = n_\textrm{s} \bigg( \occf_{i}^{(0)} \langle \psi_{i}^{(0)} | \hat{H}^{(1)} | \psi_{i}^{(0)} \rangle + \rho_{ii}^{(1)} \epsilon_{i}^{(0)} - \Lambda_{ii}^{(1)} \bigg). \label{eq:117}
\end{align}

This is indeed the derivative of the relation Eq.~(19) in the main text.

Also, the projection out of the potentially occupied space is relevant:

\begin{align}
    \hat{P}_{\perp} \Bigg| \frac{\partial F^{+(2)}}{\partial \langle \psi^{(1)}|} \Bigg\rangle = n_\textrm{s} \occf_{i}^{(0)} \bigg[ \hat{P}_{\perp} (\hat{H}^{(0)} - \epsilon_{i}^{(0)}) \hat{P}_{\perp} | \psi_{i}^{(1)} \rangle + \hat{P}_{\perp} \hat{H}^{(1)} | \psi_{i}^{(0)} \rangle \bigg]. \label{eq:118}
\end{align}

At the minimum one recovers the usual projected Sternheimer equation,

\begin{align}
    \hat{P}_{\perp} (\hat{H}^{(0)} - \epsilon_{i}^{(0)}) \hat{P}_{\perp} | \psi_{i}^{(1)} \rangle = - \hat{P}_{\perp} \hat{H}^{(1)} | \psi_{i}^{(0)} \rangle. \label{eq:118b-appendix}
\end{align}

\section{Covariance of first-order quantities}
\label{sec:covariance}

The parallel gauge choice, $A_{ij}=0$, delivers
the equations mentioned in Sec.~IV~B.
Other choices are possible, and the present section has the goal to analyze the
covariance of $\rho^{(1)}$ and $|\psi_{i}^{(1)} \rangle$.
When $A_{ij}$
does not vanish, one can compare the first-order wavefunctions $|\psi_{i}^{(1)} \rangle$ with the parallel gauge ones $|\psi_{||,i}^{(1)} \rangle$.
They differ only by their content of unperturbed eigenfunctions in the partially occupied space:

\begin{equation}
\label{eq:128}
|\psi_{i}^{(1)} \rangle = |\psi_{||,i}^{(1)} \rangle + \sum_{k}^{{pocc}} c_{ik}^{(1)} |\psi_{k}^{(0)} \rangle.
\end{equation}

One can check that
\begin{equation}
 c_{ij}^{*(1)} + c_{ji}^{(1)} = 0, \label{eq:129a} 
\end{equation}
and
\begin{equation}
c_{ij}^{*(1)} - c_{ji}^{(1)} = A_{ij}. \label{eq:129b}
\end{equation}
Also
\begin{equation}
\label{eq:133}
A_{ij} = -A_{ji}^{*}.
\end{equation}

Due to Eq.~(53) in the main text, taken to be zero at the minimum, there is no change in the projected $\hat{P}_{\perp}|\psi_{i}^{(1)} \rangle$
outside of the potentially occupied space. 
   
The expression Eq.~(38) in the main text, for the 
first-order $\rho^{(1)}(\textbf{r})$, becomes

\begin{align}
\rho^{(1)}(\textbf{r})
& = \rho_{||}^{(1)}(\textbf{r}) + n_\textrm{s} \Bigg[ \sum_{ij}^{{pocc}} 
(\rho_{ji}^{(1)} - \rho_{||,ji}^{(1)}) \psi_{i}^{*(0)}(\textbf{r}) \psi_{j}^{(0)}(\textbf{r}) + \bigg( \sum_{ik}^{{pocc}} \occf_{i}^{(0)} c_{ik}^{*(1)} \psi_{k}^{*(0)}(\textbf{r}) \psi_{i}^{(0)}(\textbf{r}) + \textrm{(c.c.) } \bigg) \Bigg].
\label{eq:130b}
\end{align}
Invariance for each ordered pair $ji$ is obtained when
\begin{equation}
    \label{eq:132}
\rho_{ji}^{(1)} = \rho_{||,ji}^{(1)} - \frac{1}{2} A_{ji} (\occf_{j}^{(0)} - \occf_{i}^{(0)}).
\end{equation}
This describes the covariance between $\rho_{ji}^{(1)}$ and $| \psi_{i}^{(1)} \rangle$, through $A_{ji}$.
From this equation, the diagonal occupation factor does not change,

\begin{equation}
    \label{eq:134}
    \rho_{ii}^{(1)} =  \rho_{||,ii}^{(1)}.
\end{equation}
Finally, one can check that

\begin{equation}
\label{eq:135a}
    F^{+(2)} [T, \{ \psi_{i}^{(1)} \}, \{ \rho_{ij}^{(1)} \}] = F^{+(2)} [T, \{ \psi_{||,i}^{(1)} \}, \{ \rho_{||,ij}^{(1)} \}].
\end{equation}

\section{The first-order density obtained from the modified first-order wavefunctions}
\label{sec:mod_check}

In this section, it is
checked that
the condition expressed by Eq.~(74) in the main text insures that the computation of Eq.~(72) in the main text delivers the correct $\rho^{(1)}$, equal to the one obtained in the parallel gauge.

Indeed, combining Eqs.~(72-74) of the main text, one obtains
\begin{align}
& \rho^{(1)}(\textbf{r}) = n_\textrm{s} \Bigg[ \sum_{i}^{{pocc}} \occf_{i}^{(0)} \big( \psi_{||,i}^{*(1)}(\textbf{r}) \psi_{i}^{(0)}(\textbf{r}) + \psi_{i}^{*(0)}(\textbf{r}) \psi_{||,i}^{(1)}(\textbf{r}) \big) \nonumber \\
& + \sum_{ij}^{{pocc}} \Theta(\occf_{i}^{(0)},\occf_{j}^{(0)}) \frac{\occf_{j}^{(0)} - \occf_{i}^{(0)}}{\epsilon_{j}^{(0)} - \epsilon_{i}^{(0)}} \bigg( \langle \psi_{i}^{(0)} | \hat{H}^{(1)} | \psi_{j}^{(0)} \rangle  \psi_{j}^{*(0)}(\textbf{r}) \psi_{i}^{(0)}(\textbf{r}) + \langle \psi_{j}^{(0)} | \hat{H}^{(1)} | \psi_{i}^{(0)} \rangle \psi_{i}^{*(0)}(\textbf{r}) \psi_{j}^{(0)}(\textbf{r}) \bigg) \Bigg]. \label{eq:158}
\end{align}

Exchange of the $i \leftrightarrow j$ indices in the second line of this equation, combined with Eq.~(75) and Eq.~(63) in the main text gives 

\begin{align}
   \sum_{ij}^{{pocc}} \frac{\occf_{j}^{(0)} - \occf_{i}^{(0)}}{\epsilon_{j}^{(0)} - \epsilon_{i}^{(0)}} & \langle \psi_{j}^{(0)} | \hat{H}^{(1)} | \psi_{i}^{(0)} \rangle \psi_{i}^{*(0)}(\textbf{r}) \psi_{j}^{(0)}(\textbf{r}) \big( \Theta(\occf_{j}^{(0)},\occf_{i}^{(0)}) + \Theta(\occf_{i}^{(0)},\occf_{j}^{(0)}) \big)  =
   \nonumber\\
   &\sum_{ij}^{{pocc}} \frac{\occf_{j}^{(0)} - \occf_{i}^{(0)}}{\epsilon_{j}^{(0)} - \epsilon_{i}^{(0)}} \langle \psi_{j}^{(0)} | \hat{H}^{(1)} | \psi_{i}^{(0)} \rangle \psi_{i}^{*(0)}(\textbf{r}) \psi_{j}^{(0)}(\textbf{r}) = n_\textrm{s} \sum_{ij}^{{pocc}} \rho_{||,ji}^{(1)} \psi_{i}^{*(0)}(\textbf{r}) \psi_{j}^{(0)}(\textbf{r}). \label{eq:161} 
\end{align}

This finally yields the expected 

\begin{align}
\rho^{(1)}(\textbf{r}) & = n_\textrm{s} \Bigg[ \sum_{ij}^{{pocc}} \rho_{||,ji}^{(1)} \psi_{i}^{*(0)}(\textbf{r}) \psi_{j}^{(0)}(\textbf{r}) + \sum_{i}^{{pocc}} \occf_{i}^{(0)} \bigg( \psi_{||,i}^{*(1)}(\textbf{r}) \psi_{i}^{(0)}(\textbf{r}) + \psi_{i}^{*(0)}(\textbf{r}) \psi_{||,i}^{(1)}(\textbf{r}) \bigg) \Bigg], \label{eq:162}
\end{align}

\indent that is, the formula for $\rho^{(1)}(\textbf{r})$ in the parallel gauge.

\section{Non-variational expressions}
\label{sec:non-var}
As in the DFPT for gapped systems, simpler expressions for $F^{+(2)}$ can be obtained, although for such expressions the variational property is lost.
Let us examine some parts of Eq.~(59) in the main text. With the use Eq.~(53) in the main text,

\begin{align}
\langle \psi_{||,i}^{(1)} | \hat{H}^{(0)} - \epsilon_{i}^{(0)} | \psi_{||,i}^{(1)} \rangle + \frac{1}{2} \bigg( \langle \psi_{||,i}^{(1)} | \hat{v}_{\textrm{ext}}^{(1)} | \psi_{i}^{(0)} \rangle + (\textrm{c.c.}) \bigg) & = - \langle \psi_{||,i}^{(1)} | \hat{H}^{(1)} | \psi_{i}^{(0)} \rangle + \frac{1}{2} \bigg( \langle \psi_{||,i}^{(1)} | \hat{v}_{\textrm{ext}}^{(1)} | \psi_{i}^{(0)} \rangle + \langle \psi_{i}^{(0)} | \hat{v}_{\textrm{ext}}^{(1)} | \psi_{||,i}^{(1)} \rangle \bigg) \nonumber \\
& = -\frac{1}{2} \bigg( \langle \psi_{||,i}^{(1)} | \hat{v}_\textrm{Hxc}^{(1)} | \psi_{i}^{(0)} \rangle + (\textrm{c.c.}) \bigg). \label{eq:167}
\end{align}

On the other hand, the second part of Eq.~(60) in the main text can be modified:

\begin{align}
\frac{1}{2} \int \int K_\textrm{Hxc}(\textbf{r},\textbf{r'}) \rho^{(1)}(\textbf{r}) \rho^{(1)}(\textbf{r'}) d\textbf{r} d\textbf{r'} 
 &= \frac{1}{2} \int \hat{v}_\textrm{Hxc}^{(1)}(\textbf{r}) \rho^{(1)}(\textbf{r}) d\textbf{r} \nonumber \\
 = \frac{1}{2} \int \hat{v}_\textrm{Hxc}^{(1)}(\textbf{r}) & \Bigg\{ n_\textrm{s} \bigg[ \sum_{i}^{{pocc}} f_{i}^{(0)} \bigg( \psi_{||,i}^{*(1)}(\textbf{r}) \psi_{i}^{(0)}(\textbf{r}) + (\textrm{c.c.}) \bigg) +\sum_{ij}^{{pocc}} \rho_{||,ji}^{(1)} \psi_{i}^{*(0)}(\textbf{r}) \psi_{j}^{(0)}(\textbf{r})\bigg] \Bigg\} d\textbf{r}. \label{eq:168}
\end{align}

This delivers a non-variational expression for $F^{(2)}$:

\begin{align}
F_{\textrm{non-var}}^{(2)} & = n_\textrm{s} \sum_{i}^{{pocc}} f_{i}^{(0)} \bigg[  \langle \psi_{i}^{(0)} | \hat{v}_{\textrm{ext}}^{(2)} | \psi_{i}^{(0)} \rangle + \frac{1}{2} \big( \langle \psi_{||,i}^{(1)} | \hat{v}_{\textrm{ext}}^{(1)} | \psi_{i}^{(0)} \rangle + (\textrm{c.c.}) \big) \bigg] +n_\textrm{s} \sum_{ij}^{{pocc}} \rho_{||,ji}^{(1)} \frac{1}{2} \langle \psi_{i}^{(0)} | \hat{v}_{\textrm{ext}}^{(1)} | \psi_{j}^{(0)} \rangle \nonumber \\
& - \frac{n_\textrm{s}}{2} \sum_{i \neq j}^{{pocc}} \frac{\epsilon_{i}^{(0)} - \epsilon_{j}^{(0)}}{f_{i}^{(0)} - f_{j}^{(0)}} |\rho_{||,ij}^{(1)}|^{2} - \frac{n_\textrm{s}}{2} \sum_{i}^{{pocc}} \frac{\partial \epsilon}{\partial f} \Bigg|_{f_{i}^{(0)}} (\rho_{||,ii}^{(1)})^{2} 
-n_\textrm{s} \mu^{(1)} \sum_{i}^{{pocc}} \rho_{||,ii}^{(1)}. \label{eq:169e}
\end{align}

The last term can be suppressed, due to the constraint Eq.~(41) in the main text.
Also Eq.~(63) in the main text can be used, as well as Eq.~(48) in the main text, which yields

\begin{align}
F_\textrm{non-var}^{(2)} & = n_\textrm{s} \sum_{i}^{{pocc}} f_{i}^{(0)} \bigg[  \langle \psi_{i}^{(0)} | \hat{v}_{\textrm{ext}}^{(2)} | \psi_{i}^{(0)} \rangle + \frac{1}{2} \big( \langle \psi_{||,i}^{(1)} | \hat{v}_{\textrm{ext}}^{(1)} | \psi_{i}^{(0)} \rangle + (\textrm{c.c.}) \big) \bigg] + n_\textrm{s} \sum_{ij}^{{pocc}} \rho_{||,ji}^{(1)} \frac{1}{2} \langle \psi_{i}^{(0)} | \hat{v}_{\textrm{ext}}^{(1)} | \psi_{j}^{(0)} \rangle \nonumber \\
& - \frac{n_\textrm{s}}{2} \sum_{i \neq j}^{{pocc}} \frac{f_{i}^{(0)} - f_{j}^{(0)}}{\epsilon_{i}^{(0)} - \epsilon_{j}^{(0)}} |\langle \psi_{i}^{(0)}| \hat{H}^{(1)} | \psi_{j}^{(0)} \rangle |^{2} - \frac{n_\textrm{s}}{2} \sum_{i}^{{pocc}} \frac{\partial f}{\partial \epsilon} \Bigg|_{\epsilon_{i}^{(0)} - \mu^{(0)}} \big( \epsilon_{i}^{(1)} - \mu^{(1)} \big)^{2}, \label{eq:170d}
\end{align}

\indent where $\mu^{(1)}$ is determined from Eq.~(65) in the main text.
Finally, the non-variational expression can be written in terms of $\psi_{\textrm{mod}}^{(1)}$, see Sec.~IV~D:

\begin{align}
F_{\textrm{non-var}}^{(2)} & = n_\textrm{s} \sum_{i}^{{pocc}} f_{i}^{(0)} \bigg[ \frac{1}{2} \big( \langle \psi_{\textrm{mod},i}^{(1)} | \hat{v}_{\textrm{ext}}^{(1)} | \psi_{i}^{(0)} \rangle + (\textrm{c.c.}) \big) \langle \psi_{i}^{(0)} | \hat{v}_{\textrm{ext}}^{(2)} | \psi_{i}^{(0)} \rangle \bigg] \nonumber \\
& - \frac{n_\textrm{s}}{2} \Bigg[ \sum_{i \neq j}^{{pocc}} \frac{f_{i}^{(0)} - f_{j}^{(0)}}{\epsilon_{i}^{(0)} - \epsilon_{j}^{(0)}} |\langle \psi_{i}^{(0)}| \hat{H}^{(1)} | \psi_{j}^{(0)} \rangle |^{2} + \sum_{i}^{{pocc}} \frac{\partial f}{\partial \epsilon} \Bigg|_{\epsilon_{i}^{(0)} - \mu^{(0)}} \big( \epsilon_{i}^{(1)} - \mu^{(1)} \big)^{2} \Bigg]. \label{eq:171d}
\end{align}

\section{Temperature-dependent phonon frequencies}
\label{sec:Tdep_phonon_SI}

This section provides additional figures related to the Sec. VI of the main text.

In particular, data are presented as a function of the electronic
physical temperature, as well as for a 
42$\times$42$\times$42 wavevector grid, finer
than the 30$\times$30$\times$30  wavevector grid of the main text.

\begin{figure}[H]
	\includegraphics[width=0.45\textwidth]{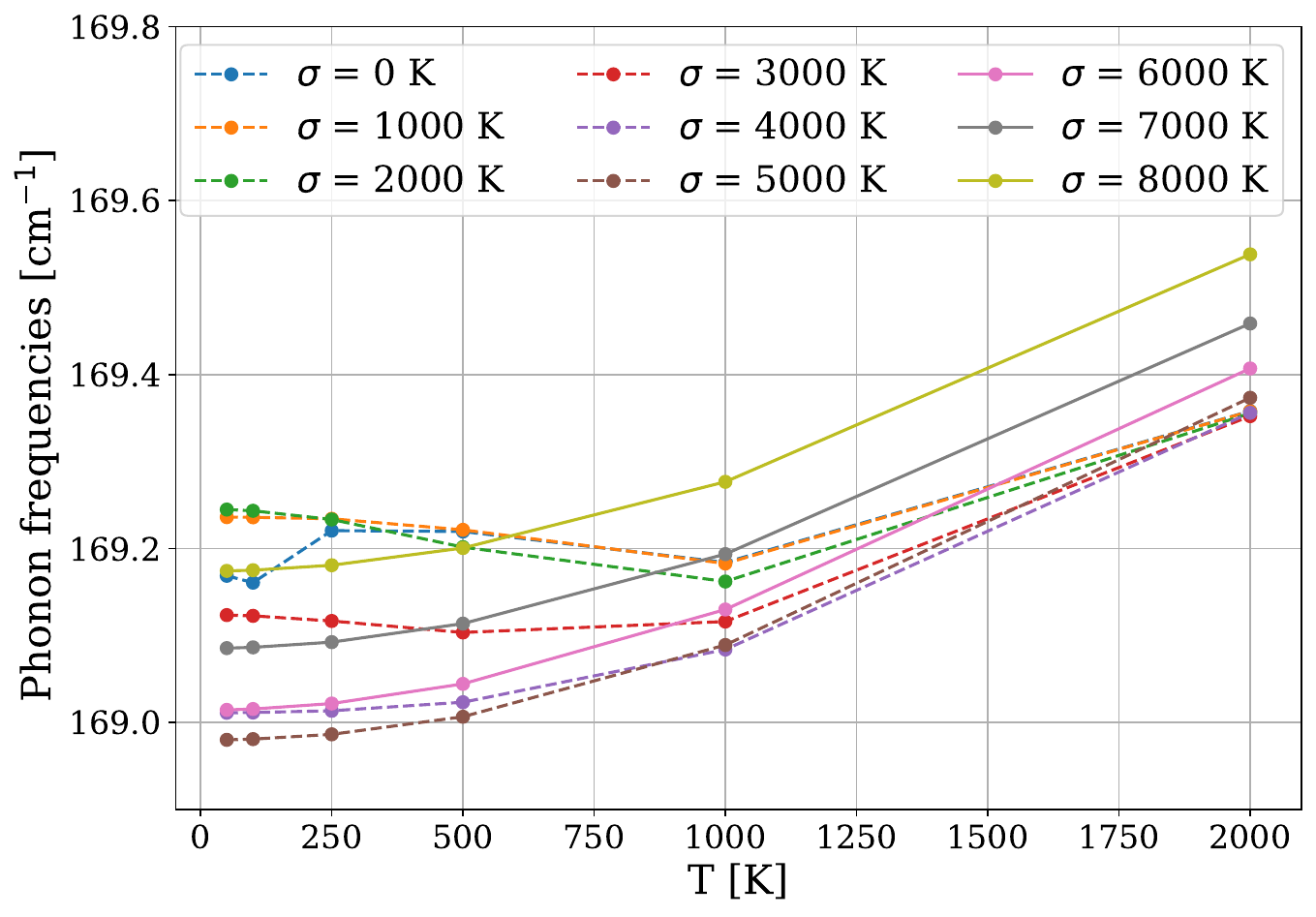} ~~\includegraphics[width=0.45\textwidth]{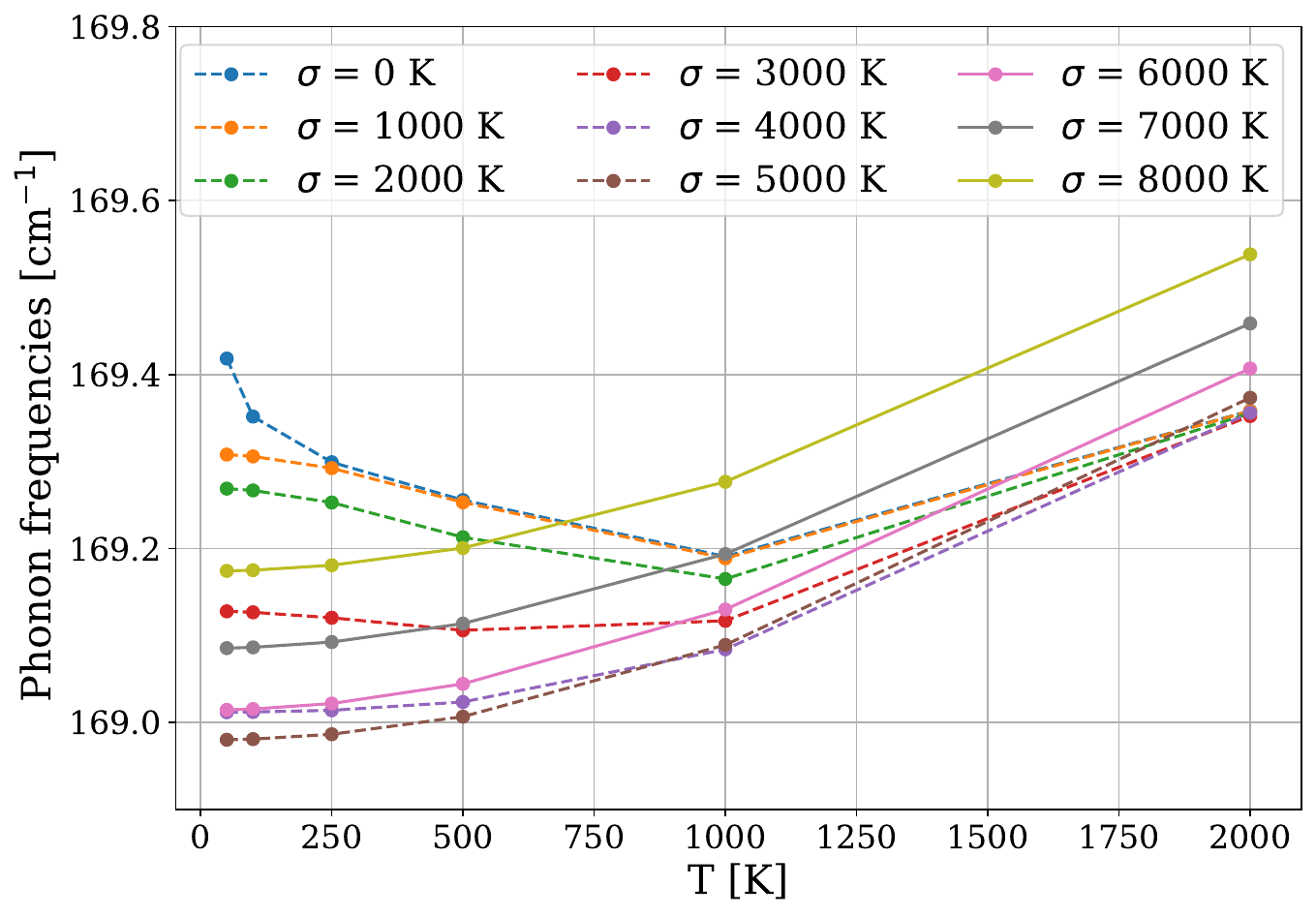}
	\caption{Transverse phonon frequencies as a function of electronic physical temperature in the range from 50 K to 2000 K, for various MP broadening values between 0 and 8000 K, with a 30$\times$30$\times$30 grid on the left and a 42$\times$42$\times$42 grid on the right. }
\end{figure}

\begin{figure}[H]
	\includegraphics[width=0.45\textwidth]{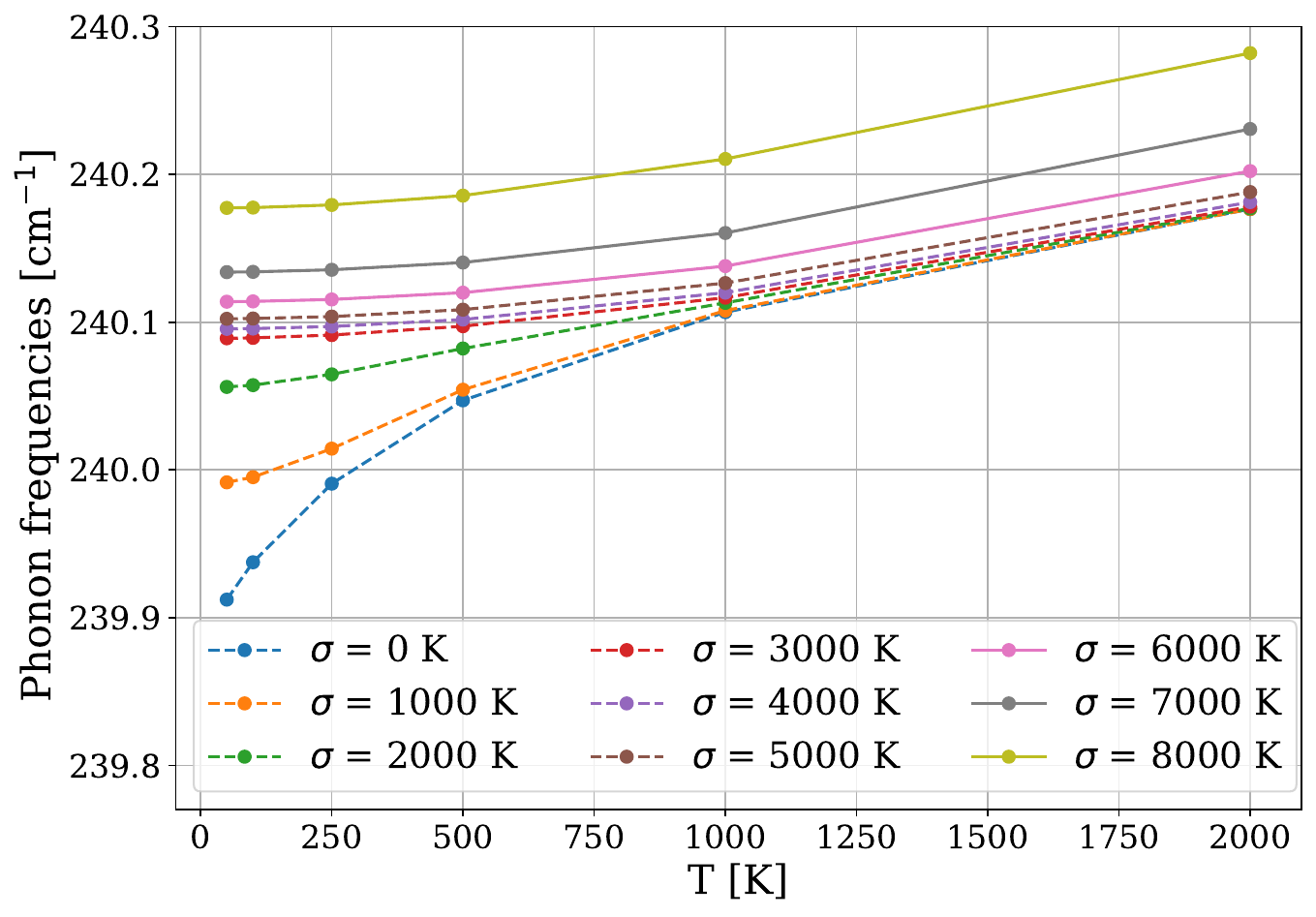} ~~\includegraphics[width=0.45\textwidth]{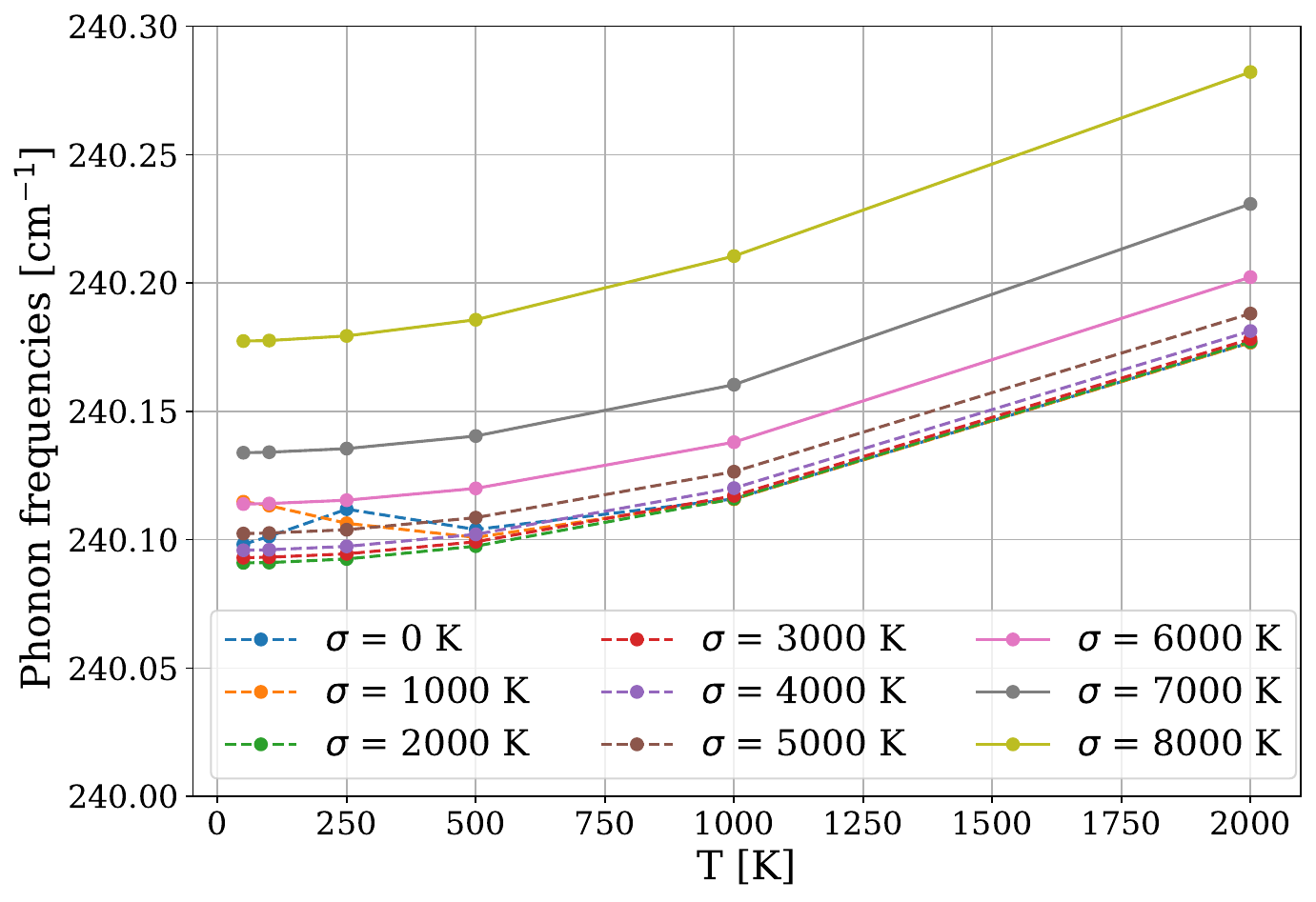}
	\caption{Longitudinal phonon frequencies as a function of electronic physical temperature in the range from 50 K to 2000 K, for various MP broadening values between 0 and 8000 K, with a 30$\times$30$\times$30 grid on the left and a 42$\times$42$\times$42 grid on the right. }
\end{figure}

\section{The effect of underconverged unperturbed wavefunctions}
\label{sec:underconvergedSI}

In this subsection, intermediate steps in the treatment of the three-level model, Sec.~VII of the main text, are provided.

The approximate eigenenergies are

\begin{align}
\tilde{\epsilon}_{2}^{(0)} & = \langle \tilde{\psi}_{2}^{(0)} | \hat{H}^{(0)} | \tilde{\psi}_{2}^{(0)} \rangle = \cos^{2}{\alpha} \cdot \epsilon_{2}^{(0)} + \sin^{2}{\alpha} \cdot \epsilon_{3}^{(0)} = \epsilon_{2}^{(0)} + \sin^{2}{\alpha} \cdot (\epsilon_{3}^{(0)} - \epsilon_{2}^{(0)}), \label{eq:195bSI}
\end{align}
and
\begin{align}
\tilde{\epsilon}_{3}^{(0)} & =  \langle \tilde{\psi}_{3}^{(0)} | \hat{H}^{(0)} | \tilde{\psi}_{3}^{(0)} \rangle = \sin^{2}{\alpha} \cdot \epsilon_{2}^{(0)} + \cos^{2}{\alpha} \cdot \epsilon_{3}^{(0)} = \epsilon_{3}^{(0)} + \sin^{2}{\alpha} \cdot (\epsilon_{2}^{(0)} - \epsilon_{3}^{(0)}). \label{eq:199SI}
\end{align}
For the second approximate eigenvector, the residual is
\begin{align}
| R_{2} \rangle  = (\hat{H}^{(0)} - \tilde{\epsilon}_{2}^{(0)}) |\tilde{\psi}_{2}^{(0)}= (\epsilon_{2}^{(0)} - \tilde{\epsilon}_{2}^{(0)}) \cos{\alpha} |\psi_{2}^{(0)} \rangle + (\epsilon_{3}^{(0)} - \tilde{\epsilon}_{2}^{(0)}) \sin{\alpha} |\psi_{3}^{(0)} \rangle, \label{eq:196SI}
\end{align}
with squared norm $R^{2}$:
\begin{align}
\langle R_{2} | R_{2} \rangle  & = \cos^{2}{\alpha} \cdot (\epsilon_{2}^{(0)} - \tilde{\epsilon}_{2}^{(0)})^{2} + \sin^{2}{\alpha} \cdot (\epsilon_{3}^{(0)} - \tilde{\epsilon}_{2}^{(0)})^{2} 
\nonumber \\
& = \cos^{2}{\alpha} \cdot \sin^{4}{\alpha} (\epsilon_{3}^{(0)} - \epsilon_{2}^{(0)})^{2} + \sin^{2}{\alpha} \cdot \cos^{4}{\alpha} (\epsilon_{3}^{(0)} - \epsilon_{2}^{(0)})^{2} = \cos^{2}{\alpha} \cdot \sin^{2}{\alpha} (\epsilon_{3}^{(0)} - \epsilon_{2}^{(0)})^{2}. \label{eq:197SI}
\end{align}
Hence, Eq.~(101) in the main text.

Concerning the approximate second-order derivative of the energy, one starts from Eq.~(102) in the main text, with actual eigenenergies and eigenfunctions for the case of the three-level system:
\begin{align}
     \tilde{E}^{(2)} = \frac{n_\textrm{s}}{2} \Bigg[ \frac{f_{2}^{(0)} - f_{1}^{(0)}}{\tilde{\epsilon}_{2}^{(0)} - \epsilon_{1}^{(0)}} |\langle \tilde{\psi}_{2}^{(0)}| \hat{H} | \psi_{1}^{(0)} \rangle |^{2} + \frac{f_{3}^{(0)} - f_{1}^{(0)}}{\tilde{\epsilon}_{3}^{(0)} - \epsilon_{1}^{(0)}} |\langle \tilde{\psi}_{3}^{(0)}| \hat{H} | \psi_{1}^{(0)} \rangle |^{2}   + \frac{f_{3}^{(0)} - f_{2}^{(0)}}{\tilde{\epsilon}_{3}^{(0)} - \tilde{\epsilon}_{2}^{(0)}} |\langle \tilde{\psi}_{3}^{(0)}| \hat{H} | \psi_{2}^{(0)} \rangle |^{2}  \Bigg].
\end{align}
This gives
\begin{align}
\tilde{E}^{(2)} =
     \frac{n_\textrm{s}}{2} \Bigg[ & \frac{\delta f - (1 - \delta f)}{\epsilon_{2}^{(0)} + \sin^{2}{\alpha} (\epsilon_{3}^{(0)} - \epsilon_{2}^{(0)}) - \epsilon_{1}^{(0)}} \cdot |\cos{\alpha} H_{12} + \sin{\alpha} H_{13}|^{2} \nonumber \\
    & - \frac{1 - \delta f}{\epsilon_{3}^{(0)} + \sin^{2}{\alpha} (\epsilon_{2}^{(0)} - \epsilon_{3}^{(0)}) - \epsilon_{1}^{(0)}} |\cos{\alpha} H_{13} - \sin{\alpha} H_{12}|^{2} \nonumber \\
    & - \frac{\delta f}{\epsilon_{3}^{(0)} + \sin^{2}{\alpha} (\epsilon_{2}^{(0)} - \epsilon_{3}^{(0)}) - (\epsilon_{2}^{(0)} + \sin^{2}{\alpha} (\epsilon_{3}^{(0)} - \epsilon_{2}^{(0)})} \cdot |\cos^{2}{\alpha} H_{32} - \sin^{2}{\alpha} H_{23}|^{2} \Bigg]. \label{eq:200SI}
\end{align}

A Taylor expansion of $\tilde{E}^{(2)}$ in term of small parameter $\sin{\alpha}$ is now performed, and the quadratic contributions are discarded.
This means $\cos{\alpha} \simeq 1$ and $\cos^{2}{\alpha} \simeq 1$.

\begin{align}
     \tilde{E}^{(2)} = - n_\textrm{s} \Bigg[ & \frac{1 - 2 \delta f}{\epsilon_{2}^{(0)} -\epsilon_{1}^{(0)}} (|H_{12}|^{2} + 2 \sin{\alpha} \Re(H_{12}^{*} H_{13}) ) - \frac{1 - \delta f}{\epsilon_{3}^{(0)} - \epsilon_{1}^{(0)}} (|H_{13}|^{2} - 2 \sin{\alpha} \Re(H_{12}^{*} H_{13}) ) \nonumber \\
     & - \frac{\delta f}{\epsilon_{3}^{(0)} - \epsilon_{2}^{(0)}} |H_{23}|^{2} \Bigg] + O(\sin^{2}{\alpha}) \label{eq:201SI}
\end{align}

The difference between the contaminated $\tilde{E}^{(2)}$ and $E^{(2)}$ is thus Eq.~(103) in the main text.
\\
\end{widetext}

\bibliography{main}